%% file: costa2018.tex
\documentclass[preprint]{aastex}
\usepackage{natbib}
\usepackage{multirow}
\usepackage[caption=false]{subfig}
\include{defs}
\usepackage{xcolor}
\usepackage{graphicx}
\usepackage{amsmath}
\usepackage{ mathrsfs }
\usepackage{ulem}

\newcommand{\pdr}{Photodissociation Region (PDR)}
\usepackage{threeparttable}

\begin{document}
\title{A Faraday Rotation Study of the Stellar Bubble and \HII Region Associated with the W4 Complex}
\author{Allison H. Costa\altaffilmark{1} and Steven R. Spangler\altaffilmark{1}}
\altaffiltext{1}{Department of Physics and Astronomy, University of Iowa, Iowa City, Iowa 52242}

\begin{abstract} 

We utilized the Very Large Array to make multifrequency polarization measurements of 20 radio sources viewed through the IC 1805 HII region and ``Superbubble''.  The measurements at frequencies between 4.33 and 7.76 GHz yield Faraday rotation measures (RMs) along 27 lines of sight to these sources. The RMs are used to probe the plasma structure of the IC 1805 HII region and to test the degree to which the Galactic magnetic field is heavily modified (amplified) by the dynamics of the HII region.  We find that IC 1805 constitutes a ``Faraday rotation anomaly'', or a region of increased RM relative to the general Galactic background value. The $|$RM$|$ due to the nebula is commonly 600 -- 800 rad m$^{-2}$. However, the observed RMs are not as large as predicted by simplified analytic models that include substantial amplification of the Galactic magnetic field within the shell. The magnitudes of the observed RMs are consistent with shells in which the Galactic field is unmodified, or increased by a modest factor, such as due to magnetic flux conservation. We also find that with one exception, the sign of the RM is that expected for the polarity of the Galactic field in this direction.  Finally, our results show intriguing indications that some of the largest values of $|$RM$|$ occur for lines of sight that pass outside the fully ionized shell of the IC 1805 HII region but pass through the Photodissociation Region associated with IC 1805.

\end{abstract}

\keywords{ISM: bubbles, ISM: \HII regions, ISM: magnetic fields, plasmas}
\section{Introduction\label{intro}}

Young massive stars in OB associations photoionize the surrounding gas, creating an \HII region, and their powerful stellar winds can inflate a bubble around the star cluster. Magnetic fields are important to the dynamics of these structures  \citep{Tomisaka:1990,Ferriere:1991,Vallee:1993,Tomisaka:1998,Haverkorn:2004,Sun:2008,Stil:2009}, and they can elongate the cavity preferentially in the direction of the magnetic field and thicken the shell perpendicular to the field \citep{Ferriere:1991,deAvillez:2005,Stil:2009}, causing deviations from the classical structure of the \citet{Weaver:1977} wind-blown bubble. Knowledge of the magnitude and direction of the magnetic field within stellar bubbles and \HII regions is important for simulations and for understanding how the magnetic field interacts with and modifies these structures.

In previous work (i.e., \citealt{Savage:2013} and \citealt{Costa:2016}), we investigated whether the Galactic magnetic field is amplified in the shell of the Rosette Nebula, an \HII region and stellar bubble associated with NGC 2244 ($\ell$ = 206.5\ddeg, $b$ = --2.1\ddeg). Other similar work investigating magnetic fields near massive star clusters has been done by \citet{Harvey:2011} and \citet{Purcell:2015}. In this work, we continue our investigation of how \HII regions and stellar bubbles modify the ambient Galactic magnetic field by considering another example of a young star cluster and an \HII region that appears to be formed into a shell by the effect of stellar winds.
\subsection{Faraday Rotation and Magnetic Fields in the Interstellar Medium}
Faraday rotation measurements probe the line of sight (LOS) component of the magnetic field in ionized parts of the interstellar medium (ISM), provided there is an independent estimate of the electron density. Faraday rotation is the rotation in the plane of polarization of a wave as it passes through magnetized plasma and is described by the equation 

\begin{equation}
\chi=\chi_{0}+\left[\left(\frac{e^{3}}{2\pi m_{e}^{2}c^{4}}\right)\int{n_e\ \mathbf{B}\cdot \textrm{d}\mathbf{s}}\right]\lambda^{2},
\label{eq:rmorg}
\end{equation}
where $\chi$ is the polarization position angle, $\chi_0$ is the intrinsic polarization position angle, the quantities in the parentheses are the usual standard physical constants in cgs units, \textit{n$_{\textrm{e}}$} is the electron density, \textbf{B} is the vector magnetic field, d\textbf{s} is the incremental path length interval along the LOS, and $\lambda$ is the wavelength. We define the terms in the square bracket as the rotation measure, RM, and we can express the RM in mixed but convenient interstellar units as 
 \begin{equation}
\textnormal{RM}=0.81\int
n_{e} \ (\text{cm$^{-3}$}) \ \mathbf{B} \ (\mu\text{G})\cdot \textrm{d}\mathbf{s} \text{ (pc)  rad m$^{-2}$.}
\label{eq:rmprat}
 \end{equation}


\subsection{The \HII Region and Stellar Bubble Associated with the W4 Complex\label{sec:structure}}

The \HII region and stellar bubble of interest for the present study is IC 1805, which is located in the Perseus Arm. The star cluster responsible for the \HII region and stellar bubble is OCl 352, which is a young cluster (1--3 Myr) \citep{Basu:1999}. OCl 352 has 60 OB stars \citep{Shi:1999}. Three of these are the O stars HD 15570, HD 15558, and HD 15629, and they have mass loss rates between 10$^{-6}$ and 10$^{-5}$ \MLu~\citep{Massey:1995} and terminal wind velocities of 2200 -- 3000 \kms~\citep{Garmany:1988,Groenewegen:1989,Bouret:2012}. We adopt the nominal center of the star cluster to be R.A.(J2000) = 02$^h$ 23$^m$ 42$^s$, decl.(J2000) = +61\ddeg 27$'$ 0$''$ ($\ell$ = 134.73, \textit{b} = +0.92) \citep{Guetter:1989} and a distance of 2.2 kpc to IC 1805 to conform with previous studies of the region (e.g., \citealt{Normandeau:1996,Dennison:1997,Reynolds:2001,Terebey:2003,Gao:2015}). In the literature, other distance values include: 2.35 kpc \citep{Massey:1995,Basu:1999,West:2007,Lagrois:2012}, 2 kpc \citep{Dickel:1980}, 2.04 kpc \citep{Feigelson:2013,Townsley:2014}, and 2.4 $\pm$ 0.1 kpc \citep{Guetter:1989}. 

\begin{figure}[htb!]
\centering
\includegraphics[width = 0.6\textwidth]{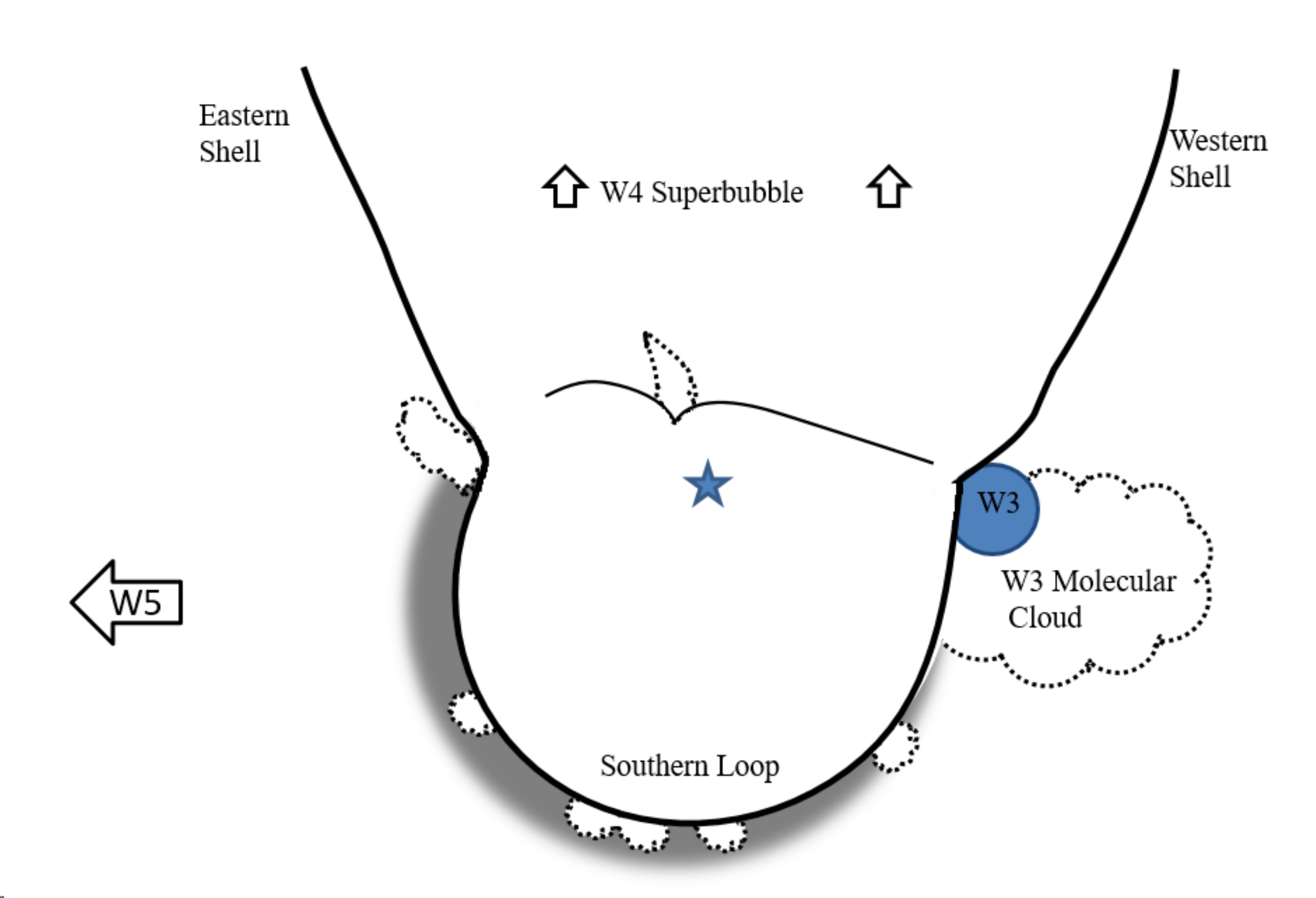}
\caption[]{Cartoon of structure of IC 1805 and the surrounding region, which includes the W3 molecular cloud, W3 (blue filled circle), and the W4 Superbubble. The solid black lines represent the bright ionized shell in Figure \ref{fig:w4}, the dotted lines show molecular material from \citet{Lagrois:2009}, and the gray shading represents the extended halo or PDR. The star represents the center of the exciting star cluster, OCl 352.}
\label{fig:cartoon}
\end{figure}

We refer to the \HII region between --0.2\ddeg~$<$ \textit{b} $<$ 2\ddeg~as IC 1805. This structure is also known as the Heart Nebula for its appearance at optical wavelengths. We differentiate this region from the northern latitudes that constitute the W4 Superbubble \citep{Normandeau:1996,West:2007,Gao:2015}, and we use the nomenclature of W4 to describe the entire region, which includes IC 1805 and the W4 Superbubble. Below we summarize the structure of IC 1805 and Figure \ref{fig:cartoon} is a cartoon diagram of the structure described here.
 
\begin{itemize}

\item[--] \textit{South.} On the southern portion of IC 1805, there is a loop structure of ionized material at 134\ddeg $<$  $\ell$ $<$ 136\ddeg, \textit{b} $<$ 1\ddeg, which we call the southern loop. \citet{Terebey:2003} find that at far infrared and radio wavelengths, the shell structure is well defined and ionization bounded, since the ionized gas lies interior to the dust shell. However, they also find that there is warm dust that extends past the southern loop and a faint ionized halo (see their Figure 6). \citet{Terebey:2003} argue that the shell is patchy and inhomogeneous in density, which allows ionizing photons to escape. \citet{Gray:1999} discuss extended emission surrounding IC 1805 and suggest that it may be evidence of an extended \HII region \citep{Anantharamaiah:1985}. Also surrounding IC 1805 are patchy regions of \hi \citep{Braunsfurth:1983,Hasegawa:1983,Sato:1990} and CO \citep{Heyer:1998,Lagrois:2009}. 

\citet{Terebey:2003} model the structure of the southern loop using radio continuum data. They assume a spherical shell and place OCl 352 at the top edge of the bubble instead of at the center to accommodate spherical symmetry (see their Figures 4 and 5). The center of their shell model is at ($\ell$, $b$) = (135.02\ddeg, 0.42\ddeg). They find an inner radius of 30 arcmin (19 pc)  and a shell thickness of 10 arcmin (6 pc) and 2.5 arcmin (2 pc) for a thick and thin shell model, respectively. \citet{Terebey:2003} report electron densities of 10 \cm~and 20 \cm~for the thick and thin shell models, respectively (see Section 3.5 and Table 3 of \citealt{Terebey:2003}). While we utilize and discuss these models in the following sections, the center position of the shell in \citet{Terebey:2003} was selected to fit the ionized shell, and as such, the shell parameters should only be used to describe the bottom of IC 1805. For latitudes near the star cluster, the model fails, as the star cluster is at the top edge of the bubble instead of at the center.

\item[--] \textit{East.} On the eastern edge of IC 1805 ($\ell$ $>$ 134.6\ddeg, \textit{b} $<$ 0.9\ddeg), \citet{Terebey:2003} find that warm dust extends outside the loop boundary and suggest that if the warm dust is associated with the ionized gas, then the bubble has blown out on the eastern side of IC 1805. 
At the Galactic latitude equal to the star cluster, the ionized gas appears to be pinched \citep{Basu:1999}, which is usually caused by higher densities. There is a clump of CO emission in the vicinity of the eastern pinch at ($\ell$, $b$) = (135.2\ddeg, 1.0\ddeg) \citep{Lagrois:2009}, and there is \hi emission on the eastern edge at ($\ell$, $b$) $\geq$ (136\ddeg, 0.5\ddeg) (see Figure 1 of \citealt{Sato:1990}).

\item[--] \textit{West.} On the western edge of IC 1805 is the W3 molecular cloud and the W3 complex, which hosts a number of compact \HII regions and young stellar objects (see \citealt{Bik:2012} and their Figure 1). \citet{Dickel:1980} modeled the structure of W3, which is thought to be slightly in front of W4, and they argue that the advancement of the IC 1805 ionization front and shock front into the W3 molecular cloud may have triggered star formation. \citet{Moore:2007} similarly conclude that the W3 molecular cloud has been compressed on one side by the expansion of IC 1805. While infrared sources nestled between the western edge of IC 1805 and eastern edge of the W3 molecular cloud are thought to be the product of this interaction , W3 Main, W3 (OH), and W3 North are thought to be sites of triggered star formation from IC 1795, which is part of W3 as well and not from the expansions of the ionization front \citep{Nakano:2017, Jose:2016,Kiminki:2015}. There is therefore uncertainty regarding a physical connection between W3 and IC 1805.

\item[--] \textit{North.} North of OCl 352, the bubble opens up into what is called the W4 Superbubble \citep{Normandeau:1997,Dennison:1997,West:2007,Gao:2015}, which is a sealed ``egg-shaped'' structure that extends up to \textit{b} $\sim$ 7\ddeg~ \citep{Dennison:1997,West:2007}. At the latitude of the star cluster, \citet{Lagrois:2009} estimate the distance between the eastern and western shell to be $\sim$ 1.2\ddeg~(46 pc) that increases in size up to 1.6\ddeg~(61 pc) at \textit{b} = 1.8\ddeg~(see Figure 11 of \citealt{Lagrois:2009}). At higher latitudes, \citet{Dennison:1997} model the thickness of the shell to be between 10--20 pc (16 -- 31 arcminutes) from H$\alpha$ observations.

The ``v''-shaped feature seen in Figure \ref{fig:w4} at ($\ell$, $b$) $\sim$ (134.8\ddeg, 1.35\ddeg) is prominent in the ionized emission, and \citet{Heyer:1996} report a cometary-shaped molecular cloud near ($\ell$, $b$) $\sim$ (134.8\ddeg, 1.35\ddeg). The alignment of the cometary cloud, as it is pointed towards IC 1805, suggests that the UV photons from the star cluster are responsible for the ``v'' shaped feature in the ionized emission on the side closest to the star cluster \citep{Dennison:1997,Taylor:1999}. \citet{Lagrois:2009} argue, from radial velocity measurements, that the cloud is located on the far side of the bubble wall, and while it may appear to be a cap to the bubble connecting to the southern loop, it is simply a projection effect. As such, the ridge of ionized material directly north of OCl 352 is not the outer radius of the shell but is part of the rear bubble wall.

\item [--] \textit{PDR.} The \hi and molecular emission near the southern ($\ell$ $<$ 0.9\ddeg) portions of IC 1805 suggest that a \pdr{} has formed exterior to the \HII region. PDRs are the transition layer between the fully ionized \HII region and molecular material, where far UV photons can propagate out and photodissociate molecules. We discuss the importance and observational evidence of a PDR in Section \ref{sec:pdr}.
\end{itemize}

There is an extensive literature on the W4 region and its relationship to W3, dealing with the morphology \citep{Dickel:1980,Dickel:1980b,Braunsfurth:1983,Normandeau:1996,Dennison:1997,Heyer:1998,Taylor:1999,Basu:1999, Terebey:2003,Lagrois:2009,Lagrois:2009b,Stil:2009} and star formation history \citep{Carpenter:2000,Oey:2005}. In the following paragraphs, we summarize those results from the literature that are most relevant to our polarimetric study and inferences on magnetic fields in this region. 

Measurements of the total intensity and polarization of the Galactic nonthermal emission in the vicinity of \HII regions are of interest because the \HII regions and environs act as a Faraday-rotating screen inserted between the Galactic emission behind the \HII region and that in front. Few radio polarimetric studies exist in the literature to date of the IC 1805 stellar bubble. \citet{Gray:1999} present their polarimetric results of the W3/W4 region at 1420 MHz with the Dominion Radio Astrophysical Observatory (DRAO) Synthesis Telescope. They find zones of strong depolarization near the \HII regions, particularly in the south, where there is a halo of extended emission around IC 1805. 
They conclude that RM values on order 10$^3$ \radm~and spatial RM gradients must exist to explain the depolarization near the \HII region. More recently, \citet{Hill:2017} present results of their polarimetric study of the Fan region ($\ell$ $\sim$ 130\ddeg, --5\ddeg $\leq$ \textit{b} $\leq$ +10\ddeg), which is a large structure in the Perseus arm that includes W3/W4. While the focus of their study was not on W4 specifically, they find similar results to \citet{Gray:1999} in that there is sufficient Faraday rotation to cause beam depolarization in the regions of extended emission.

In the W4 Superbubble, \citet{West:2007} determined the LOS magnetic field strength by estimating                                                                                                                                              depolarization effects along adjacent lines of sight. Using estimates of the shell thickness and the electron density from \citet{Dennison:1997}, \citet{West:2007} estimate \blos~$\sim$ 3.4 -- 9.1 $\mu$G for lines of sight at \textit{b} $>$ 5\ddeg. \citet{Gao:2015} also report \blos~estimates in the W4 Superbubble by assuming a passive Faraday screen model \citep{Sun:2007} and measuring the polarization angle for lines of sight interior and exterior to the screen. For the western shell ($\ell$ $\sim$ 132.5\ddeg, 4\ddeg $<$ \textit{b} $<$ 6\ddeg) and the eastern shell ($\ell$ $\sim$ 136\ddeg, 6\ddeg $<$ \textit{b} $<$ 7.5\ddeg) in the superbubble, \citet{Gao:2015} report negative RMs between --70 and --300 \radm~in the western shell and positive RMs on order +55 \radm~ in the eastern shell. \citet{Gao:2015} conclude that the sign reversal is expected in the case of the Galactic magnetic field being lifted out of the plane by the expanding bubble. With H$\alpha$ estimates from \citet{Dennison:1997} for the electron density and geometric arguments for the shell radii of the W4 Superbubble, \citet{Gao:2015} estimate $|$\blos$|$  $\sim$ 5 $\mu$G.

\citet{Stil:2009} compare their magnetohydrodynamic simulations of superbubbles to the W4 Superbubble. In general, they find that the largest Faraday rotation occurs in a thin region around the cavity, and inside the cavity, it would be smaller. They also present two limiting cases for the orientation of the Galactic magnetic field with respect to the line of sight, and the consequences for the RMs through the shell. If the Galactic magnetic field is perpendicular to the observer's line of sight, then the contributions to the RM from the front and rear bubble wall would be of equal but opposite magnitude, except for small asymmetries which would lead to low RMs (\about~20 \radm) through the cavity. This requires the magnetic field to be bent by the bubble to have a non-zero line of sight component. If the Galactic magnetic field is parallel to the line of sight, then the RMs through the front and rear bubble wall reinforce each other, and there are high RMs for lines of sight through the shell. In this case, there are higher RMs (\about{} 3 $\times$ 10$^3$ \radm) everywhere.

There are also studies of the magnetic field for W3. From \hi Zeeman observations, \citet{vanderWerf:1990} conclude that the \blos~has small-scale structures that can vary on order of 50 $\mu$G over $\sim$ 9 arcsec scales.  \citet{Roberts:1993} report values of the LOS magnetic field from \hi Zeeman observations towards three resolved components of W3. The three components are near ($\ell$, $b$) $\sim$ (133.7\ddeg, 1.21\ddeg), with a maximum separation of 1.5 arcmin, and the LOS magnetic field is between --50 $\mu$G and +100 $\mu$G. \citet{Balser:2016} observed carbon radio recombination line (RRL) widths to estimate the total magnetic field strength in the photodissociation region (see \citealt{Roshi:2007} for details). They report B$_{\textrm{tot}}$ = 140 -- 320 $\mu$G near W3A (133.72\ddeg, 1.22\ddeg) and argue that for a random magnetic field, B$_{\textrm{tot}}$ = 2 $|$\blos$|$, which would then be consistent with the \citet{Roberts:1993} estimates of the \blos. It should be noted that these magnetic field strengths are substantially larger than those inferred for the W4 Superbubble on the basis of polarimetry of the Galactic background (see text above).

In this paper, we present new Faraday rotation results for IC 1805 to investigate the role of the magnetic field in the \HII region and stellar bubble. As in \citet{Savage:2013} and \citet{Costa:2016}, we utilize an arguably simpler and more direct method of inferring the LOS component of the magnetic field in \HII regions.  This is the measurement of the Faraday rotation of nonthermal background sources (usually extragalactic radio sources) whose lines of sight pass through the \HII region and its vicinity. In Section \ref{sec:obs}, we describe the instrumental configuration and observations, including source selection. Section \ref{sec:dataredux} details the data reduction process, including the methods used to determine RM values. In Section \ref{sec:obsres}, we report the results of the RM analysis and discuss Faraday rotation through the W4 complex in Section \ref{sec:fr}. We present models for the RM within the \HII region and stellar bubble in Section \ref{sec:models}. We discuss our observational results and their significance for the nature of IC 1805 in Section \ref{sec:results} and compare the results of this study with our previous study of the Rosette nebula in Section \ref{sec:rosette}. We discuss future research in Section \ref{sec:fut}, and present our conclusions and summary in Section \ref{sec:sum}. 


%
%
%
%
%

\section{Observations}\label{sec:obs}

\subsection{Source Selection\label{sec:sourceselect}}

\begin{figure}[htb!]
\centering
\includegraphics[width=0.9\textwidth]{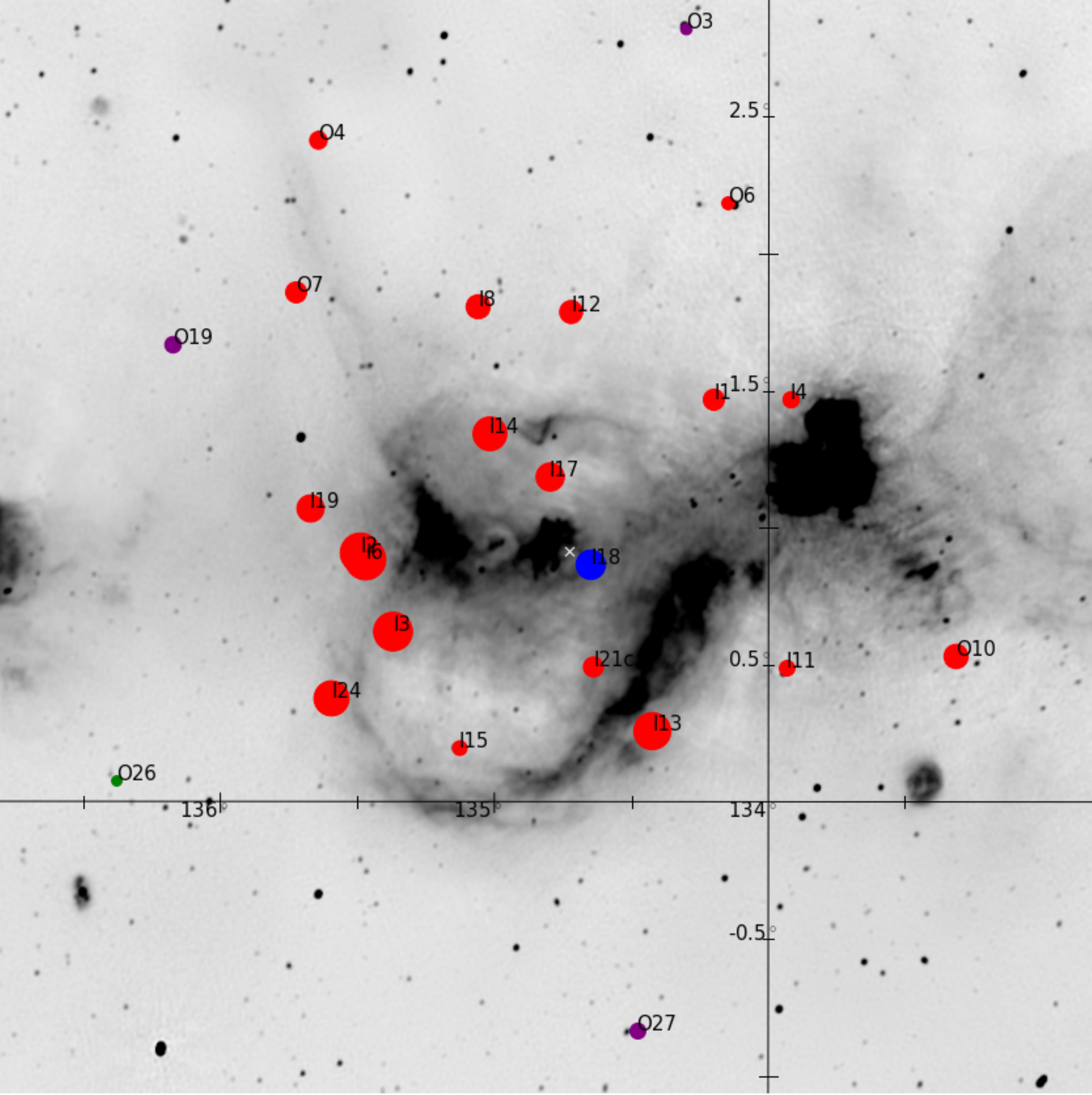}
\caption[Radio Continuum map at 1.42GHz of W4]{Mosaic of IC 1805 from the Canadian Galactic Plane Survey at 1.42 GHz, with Galactic longitude and latitude axes. The lines of sight listed in Table \ref{tab:sources} are the red and blue symbols, where positive RMs are blue and negative RMs are red. The green and purple symbols are RM values from \citet{Taylor:2009} or \citet{Brown:2003}, where positive RMs are green and negative RMs are purple. We utilize the naming scheme from Table \ref{tab:sources} for the RM values from the literature for ease of reference, but we omit the ``W4-'' prefix in this image for clarity. The size of the plotted symbols is proportional to the $|$RM$|$ value.}
\label{fig:w4}
\end{figure}

Our criteria for source selection were identical to \citet{Savage:2013} and \citet{Costa:2016} in that we searched the National Radio Astronomy Observatory Very Large Array Sky Survey (NVSS, \citealt{Condon:1998}) database for point sources within 1\ddeg~of OCl 352 (the ``I'' sources) with a minimum flux density of 20 mJy. We also searched in an annulus centered on the star cluster with inner and outer radii of 1\ddeg~and 2\ddeg~for outer sources (``O'') to measure the background RM due to the general ISM. We identified 31 inner sources and 26 outer sources in the region. We then inspected the NVSS postage stamps to ensure that they were point sources at the resolution of the NVSS (\about{} 45 arcseconds). We discarded sources that showed extended structure similar to Galactic sources. We selected 24 inner sources and 8 outer sources from this final list.
\begin{table}
\centering
\caption{ List of Sources Observed}
\begin{threeparttable}
\centering
\small
\begin{tabular}{ccccccccc}
\hline
Source & $\alpha$(J2000) & $\delta$(J2000) & $\emph{l}$  & $\emph{b}$ & $\xi$\tnote{a} & S$_{4.33\textrm{GHz}}$ & m \\
Name & h m s & $^o$ $'$ $''$  &  ($^o$) & ($^o$) & (arcmin) & (mJy) & ($\%$)\\
\hline
W4-I1 & 02 30 16.2 & +62 09 37.9 & 134.19 & 1.47 & 46.0 & 77 & 4 \\ 
W4-I2 & 02 38 34.2 & +61 08 46.6 & 135.49 & 0.91 & 46.2 & 5 & 12 \\ 
W4-I3 & 02 36 45.5 & +60 55 48.8 & 135.38 & 0.63 & 42.8 & 82 & 3 \\ 
W4-I4 & 02 27 59.8 & +62 15 44.0 & 133.91 & 1.47 & 58.9 & 46 & 10 \\ 
W4-I5\tnote{b} & 02 28 01.6 & +62 02 16.7 & 133.99 & 1.26 & 48.4 & --- & --- \\ 
W4-I6 & 02 38 19.9 & +61 08 03.5 & 135.47 & 0.89 & 44.8 & 12 & 11 \\
W4-I7\tnote{b} & 02 27 33.8 & +61 55 58.1 & 133.98 & 1.14 & 46.6 & -- &  --- \\ 
W4-I8 & 02 38 10.1 & +62 08 57.0 & 135.05 & 1.81 & 57.1 & 47 & 9 \\ 
W4-I9 & 02 28 21.6 & +61 28 36.5 & 134.23 & 0.75 & 31.1 & --- & --- \\ 
W4-I10\tnote{b} & 02 29 13.0 & +61 00 53.4 & 134.50 & 0.36 & 36.2 & --- & --- \\ 
W4-I11 & 02 25 15.2 & +61 19 14.4 & 133.94 & 0.47 & 54.0 & 39 & 3 \\ 
W4-I12 & 02 35 20.6 & +62 16 02.3 & 134.70 & 1.79 & 52.5 & 77 & 2 \\ 
W4-I13 & 02 28 25.1 & +60 56 20.2 & 134.44 & 0.25 & 43.5 & 16 & 6 \\ 
W4-I14 & 02 36 19.2 & +61 44 05.5 & 135.01 & 1.35 & 31.4 & 2 & 16 \\ 
W4-I15 & 02 33 36.1 & +60 37 40.4 & 135.14 & 0.20 & 49.8 & 27 & 4 \\ 
W4-I16 & 02 36 56.8 & +61 57 58.6 & 134.99 & 1.59 & 43.3 & 35 & 0 \\ 
W4-I17 & 02 34 08.8 & +61 40 35.5 & 134.80 & 1.19 & 17.1 & 9 & 16 \\ 
W4-I18 & 02 31 56.3 & +61 25 50.9 & 134.65 & 0.87 & 5.6 & 24 & 2 \\ 
W4-I19\tnote{c} & 02 40 31.7 & +61 13 45.9 & 135.68 & 1.09 & 57.8 & 11 & 3 \\ 
W4-I20 & 02 27 03.9 & +61 52 24.9 & 133.94 & 1.07 & 47.6&  457 & 0 \\  
W4-I21 & 02 30 44.5 & +61 05 30.2 & 134.64 & 0.50 & 25.9 & 5 & 7 \\ 
W4-I22\tnote{b} & 02 26 07.8 & +61 56 43.7 & 133.82 & 1.09 & 55.4 & --- & --- \\ 
W4-I23 & 02 40 30.9 & +61 47 10.1 & 135.45 & 1.59 & 59.3 & 3 & 0 \\ 
W4-I24 & 02 37 45.1 & +60 37 31.4 & 135.61 & 0.40 & 61.5 & 20 & 10 \\ 
W4-O1 & 02 41 33.9 & +61 26 29.5 & 135.70 & 1.33 & 63.5 & 377 & 0 \\ 
W4-O2 & 02 35 37.8 & +59 56 29.5 & 135.64 & -0.33 & 93.0 &  107 & 0 \\ 
W4-O4 & 02 44 57.7 & +62 28 06.5 & 135.64 & 2.43 & 105.8 & 747 & 0 \\ 
W4-O5 & 02 21 52.6 & +60 10 03.2 & 133.96 & -0.75 & 110.3 & 94 & 0 \\ 
W4-O6 & 02 31 59.2 & +62 50 34.1 & 134.12 & 2.18 & 83.7 & 120 & 4 \\ 
W4-O7 & 02 43 35.6 & +61 55 54.6 & 135.72 & 1.88 & 82.7 & 52 & 0 \\ 
W4-O8 & 02 23 04.5 & +60 58 19.6 & 133.82 & 0.05 & 75.2 & 40 & 0 \\ 
W4-O10 & 02 20 26.2 & +61 34 46.2 & 133.31 & 0.51 & 88.0 & 64 & 3 \\ 
\hline
\end{tabular}
\label{tab:sources}
\begin{tablenotes}
\item[a] Angular distance between the line of sight and a line of sight through the center of the star cluster.
\item[b] NVSS position. No source detected in the Stokes I map in any frequency bin.
\item[c] High Mass X-Ray Binary \lsi.
\end{tablenotes}
\end{threeparttable}
\end{table}

The sources are listed in Table \ref{tab:sources}, where the first column lists the source name in our nomenclature. The second and third columns list the right ascension ($\alpha$) and declination ($\delta$) of the observed sources. The positions are determined with the {\sc imfit} task in CASA, which fits a 2D Gaussian to the intensity distribution at 4.33 GHz. Columns four and five give the Galactic longitude ($\ell$), Galactic latitude (\textit{b}), which is converted from $\alpha$ and $\delta$ using the Python \textit{Astropy} package, and the angular separation from the center of the nebula ($\xi$) is given in column six. Column seven lists the flux density at 4.33 GHz calculated with {\sc imfit}, and column eight gives the mean percent linear polarization (m = \textit{P}/\textit{I}) as measured across the eight 128 MHz maps and assuming a \fsimp{} source. Figure \ref{fig:w4} is a radio continuum mosaic from the Canadian Galactic Plane Survey (CGPS) \citep{Taylor:2003,Landecker:2010} with the location of the sources, along with the names, indicated with filled circles.


\subsection{VLA Observations\label{sec:vlaobs}}
\begin{table}[!htb]
\centering
\begin{threeparttable}
\caption{Log of Observations \label{tab:logofobs}}

\begin{tabular}{p{0.54\linewidth}
				p{0.3\linewidth}}
\hline
VLA Project Code & 13A-035 \\
Date of Observations &  2013 July 10, 13, 16, and 17\\

Number of Scheduling Blocks & 4 \\

Duration of Scheduling Blocks (h) & 4\\

Frequencies of Observation\tnote{a}~ (GHz) & 4.850; 7.250\\

Number of Frequency Channels per IF & 512\\


Channel Width (MHz) & 2 \\

VLA array & C \\

Restoring Beam (diameter) & 4\farcs81\\

Total Integration Time per Source & 18--25 minutes\tnote{b}\\

RMS Noise in Q and U Maps ($\mu$Jy/beam) & 39\tnote{c}\\

RMS Noise in RM Synthesis Maps ($\mu$Jy/beam) & 23\tnote{d} \\

\hline
\end{tabular}
\begin{tablenotes}
\item[a] The observations had 1.024 GHz wide intermediate frequency bands (IFs) centered on the frequencies listed, each composed of eight 128 MHz wide subbands.
\item [b] The ``O'' sources (see Table \ref{tab:sources}) averaged 18 minutes, and the ``I'' sources, being weaker, were between 22--25 minutes.
\item [c] This number represents the average rms noise level for all the Q and U maps.
\item[d] Polarized sensitivity of the combined RM Synthesis maps.
\end{tablenotes}
\end{threeparttable}
\end{table}

We observed 32 radio sources with the NSF's Karl G. Jansky Very Large Array (VLA)\footnote{\footnotesize{The Karl G. Jansky Very Large Array is an instrument of the National Radio Astronomy
Observatory (NRAO). The NRAO is a facility of the National Science Foundation,
operated under cooperative agreement with Associated Universities, Inc.}} whose lines of sight pass through or near to the shell of the IC 1805 stellar bubble. Table \ref{tab:logofobs} lists details of the observations. Traditionally, polarization observations require observing a polarization calibrator source frequently over the course of an observation to acquire at least 60\ddeg~of parallactic angle coverage. This is done to determine the instrumental polarization (D-factors, leakage solutions). Since the completion of the upgraded VLA, shorter scheduling blocks, typically less than 4 hours in duration, have become a common mode of observation. It is difficult, if not impossible, with very short scheduling blocks to acquire enough parallactic angle coverage to measure the instrumental calibration with a polarized source. Another method of determining the instrumental polarization is to observe a single scan of an unpolarized source. This technique can be used with shorter scheduling blocks.

In this project we calibrated the instrumental polarization using both techniques. We used the source J0228+6721, observed over a wide range of parallactic angle, and also made a single scan of the unpolarized source 3C84. Use of the CASA task \polcal~on the J0228+6721 data solved for the instrumental polarization, determined by the antenna-specific D factors \citep{Bignell:1982}, which are complex, as well as the source polarization (\textit{Q} and \textit{U} fluxes). In the case of 3C84, \polcal~solves only for the D factors. 
We find no significant deviations between these two calibration methods, indicating accurate values for the instrumental polarization parameters.

3C138 and 3C48 functioned as both flux density and polarization position angle calibrators. J0228+6721 was used to determine the complex gain of the antennas as a function of time as well to as serve as a check, as described above, for the D-factors. We observed the program sources for 5 minute intervals and interleaved the observations of J0228+6721. There was one observation of 3C138, 3C48, and 3C84 each. For our final data products, we utilized 3C84 as the primary leakage calibrator and 3C138 as the flux density and polarization position angle calibrator.




\section{Data Reduction\label{sec:dataredux}}

The data were reduced and imaged using the NRAO Common Astronomy Software Applications (CASA)\footnote{\footnotesize{For further reference on data reduction, see the NRAO Jansky VLA tutorial ``EVLA Continuum Tutorial 3C391'' (http$://$casaguides.nrao.edu$/$index.php$?$title$=$EVLA$_{-}$Continuum$_{-}$Tutorial$_{-}$3C391)}} version 4.5. The procedure for the data reduction as described in Section 3 of \citet{Costa:2016} is identical to the procedure we employed in this study. The only difference for the current data set is that in the CASA task \clean, we utilized \textit{Briggs} weighting with the ``robust'' parameter set to 0.5, which adjusts the weighting to be slightly more \textit{natural} than \textit{uniform}. \textit{Natural} weighting has the best signal/noise ratio at the expense of resolution, while \textit{uniform} is the opposite. \textit{Briggs} weighting allows for intermediate options. As in our previous work, we also implemented a cutoff in the (\textit{u}, \textit{v}) plane for distances $<$ 5000 wavelengths to remove foreground nebular emission.

\begin{figure}[!htb]
\centering
\subfloat[\label{fig:I18}]{
\includegraphics[width=0.4\textwidth]{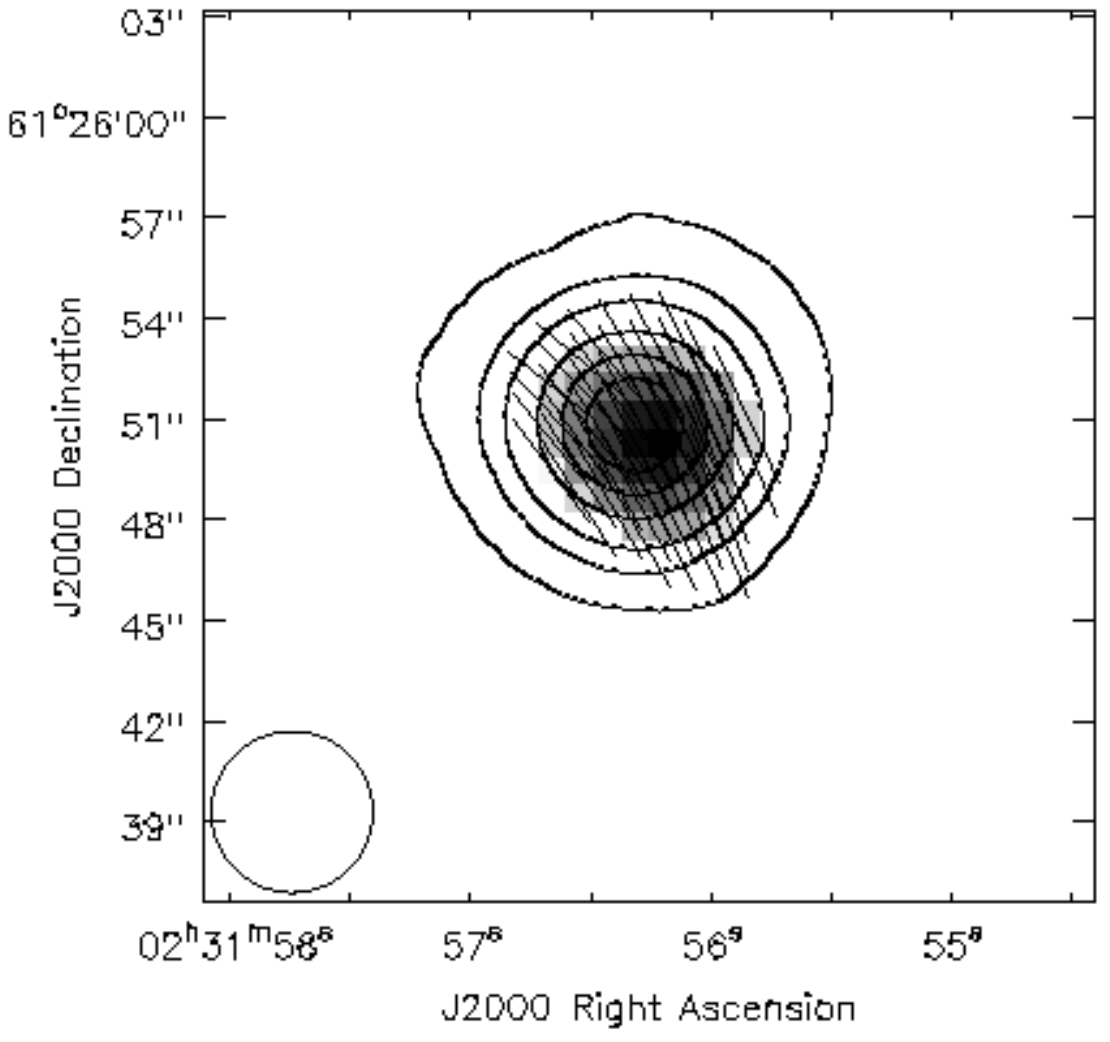}}
\quad
\subfloat[\label{fig:I24}]{
\includegraphics[width=0.42\textwidth]{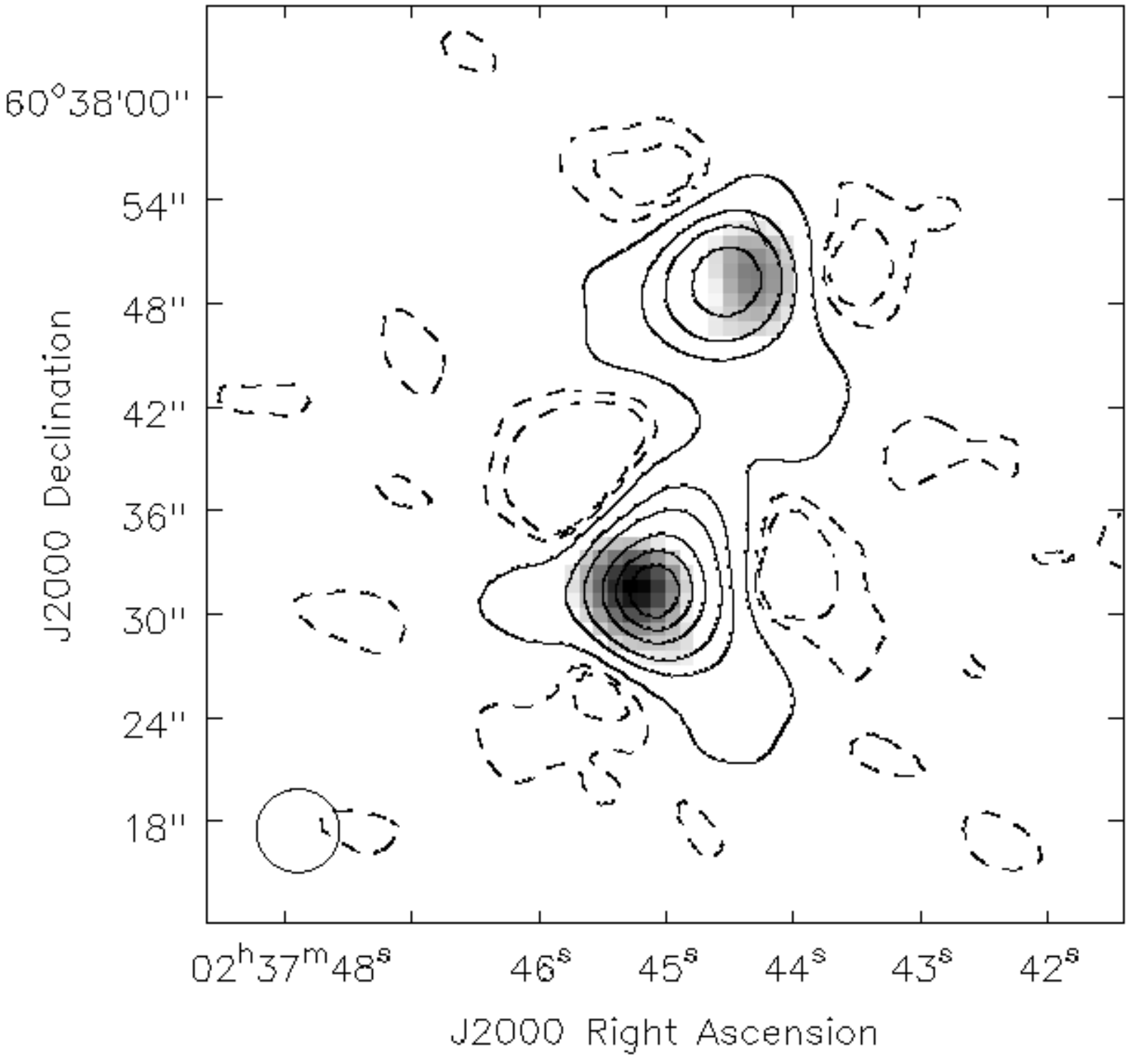}}
\caption[CASA Map]{Map of (a) W4-I118 and (b) W4-I24 at 4913 MHz. The circle in the lower left is the restoring beam. The gray scale is the linear polarized intensity, \textit{P}, the vectors show the polarization position angle, $\chi$, and the contours are the Stokes \textit{I} intensity with levels of -2, -1 , 2, 10, 20, 40, 60, and 80$\%$ of the peak intensity,  21.5 mJy beam$^{-1}$ and 11.7 mJy beam$^{-1}$ for W4-I18 and W4-I24a, respectively. }
\label{fig:MAP}
\end{figure}

Similar to \citet{Costa:2016}, we had two sets of data products after calibration and imaging. The first set of images consisted of radio maps (see Figures \ref{fig:I18} and \ref{fig:I24}) of each Stokes parameter, formed over a 128 MHz wide subband for each source. These images were inputs to the \chilam~analysis (Section \ref{sec:chilam}), and there were typically 14 individual maps for each source per Stokes parameter. 

The second set of images consisted of maps of \textit{I}, \textit{Q}, and \textit{U} in 4 MHz wide steps across the entire bandwidth using the \clean~mode ``channel'', which averages two adjacent 2 MHz channels. 
 Ideally, changes in \Q{} and \U{} should only be due to Faraday rotation; however, the spectral index can affect \Q{} and \U{} independently of the RM, which can be interpreted as depolarization. RM Synthesis does not, by default, account for the spectral index, so a correction must be applied prior to performing RM Synthesis (see Section 3 of \citealt{Brentjens:2005}). We first determine the spectral index, $\alpha$, of each source from a least-squares fit to the log of the flux density, $S_\nu$, and the log of the frequency, $\nu$. We adopt the convention that $S_\nu$ \about~ $\nu^{-\alpha}$. We use the center frequency, $\nu_c$, of the band and the measured value of \textit{Q} and \textit{U} at each frequency, $\nu$, to find \textit{Q$_o$} and \textit{U$_o$} using the relationship \[Q = Q_o\left(\frac{\nu}{\nu_c}\right)^{-\alpha} \textrm{ and } \ U = U_o\left(\frac{\nu}{\nu_c}\right)^{-\alpha}. \] The final images consisted of approximately 336 maps per source, per Stokes parameter, as inputs for the RM Synthesis analysis (Section \ref{sec:rmsyn}). 

\subsection{Rotation Measure Analysis via a Least-Squares Fit to $\chi$ vs $\lambda^2$ \label{sec:chilam}}
The output of the CASA task \clean~produces images in Stokes \textit{I, Q, U,} and \textit{V}. From these images, we generated maps with the task \immath~of the linear polarized intensity \textit{P}, \[P = \sqrt{Q^2+U^2} \] and the polarization position angle $\chi$,
\begin{equation*}
\chi=\frac{1}{2}\tan^{-1}{\left(\frac{U}{Q}\right)}
\end{equation*}
for each source over a 128 MHz subband. Data that are below the threshold of 5$\sigma_{\textrm{Q}}$ are masked in the \textit{P} and $\chi$ maps, where $\sigma_{\textrm{Q}}$ = $\sigma_{\textrm{U}}$  is the rms noise in the \textit{Q} data. This threshold prevents noise in the \textit{Q} and \textit{U} data from generating false structure in the \textit{P} and $\chi$ maps. Examples of images are shown in Figure \ref{fig:MAP}, which displays the total intensity, polarized intensity, and polarization position angle for sources W4-I18 and W4-I24. W4-I18 is an example of a point  source, or slightly resolved source. Twelve of the sources in Table \ref{tab:sources} were of this type and unresolved to the VLA in C array. Eight sources were like W4-I24, showing extended structure in the observations and potentially yielding RM values on more than one line of sight.

\begin{figure}[hbt!]
\centering
\subfloat[][\label{fig:I1chi}]{
\includegraphics[width=0.48\textwidth]{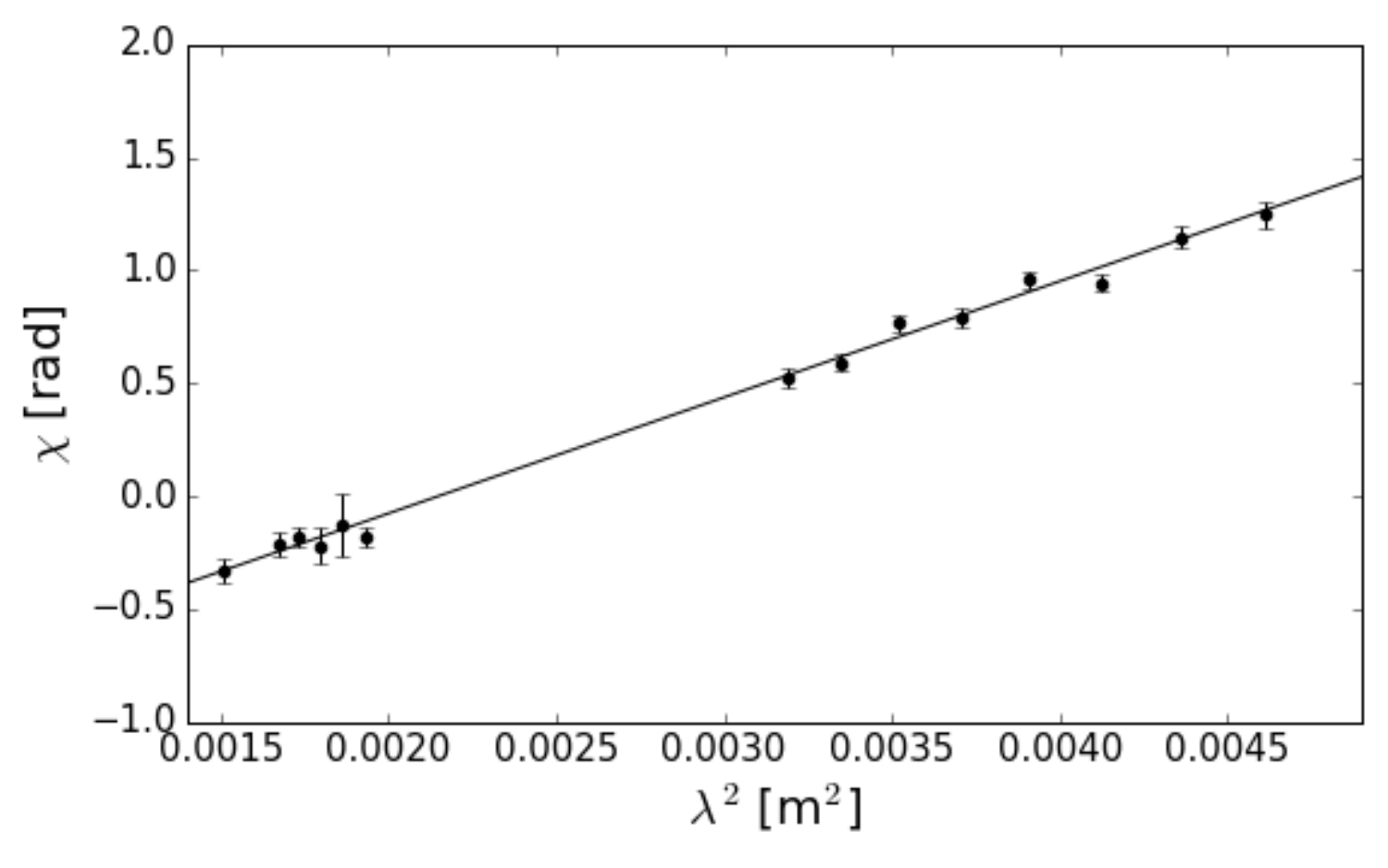}} \quad
\subfloat[][\label{fig:I18chi}]{
\includegraphics[width=0.48\textwidth]{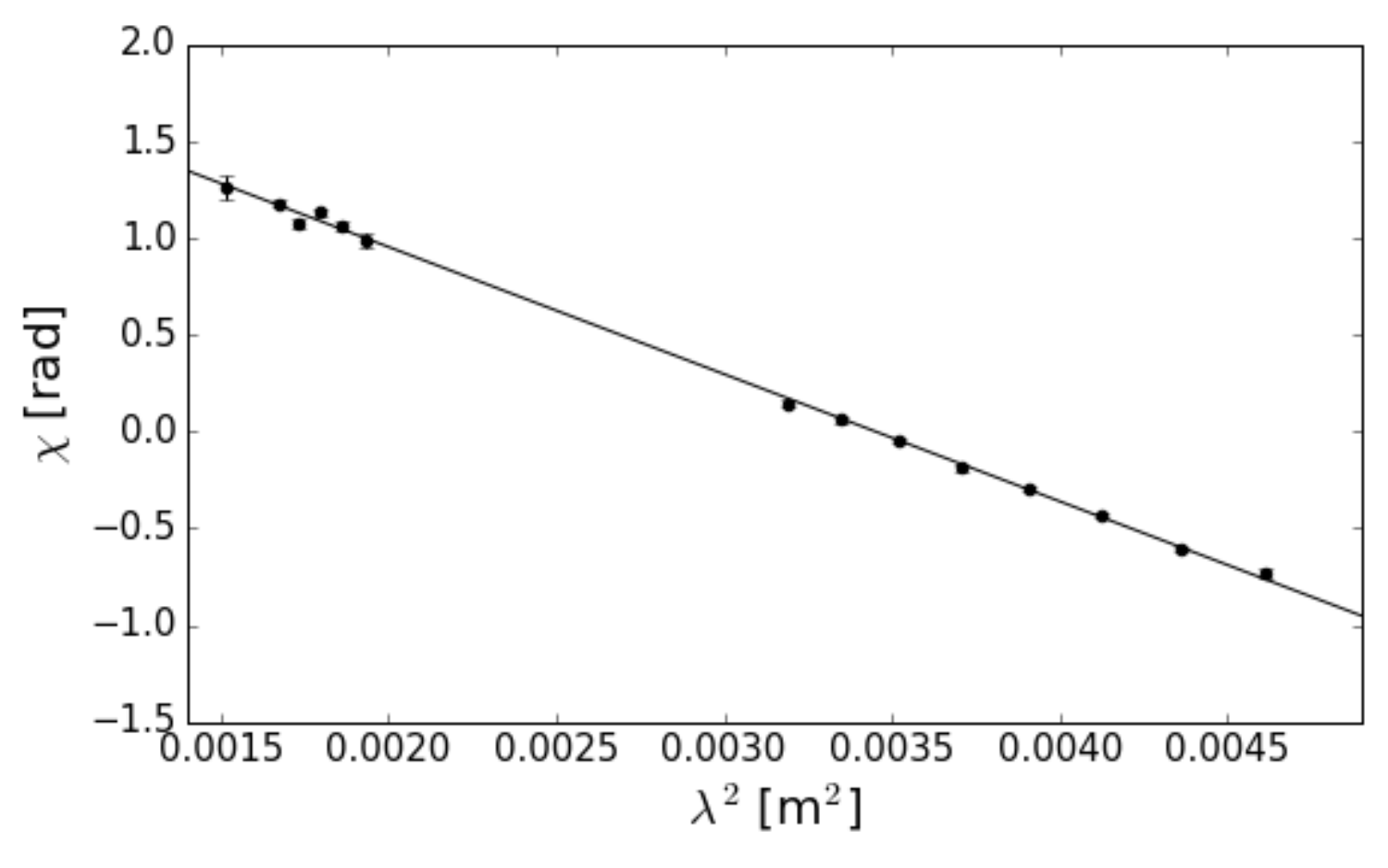}}\\
\caption[Plot of $\chi(\lambda^{2})$]{Plot of the polarization position angle as a function of the square of the wavelength, $\chi(\lambda^{2})$, for the source (a) W4-I18, RM= +514 $\pm$ 12 \radm, and (b) W4-I24a, RM = --658 $\pm$ 5 \radm. Each plotted point results from a measurement in a single 128 MHz-wide subband. The gap in \lamsq{} coverage is due to the observation configuration that consisted of two 1.024 GHz wide IFs separated by \about{} 1.4 GHz.}
\label{fig:newpol}
\end{figure}
In the case of a single foreground magnetic-ionic medium responsible for the rotation of an incoming radio wave, the relation between $\chi$ and $\lambda^2$ is linear, and we calculate the RM through a least-squares fit of \chilam. To measure $\chi$, we select the pixel that corresponds to the highest value of \textit{P} on the source in the 4338 MHz map, and we then measure $\chi$ at that location in each subsequent 128 MHz wide subband. Figure \ref{fig:newpol} shows two examples of the least-squares fit to \chilam. The $\chi$ errors are \(\sigma_{\chi} = \frac{\sigma_{Q}}{2P},\) (\citealt{Everett:2001}, Equation 12).

\subsection{Rotation Measure Synthesis \label{sec:rmsyn}}

In additional to the least-squares fit to \chilam, we performed Rotation Measure Synthesis \citep{Brentjens:2005}. The inputs to RM Synthesis are images in Stokes \textit{I}, \textit{Q}, and \textit{U} across the entire observed spectrum in 4 MHz spectral intervals. We refer the reader to Section 3.1.2 of \citet{Costa:2016} for a detailed account of our procedure, which follows the implementation of RM Synthesis as developed by \citet{Brentjens:2005}. The goal of RM synthesis is to recover the Faraday dispersion function $F(\phi)$.  Here $\phi$, the Faraday depth, is a variable which is Fourier-conjugate to $\lambda^2$ (see \citealt{Costa:2016}, Equations 3 and 4), and has units of \radm.  We also refer to $F(\phi)$ as the ``Faraday spectrum''.


\begin{table}[htb!]
\centering
\caption{Rotation Measure Synthesis Parameters\label{tab:rmpar}}
\begin{threeparttable}
\centering
\begin{tabular}{p{2cm} p{4cm} p{9.5cm}}
\hline 
$\Delta \lambda^2$ & 3.2 $\times$ 10$^{-3}$ (m$^2$) & Total bandwidth\tnote{a}. \\ 

$\lambda^2_{min}$ & 1.5 $\times$ 10$^{-3}$ (m$^2$) & Shortest observed wavelength squared. \\ 
$\delta \lambda^2$ & 4.8 $\times$ 10$^{-6}$ (m$^2$) & Width of a channel; Eq (35) \citet{Brentjens:2005}. \\ 

$\delta \phi$ & 1072\tnote{b}~  (\radm) & FWHM of RMSF; Eq (61) \citet{Brentjens:2005}. \\ 
\\
\multirow{2}{*}{max-scale} & \multirow{2}{*}{2098 (\radm)} & Sensitivity to extended Faraday structures; Eq (62) \citet{Brentjens:2005}. \\ 
\\
\multirow{3}{*}{$|\phi_{max}|$} & \multirow{3}{*}{3.6 $\times$ 10$^{5}$ (\radm)} & Maximum detectable Faraday depth before bandwidth depolarization;  Eq (63) \citet{Brentjens:2005}. \\  \\ \hline
\end{tabular}
\begin{tablenotes}
\item[a] This bandwidth includes the frequencies not observed that lie between our two IFs. They are set to 0 via the weighting function, W($\lambda^2$).
\item[b] Since flagging for RFI and bad antennas were done individually for each scheduling block, the FWHM of the RMSF can vary slightly from source to source. However, these slight variations are not significant in our interpretation of the RM values report in this paper.
\end{tablenotes}
\end{threeparttable}
\end{table}

\fphi~is recovered via an \rmclean~algothrim \citep{Heald:2009,Bell:2012}, and we applied a 7$\sigma$ cutoff, which is above the amplitude at which peaks due to noise are likely to arise \citep{Brentjens:2005,Macquart:2012,Anderson:2015}.  The \textsc{rmsynthesis} algorithm initially searched for peaks in the Faraday spectrum using a range of $\phi$ $\pm$ 10,000 \radm~at a resolution of 40 \radm~to determine if there were significant peaks at large values of $|\phi|$. Then, we performed a finer search at $\phi$ = $\pm$ 3000 \radm~at a resolution of 10 \radm. The RM Synthesis parameters, such as the full-width-at-half-maximum (FWHM) of the rotation measure spread function (RMSF) and the maximum detectable Faraday depth, are given in Table \ref{tab:rmpar}. 

\begin{figure}[hbt!]
\centering
\subfloat[\label{fig:psynmap}]{
\includegraphics[width=0.48\textwidth]{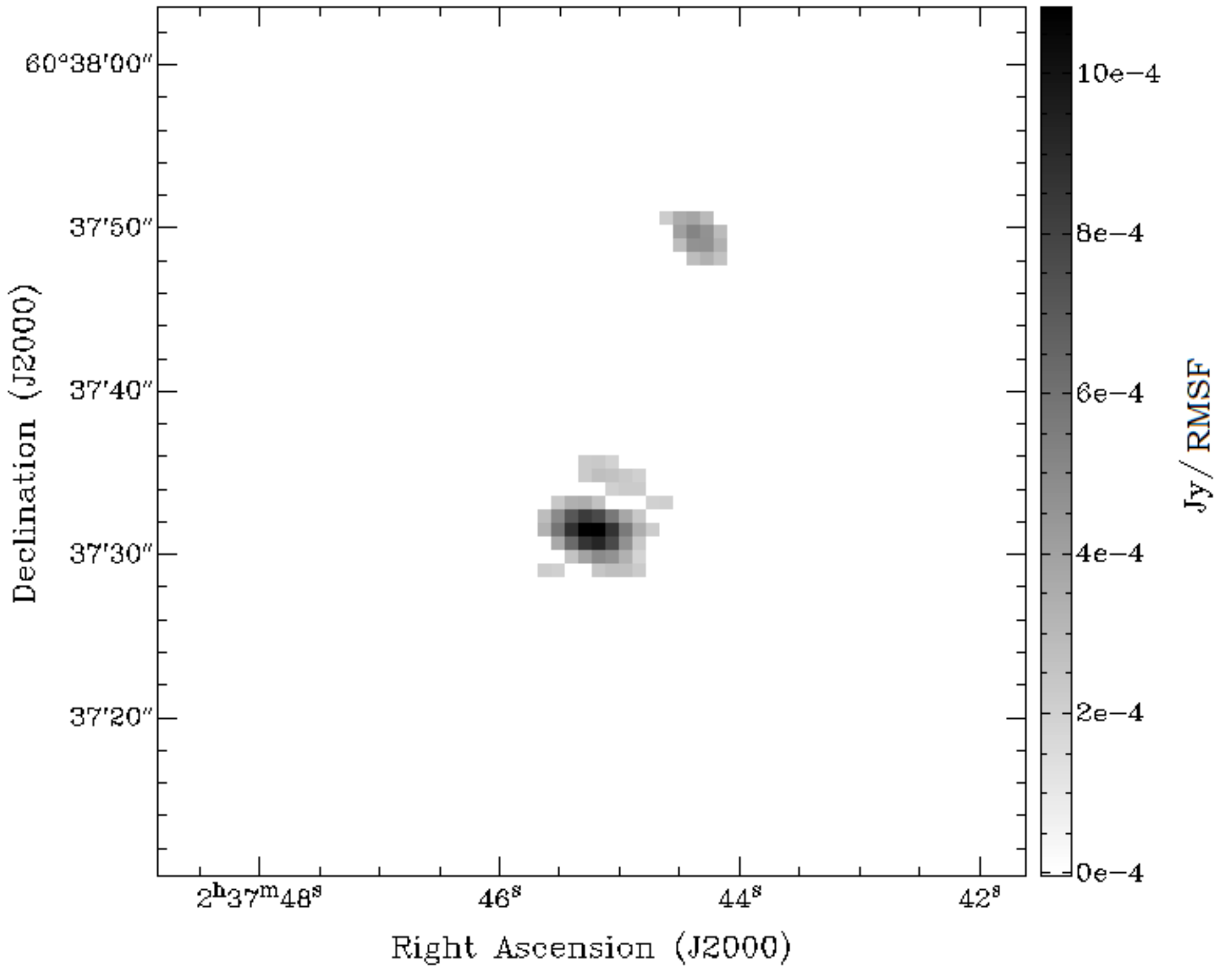}}
\quad
\subfloat[\label{fig:rmsynmap}]{
\includegraphics[width=0.48\textwidth]{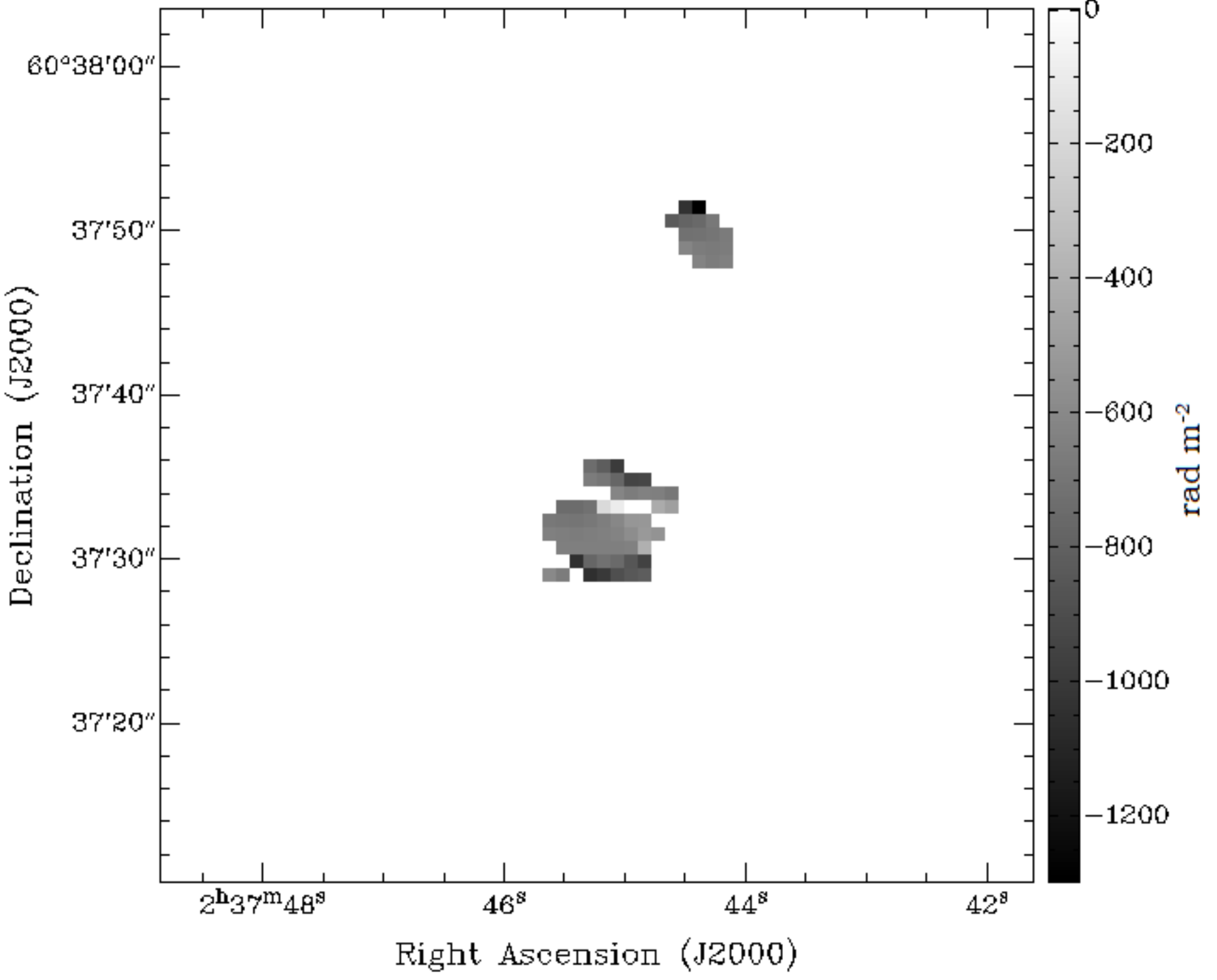}}
\caption[Plot of RM Synthesis Map]{(a) Linear polarization map and (b) RM map of W4-I24 from the RM Synthesis analysis. In both images, the data cube was flattened over the Faraday depth axes, and a threshold of 7$\sigma$ was applied. W4-I24 has extended structure, so there are two peaks, which are also present in the CASA maps (Figure \ref{fig:I24}).}
\end{figure}

As in \citet{Costa:2016}, we utilized an IDL code for the \textsc{rmsynthesis} and \rmclean~algorithms. The output of the IDL code is a data cube in Faraday depth space that is equal in range to the range of $\phi$ that was searched over in the \textsc{rmsynthesis} algorithm. The data cube contains, for example, 500 maps of the polarized intensity as a function of spatial coordinates and $\phi$, which ranges between $\pm$ 10,000 \radm~at intervals of 40 \radm. Initially, we generated these maps for a 1024 $\times$ 1024 pixel image. We then used the Karma package \citep{Gooch:1995} tool \textsc{kvis} to review the maps to search for sources or source components away from the phase center that, while being too weak to detect in the 128 MHz maps, may be detectable in the RM Synthesis technique since it uses the entire bandwidth to determine the Faraday spectrum\footnote{private communication, L. Rudnick}. However, no such sources were identified above the cutoff. From the 1024 $\times$ 1024 maps of the Faraday spectrum, we identified the \textit{P$_{\textrm{max}}$} for the observed sources and extracted the Faraday spectrum at that location. Figure \ref{fig:psynmap} shows an example of a \textit{P$_{\textrm{max}}$} map that has been flattened along the $\phi$ axis, i.e., the gray scale in the image represents the full range of $\phi$. From this map, it is easy to identify the spatial location of \textit{P$_{\textrm{max}}$} for the source, which agrees with the location of the peak linear polarized intensity in the \chilam~analysis. We obtained this same result in \citet{Costa:2016} for the Rosette Nebula.

\begin{figure}[htb!]
\centering
\subfloat[\label{fig:RMSYN}]{
\includegraphics[width=0.45\textwidth]{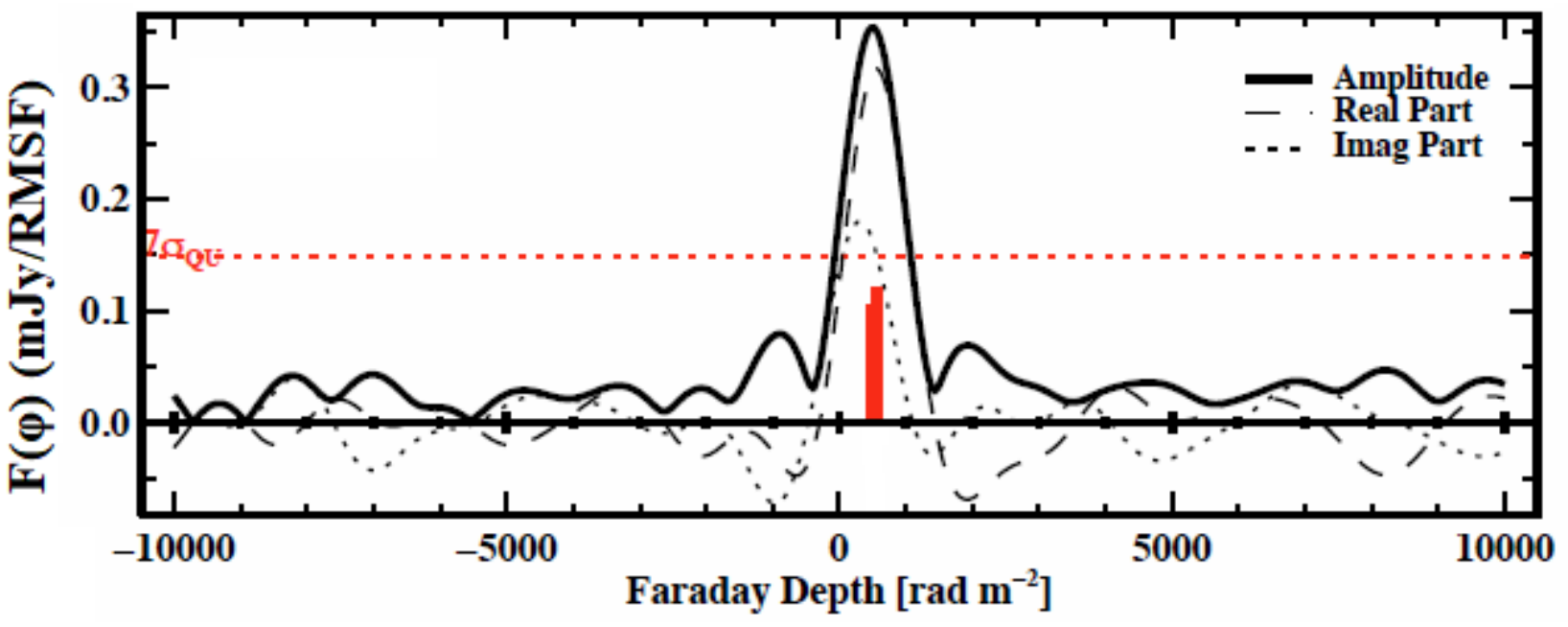}}
\quad
\subfloat[\label{fig:RMSF}]{
\includegraphics[width=0.45\textwidth]{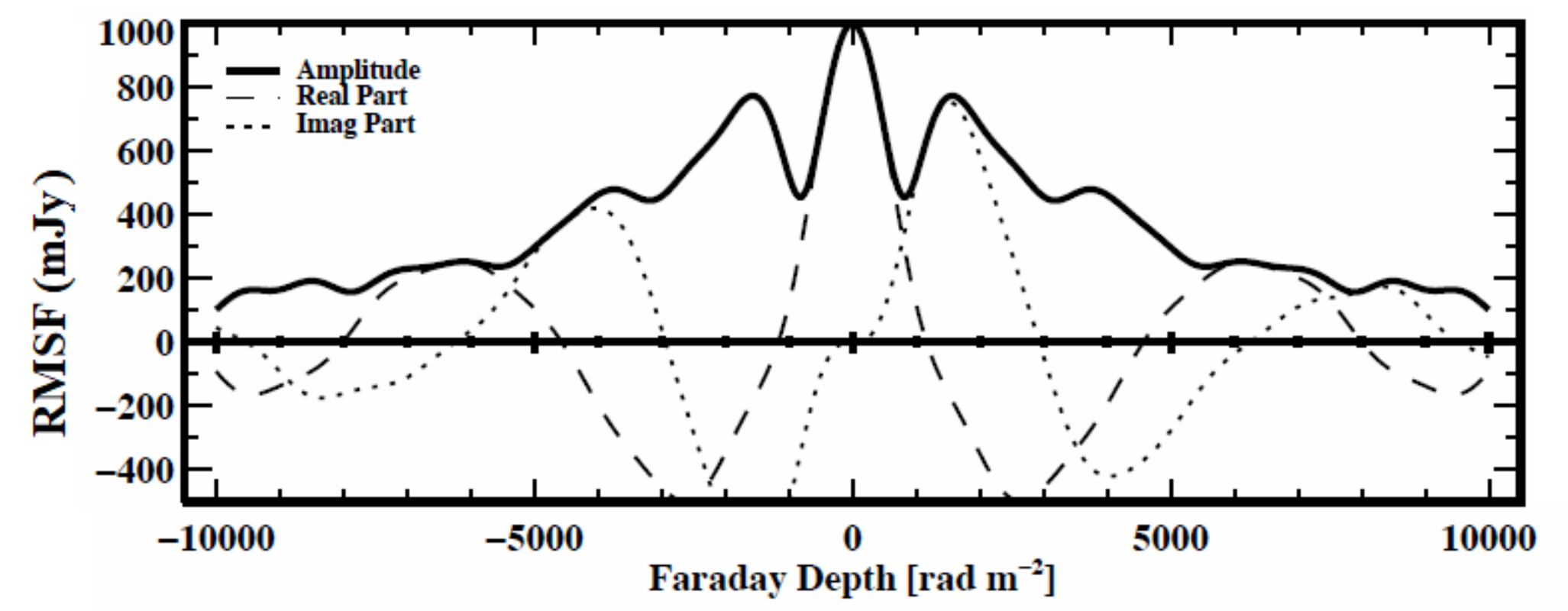}}
\caption[]{Plot of (a) cleaned Faraday dispersion function, F($\phi$), for W4-I18, where $\phi_{peak}$ = 501 $\pm$ 33 \radm~and (b) the RMSF (R($\phi$)). The 7$\sigma$ cutoff is shown in the red, dashed, horizontal line. The vertical red lines are the clean components, and the black curves represent the \textit{P} (solid), \textit{Q} (dot dashed), and U (dashed) components of the spectrum. }
\label{fig:rmsynrmsf}
\end{figure}

To determine the RM, we fit a 2 degree polynomial to the Faraday spectrum at each pixel in the 1024 x 1024 image above the 7$\sigma$ cutoff. The gray scale in Figure \ref{fig:rmsynmap} shows the RM value from the fit to each pixel. The image is zoomed and centered on the source. While we can mathematically determine the RM at each pixel, the sources are not resolved, so we only select the RM at the spatial location of \textit{P$_{\textrm{max}}$}.  Figure \ref{fig:RMSYN} plots the Faraday spectrum and \rmclean~components for W4-I18, and Figure \ref{fig:RMSF} shows the RMSF.

\citet{Anderson:2015} describe two cases for the behavior of the Faraday spectrum. A source is considered \fsimp~when \fphi~is non-zero at only one value of $\phi$, \textit{Q} and \textit{U} as a function of \lamsq~vary sinusoidally with equal amplitude, and \plam~is constant. The \fsimp~case has the physical meaning of a uniform Faraday screen in the foreground that is responsible for the Faraday rotation, and $\chi$ is linearly dependent on \lamsq. If a source is \fsimp, then \fphi~is a delta function at a Faraday depth equal to the RM. The second behavior \citet{Anderson:2015} describe for the Faraday spectrum is a \fcomp~source, which is any spectrum that deviates from the criteria set for the \fsimp~case. A \fcomp~spectrum can be the result of depolarization in form of beam depolarization, internal Faraday dispersion, multiple interfering Faraday rotating components, etc. \citep{Sokoloff:1998}.

\section{Observational Results\label{sec:obsres}}

\subsection{Measurements of Radio Sources Viewed Through the W4 Complex\label{sec:obs2}}
\begin{figure}[htb!]
\centering
\includegraphics[width=0.6\textwidth]{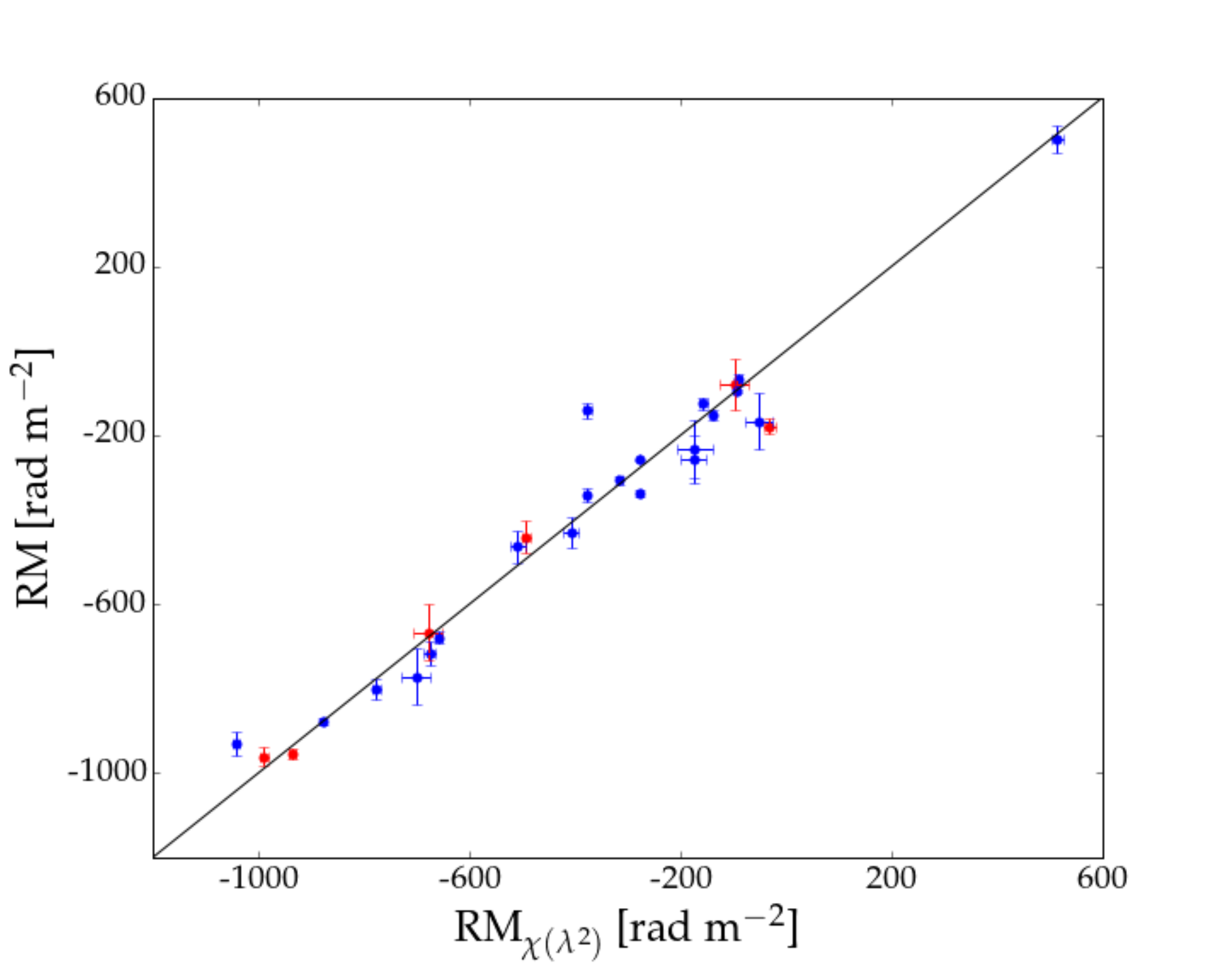}
\caption[]{ Plot of RM values derived from the \chilam~analysis vs the RM Synthesis analysis. The blue markers are the RM from primary component and the red are the secondary component. The straight line represents perfect agreement between the two sets of measurements.}
\label{fig:comprm}
\end{figure}
We measured 27 RM values for 20 lines of sight, including secondary components, through or near to IC 1805. In Table \ref{tab:results}, the first column lists the source name using our naming scheme, and column two gives the component, if the source had multiple resolved components for which we could determine a RM value. 
Columns three and four list the RM value from the least-squares method and the reduced chi-squared value, respectively. Column five lists the RM value determined from the RM Synthesis technique and the associated error (Equation 7 of \citealt{Mao:2010}).

Figure \ref{fig:comprm} shows the agreement between the two techniques for determining the RM. As in \citet{Costa:2016}, we find good agreement between the two techniques, and the good agreement between the results using the two techniques gives us confidence in our RM measurements.

\begin{table}[!hbtp]
\centering
\begin{threeparttable}
\caption{Faraday Rotation Measurement Values through the W4 Complex \label{tab:results}}
\begin{tabular}{ccccccc}
\hline
\multirow{2}{*}{Source} & \multirow{2}{*}{Component} &  RM\tnote{a} & Reduced  & RM\tnote{c} & $\xi$\tnote{d} & $\xi$\tnote{e}\\
 &  & (\radm) & $\chi^{2}$\tnote{b} & (\radm) & (pc) & (pc)\\
\hline 
W4-I1 & a & -277 $\pm$ 1 & 29 & -258 $\pm$ 3 & 29 & 53\\ \hline
\multirow{2}{*}{W4-I2} & a & -1042 $\pm$ 7 & 1.5 & -930 $\pm$ 30 & \multirow{2}{*}{30}& \multirow{2}{*}{26}\\
 & b & -935 $\pm$ 6 & 1.5 & -954 $\pm$ 11 & & \\ \hline 
W4-I3 & a & -876 $\pm$ 2 & 1.9 & -878 $\pm$ 8 & 27 & 16 \\ \hline
\multirow{2}{*}{W4-I4} & a & -139 $\pm$ 3 & 1.9 & -153 $\pm$  12 & \multirow{2}{*}{38} & \multirow{2}{*}{59}\\
& b & -91 $\pm$ 6 & 2 & -68 $\pm$ 15 & & \\ \hline
W4-I6 & a & -990 $\pm$ 8 & 1.3 & -961 $\pm$ 23 & 29 & 25\\ \hline
W4-I8 & a & -276 $\pm$ 2 & 4.4 & -337 $\pm$ 8 & 37& 54\\ \hline
W4-I11 & a & -377 $\pm$ 8 & 12 & -141 $\pm$ 18 & 35 & 41\\ \hline
W4-I12 & a & -315 $\pm$ 4 & 2.8 & -306 $\pm$ 10 &34 & 54\\ \hline
\multirow{2}{*}{W4-I13} & a & -777 $\pm$ 8 & 1.2 & -801 $\pm$ 24 & \multirow{2}{*}{28} & \multirow{2}{*}{36}\\
 & b & -701 $\pm$ 28 & 0.5 & -772 $\pm$ 66 & & \\ \hline
W4-I14 & a & -678 $\pm$ 27 & 0.6 & -666 $\pm$ 66 & 20 & 36\\ \hline
W4-I15 & a & -157 $\pm$ 9 & 0.8 & -124 $\pm$ 14 & 32 & 10\\ \hline
\multirow{2}{*}{W4-I17}& a & -492 $\pm$ 8 & 1.6 & -440 $\pm$ 40 & \multirow{2}{*}{11} & \multirow{2}{*}{31}\\ 
& b & -509 $\pm$ 15 & 0.6 & -464 $\pm$ 40 & & \\ \hline
W4-I18 & a & +514 $\pm$ 12 & 1.1 & +501 $\pm$ 33 & 4 & 22\\ \hline
W4-I19 & a & -407 $\pm$ 14 & 0.3 & -431 $\pm$ 36 & 37 & 34\\ \hline
\multirow{3}{*}{W4-I21}  & a & -53 $\pm$ 26 & 1.4 & -167 $\pm$ 67 & \multirow{3}{*}{17} & \multirow{3}{*}{15}\\ 
 & b & -98 $\pm$ 28 & 0.5 & -79 $\pm$ 62& &  \\ 
 & c & -173 $\pm$ 34 & 0.5 & -232 $\pm$ 70 & & \\ \hline
\multirow{2}{*}{W4-I24}& a & -658 $\pm$ 5 & 1.4 & -678 $\pm$ 14 & \multirow{2}{*}{39} & \multirow{2}{*}{23}\\ 
 & b & -675 $\pm$ 12 & 0.4 & -716 $\pm$ 30 & & \\ \hline
W4-O4 & a & -31 $\pm$ 12 & 24 & -178 $\pm$ 18 &68 &81 \\ \hline
W4-O6 & a & -95 $\pm$ 1 & 3.6 & -96 $\pm$ 4 & 54 & 76\\ \hline
W4-O7 & a & -175 $\pm$ 24 & 0.8 & -256 $\pm$ 56 & 53 & 62\\ \hline
W4-O10 & a & -379 $\pm$ 5 & 1.8 & -343 $\pm$ 16 & 56 & 66\\ \hline

\end{tabular}
\begin{tablenotes}
\item[a] RM value obtained from a least-squares linear fit to $\chi(\lambda^2)$. The errors are 1$\sigma$.
\item[b] Reduced $\chi^{2}$ for the \chilam~fit.
\item[c] Effective RM derived from RM Synthesis.
\item[d] Distance from center of OCl 352.
\item[e] Distance from \citet{Terebey:2003} center.
\end{tablenotes}
\end{threeparttable}
\end{table}


\subsection{Report on Faraday Complexity and Unpolarized Lines of Sight\label{sec:depol}}

In the last paragraph of Section \ref{sec:rmsyn}, we discuss Faraday complexity. If a source is \fsimp, then the RM is equal to a delta function in \fphi~at the Faraday depth. If a source is \fcomp, then the interpretation of the RM is not as straightforward. There is extensive literature (e.g. \citealt{Farnsworth:2011}, \citealt{OSullivan:2012}, \citealt{Anderson:2015}, \citealt{Sun:2015}, \citealt{Purcell:2015}) to understand Faraday complexity.

One indicator of a \fcomp~source is a decreasing fractional polarization, \textit{p} = \textit{P}/\textit{I}, as a function of \lamsq. The ways in which this can arise are discussed at the end of Section \ref{sec:rmsyn}.  Although depolarization does not necessarily lead to a net rotation of the source $\chi$, its presence indicates the potential for a $\chi$ rotation independent of the plasma medium through which the radio waves subsequently propagate.  This could result in an error in our deduced RMs. Nine of the sources, W4-I1, -I3, -I11, -I15, I21b, -O4, -O6, -O7, and  -O10 show a decreasing \textit{p} with increasing \lamsq.

A rough estimate of the potential position angle rotation associated with depolarization may be obtained using the analysis in \citet{Cioffi:1980}. These calculations assume that depolarization arises from Faraday rotation within the synchrotron radiation source, and we can estimate the effect of internal depolarization from the changes in fraction polarization. Given the fractional polarization at the shortest and longest wavelength, we obtain the corresponding polarization angle change from Figure 1 of \citet{Cioffi:1980} and then calculate a RM due to internal depolarization. If the calculated RM due to depolarization (RM$_{depol}$) is larger than the observed RM, then the RM is potentially affected by depolarization. 
W4-I1, -O6, -O7, and -O10 show RM$_{depol}$ \about~ RM$_{obs}$, which indicates that internal depolarization could affect the observed RM. The observed RM of W4-I3 is \about~3 times larger than RM$_{depol}$, so it is not affected by internal depolarization.

Depolarization due to internal Faraday rotation makes predictions for the form of $\chi (\lambda^2)$ which would not have $\chi \propto \lambda^2$ \citep{Cioffi:1980}. For all of the sources mentioned above, we compared the observed behavior of \chilam~to the predicted behavior (Equation 4b of \citealt{Cioffi:1980}). Within the errors, only W4-O7 is consistent with the non-linear behavior of a RM affected by internal depolarization. We interpret this result as meaning that our deduced RM values for most of the sources are not significantly in error due to internal depolarization, and we consider the measurement of depolarization as providing a cautionary flag.

We also considered whether our measurements could have been affected by bandwidth depolarization or beam depolarization. Bandwidth depolarization occurs when the polarization angle varies over frequency averaged bins, $\Delta\nu$. For example in this study, we use values of $\chi$ in 128 MHz wide bins (Section \ref{sec:chilam}) and 4 MHz (Section \ref{sec:rmsyn}). For a center frequency of the lowest frequency bin, we use $\nu_c$ = 4466 MHz, and the relationship between the change in polarization position angle, $\Delta\chi$, is
\begin{equation}
|\Delta\chi| = 2|\textrm{RM}| \; c^2 \; \frac{\Delta\nu}{\nu_c^3},
\label{eq:bandwidth}
\end{equation}
where c is the speed of light. This formula shows that even for $|$RM$|$ = 10$^4$ \radm, which is far larger than any RMs we measure, the Faraday rotation across the band is 0.41 radians.  This is insufficient to cause substantial bandwidth depolarization. Beam depolarization occurs when there are small scale variations of the electron density or the magnetic field within a beam. It is unlikely that the RMs are affected by beam depolarization as the beam at 6 cm for the VLA in C array is \about~5 arcseconds. 
We interpret these RMs as a characteristic value due to the plasma medium (primarily the Galactic ISM) between the source and the observer. In the analysis that follows, we choose the RM values from the RM Synthesis method.


When the data were mapped and inspected, we found that a few sources that had passed our criteria for flux density and compactness to the VLA D array at L band (1.42 GHz) were completely unpolarized. W4-I16 and W4-O8 are not polarized at any frequency, and the RM Synthesis technique does not show significant ($>$ 7$\sigma$) peaks at any $\phi$. Three of the lines of sight, W4-I5, W4-I10, and W4-I22, have no source in the field. Despite appearing to be point sources in the NVSS postage stamps (see Section \ref{sec:obs}), we do not observe a source at these locations, and they may have been clumpy foreground nebular emission that was filtered out during the imaging process.

Subsequent investigations determined that some of the selected sources were previously cataloged ultra compact \HII regions associated with the W3 star formation region. These sources are W4-I7 (W3(OH)-C), W4-I9 (AFGL 333), and W4-I20 (W3(OH)-A) \citep{Feigelson:2008,Navarete:2011,Roman:2015}. The W4-I7 field has no source at the observed $\alpha$ and $\delta$, despite it being identified as W3(OH)-C. We do not observe a source at this location in any frequency bin. W3(OH)-A, however, is observed and is a point source in our maps at all frequencies. Similarly, W4-I9 is detected in each frequency bin and is an extended source. These sources are unpolarized and do not feature in our subsequent analysis. 

\subsection{A Unique Line of Sight Through the W4 Region: LSI +61\ddeg 303\label{sec:HMXB}}
%
%
%
%
%

W4-I19 has a spectrum which is inconsistent with an optically-thin extragalactic radio source. It is linearly polarized, and we measure RM = --431 $\pm$ 36 \radm. Investigation of this source during the data analysis phase revealed that it is not an extragalactic source, although it passed our selection criteria for flux and compactness. W4-I19 is the high mass X-ray binary (HMXB) LSI +61\ddeg303 \citep{Gregory:1979,Bignami:1981}, which is notable for being one of five known gamma ray binary systems \citep{Frail:1991}. This system has been extensively studied, and as a result, much is known about the nature of the compact object \citep{Massi:2004,Massi:2004b,Dubus:2006,Paredes:2007,Massi:2017},  the stellar companion \citep{Casares:2005,Dubus:2006,Paredes:2007}, orbital period \citep{Gregory:2002}, radio structure \citep{Albert:2008}, and radial velocity \citep{Gregory:1979,Lestrade:1999}.  

The spatial location of LSI +61\ddeg303 is important for understanding the RM we determined for this source. \citet{Frail:1991} argue that since signatures of the Perseus arm shock are present in the absorption spectrum to LSI +61\ddeg303 but not the post-shock gas from the Perseus arm, LSI +61\ddeg303 must lie between the two features at a distance of 2.0 $\pm$ 0.2 kpc. They also report that they do not see absorption features due to the IC 1805 ionization front and shock front. The estimated distance to \lsi{} is consistent with distance estimates to OCl 352. The position relative to the nebula has consequences for the interpretation of the RM that we measure. The possibilities are:
\begin{enumerate}
\item LSI +61\ddeg303 is in front of the stellar bubble and \HII region, so it is exterior to a region modified by OCl 352. The RM is then an estimate of the foreground ISM between us and the nebula.
\item If LSI +61\ddeg303 is at the same distance as IC 1805 or slightly behind (greater distance), then the RM is unique among our sources in that it is not affected by Faraday rotation from material in the outer Galaxy.  The RM is then probing at least a part of the Faraday rotating material due to the nebula. 
\end{enumerate}

To further determine the position of \lsi{} with respect to IC 1805, we review the current state of knowledge on the subject from the literature. \citet{Dhawan:2006} observed \lsi{} with the Very Long Baseline Array (VLBA) and report a proper motion of ($\mu_{\alpha}$, $\mu_{\delta}$) = (-0.30 $\pm$ 0.07, -0.26 $\pm$ 0.05) mas yr$^{-1}$. \citet{Aragona:2009} report a radial velocity for \lsi~of $V_{rad}$ = --41.4 $\pm$ 0.6 \kms, which agrees with previous estimates by \citet{Casares:2005}. For OCl 352, \citet{Dambis:2001} estimate the radial velocity to be --41 $\pm$ 3 \kms, and more recent estimates by \citet{Kharchenko:2005} ($V_{rad}$ = --47 $\pm$ 18 \kms) agree within the errors. Both \lsi~and OCl 352 have similar radial velocities, and the proper motion estimates by \citet{Dhawan:2006} indicate that \lsi~is moving similarly on the plane of the sky to OCl 352, which has a proper motion of ($\mu_{\alpha}$, $\mu_{\delta}$) = (--1.0 $\pm$ 0.4, --0.9 $\pm$ 0.4) mas yr$^{-1}$ \citep{Dambis:2001}.

From proper motion and radial velocity estimates, \lsi~appears to be moving in relatively the same direction and speed as OCl 352. Using a distance of 2 kpc to \lsi~and 2.2 kpc to OCl 352, the transverse velocities are \about~3~\kms{} and \about~14 \kms, respectively.  If \lsi~originally belonged to OCl 352, then it is unlikely that it is in front of IC 1805, given that both are moving at the same radial velocity. While \lsi~appears to be outside the obvious shell structure of IC 1805, it is more likely that it is probing material modified by OCl 352. We discuss this possibility further in Section \ref{sec:pdr}. 

If \lsi~did not originate in OCl 352, then it is possible to still be in front of the nebula, despite the similar velocities. In such a case, the RM we obtained for this line of sight is due to the ISM between us and IC 1805. 
The RM value we find for \lsi~is nearly 3 times larger than the background RM, which we discuss in Section \ref{sec:bkgrm}. This would require a magneto-ionic medium between the observer and the nebula capable of producing \about~400 \radm~along this line of sight.  As may be seen from Table \ref{tab:results} and Figure \ref{fig:w4}, other lines of sight near IC 1805, but exterior to the shell, do not have as large of RM values (e.g. W4-O26, -O19, -O7, -I11). It therefore seems most probable that the RM for W-I19 (\lsi) is dominated by plasma in W4.

In summary, there is evidence in the literature that suggests \lsi~may lie within a region modified by OCl 352, particularly if \lsi~did indeed once belong to OCl 352. If this is the case, then the RM we find is unaffected by the ISM in the outer galaxy and is due to the material near IC 1805. 

\section{Results on Faraday Rotation Through the W4 Complex\label{sec:fr}}
\subsection{The Rotation Measure Sky in the Direction of W4}
\begin{figure}[!htb]
\centering
\subfloat[][\label{fig:northb}]{
\includegraphics[width=0.48\textwidth]{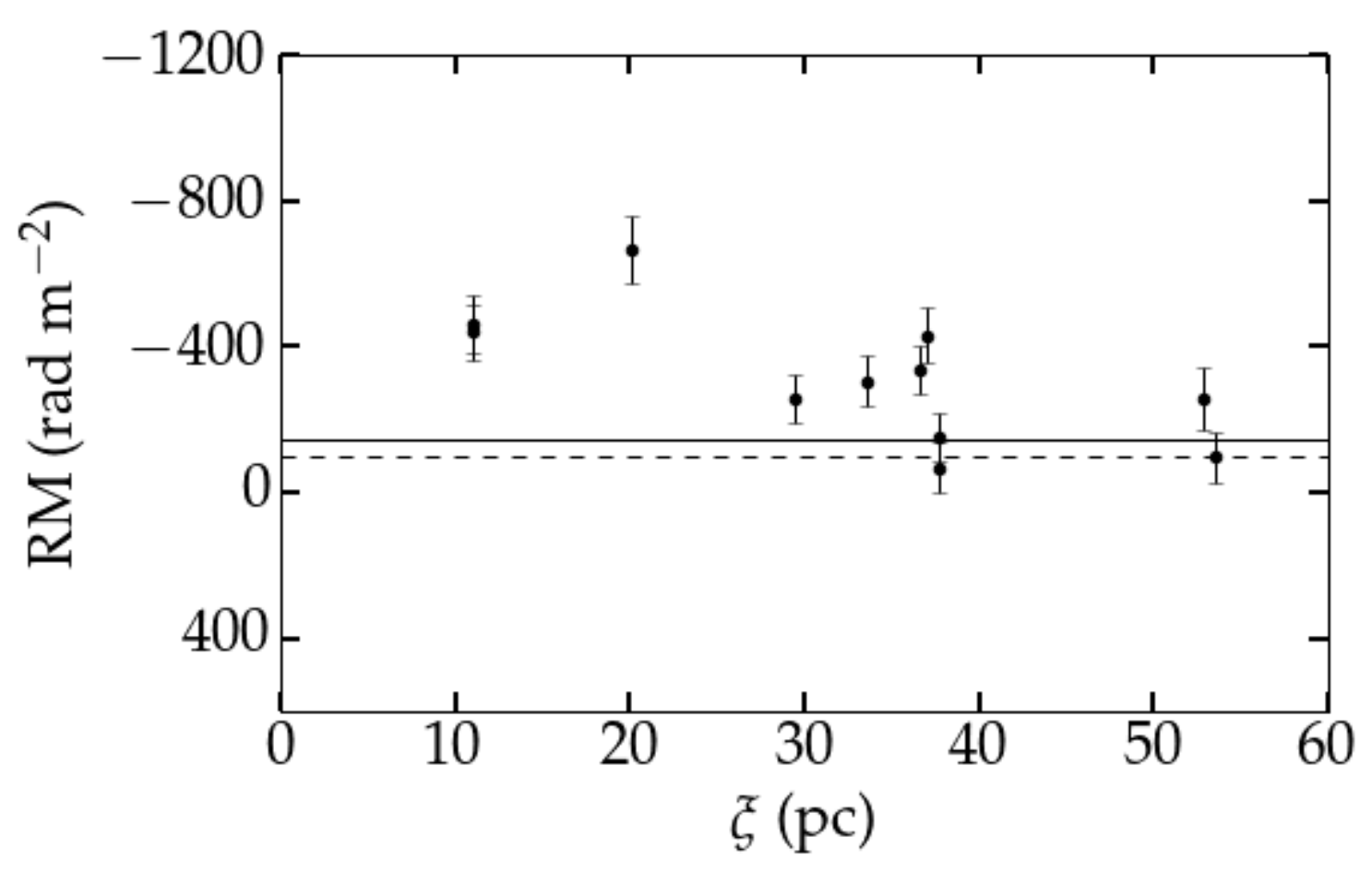}}
\quad
\subfloat[][\label{fig:southb}]{
\includegraphics[width=0.48\textwidth]{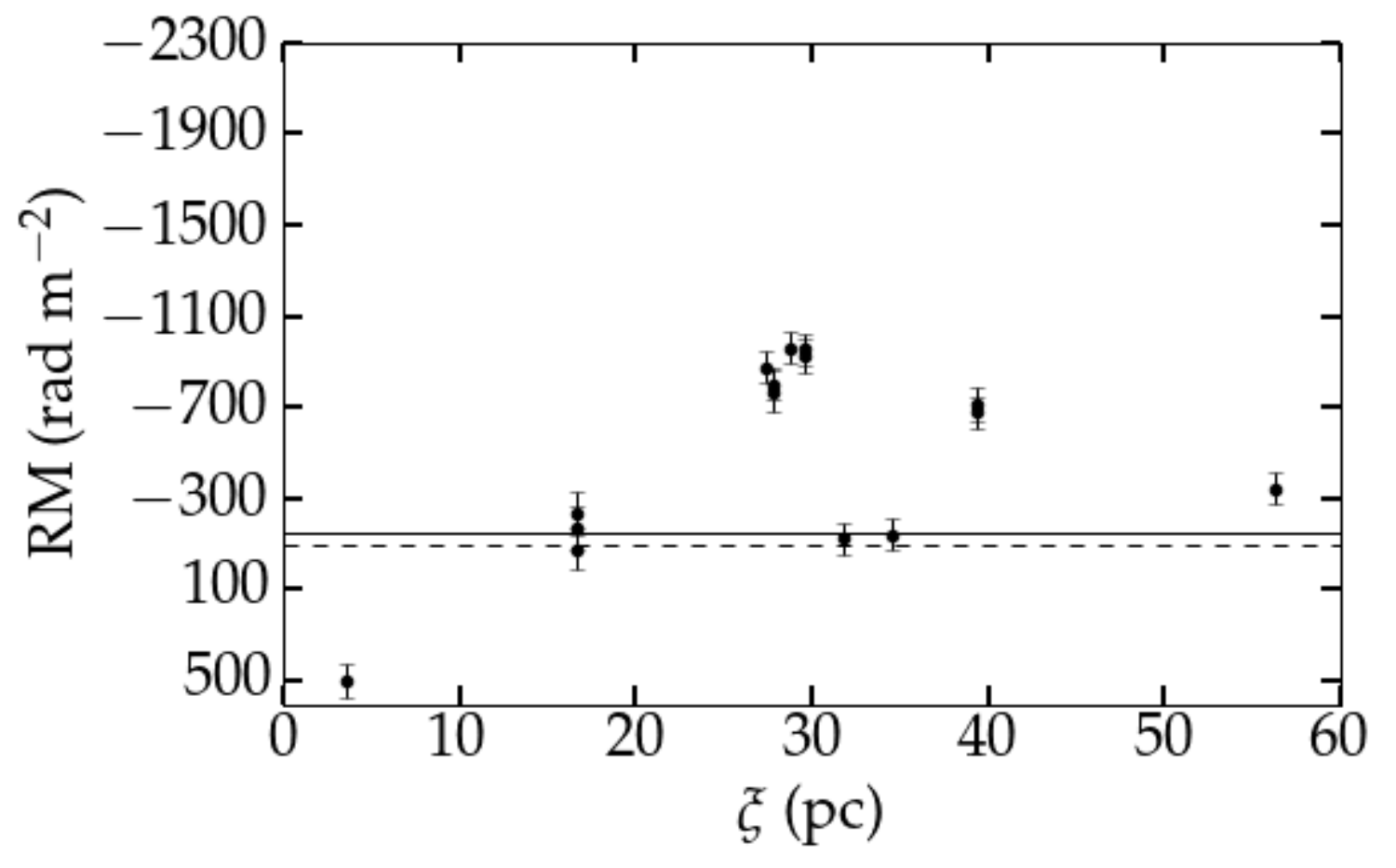}}
\caption[]{Plot of RM vs distance from the center of OCl 352 for (a) lines of sight that pass through the W4 Superbubble and (b) lines of sight near and close to the southern loop. The solid line represents the estimate of the background RM using sources in this study and in the literature, and the dashed line is the predicted background RM from the \citet{vanEck:2011} model of the Galactic magnetic field.}
\label{fig:rmvsxi}
\end{figure}

\citet{Whiting:2009}, \citet{Savage:2013}, and \citet{Costa:2016} compared observations to a model of the ionized shell in which the RM depended only on $\xi$, the impact parameter, or closest approach of a line of sight to the center of the shell. In anticipation of a similar analysis in this study, we show Figures \ref{fig:northb} and \ref{fig:southb}, which plot the RM versus distance from the center of star cluster for the lines of sight through the W4 Superbubble (W4-I1, -I4, -I8, -I12, -I14, -I17, -O4, -O6, and -O7) and the ones through or close to the southern loop (W4-I2, -I3, -I6, -I11, -I13, -I15, -I18, -I19, -I21, -I24, -O10).

In Section \ref{sec:structure}, we discussed the morphology of the region around IC 1805 and made the distinction between the southern latitudes and the northern latitudes, so in the following sections, we address each region near IC 1805 separately.




\subsection{The Galactic Background RM in the Direction of W4\label{sec:bkgrm}}

In \citet{Savage:2013}, we determined the background RM in the vicinity of the Rosette Nebula ($\ell$ $\sim$ 206\ddeg) by finding the median value of the RM for sources outside the obvious shell structure of the Rosette. Determining the background RM near IC 1805 is difficult, however, due to proximity of W3, the W3 molecular cloud, and the W4 Superbubble. Given the morphological difference between the northern and southern parts of IC 1805, we assume that sources south of OCl 352 (\textit{b} $<$ 0.9\ddeg) should be modeled independently of the northern sources, since the W4 Superbubble extends up to \textit{b} $\sim$ 7\ddeg~\citep{West:2007}. The lines of sight north of the star cluster are intersecting the W4 Superbubble and are not probing the RM due to the general ISM independent of IC 1805. Therefore, the only lines of sight that are potentially probing the RM in the vicinity of IC 1805 are those exterior to the shell structure of the southern loop.
\begin{table}[htb!]
\centering
\caption{List of Sources with RM values from Catalogs\label{tab:taylor}}
\begin{threeparttable}
\centering
\begin{tabular}{cccc}
\hline
Source & $\alpha$(J2000) & $\delta$(J2000) & RM\tnote{a} \\
Name & h m s & $^o$ $'$ $''$  &  (rad m$^2$) \\
\hline
W4-O3 & 02 35 43.0 & +63 22 33.0 & --138\tnote{b} $\pm$ 18 \\
W4-O19 & 02 46 23.9 & +61 33 19.9 & --157\tnote{c} $\pm$ 15 \\
W4-O26 & 02 42 32.3 & +60 02 31.0 & +61 $\pm$ 41  \\
W4-O27 & 02 25 48.7 & +59 53 52.0 & --145 $\pm$ 22 \\

\hline
\end{tabular}
\begin{tablenotes}
\item[a] RM values from \citet{Brown:2003} unless otherwise noted.
\item[b] \citet{Taylor:2009} give --75 $\pm$ 9 \radm~for this line of sight.
\item[c] RM value from \citet{Taylor:2009}.
\end{tablenotes}
\end{threeparttable}
\end{table}


If we apply the thick shell model from \citet{Terebey:2003} (see Section \ref{sec:structure} for details), then the lines of sight with RM values exterior to the shell are W4-I2, -I11, and -O10. For the thin shell case, W4-I6, -I13 and -I24 are also exterior sources. The mean RM value for the background using these sources is --554 \radm~and --670 \radm~for the thick or thin shell, respectively. In Table \ref{tab:taylor}, we list RM values from the literature for lines of sight near IC 1805 that we include in our estimate of the background RM. The mean RM value for these sources (excluding W4-O3 for being in the superbubble) is --80 \radm. The sources W4-I2, -I6, -I13, and -I24 are seemingly outside the obvious ionized shell structure; however, they are also the lines of sight for which we measure some of the highest RM values. This is a surprising result, and one we did not observe in the case of the Rosette Nebula. It strongly suggests that the lines of sight to W4-I2, -I6, -I13, and -I24 have RMs that are dominated by the W4 complex, despite the fact that they are outside the obvious ionized shell of IC 1805.  We discuss this further in the next section. 

For the present discussion, we exclude these sources from the estimate of the background. Using W4-I11, -O10, -O19, -O26, and -O27, we find a mean value for the background RM due to the ISM of --145 \radm. While this value is similar in magnitude to the value of the background RM we found in our studies on the Rosette Nebula, we have significantly fewer lines of sight, and only two of the lines of sight were observed in this study. 

Due to a low number of lines of sight exterior to IC 1805, we utilize the model of a Galactic magnetic field by \citet{vanEck:2011} to estimate the background RM due to the ISM. From their Figure 6, they find the Galactic magnetic field is best modeled by an almost purely azimuthal, clockwise field. \citet{vanEck:2011} use their model to predict the RM values in the Galaxy, and in the vicinity of IC 1805, their model predicts RMs of order --100 \radm. Using this as an estimate of the background RM, we find an excess RM due to IC 1805 of +600 to --860 \radm.


\subsection{High Faraday Rotation Through Photodissociation Regions\label{sec:pdr}}
The lines of sight with the highest RM values, W4-I2, -I6, and -I24, appear to be outside the obvious shell of the southern loop. These sources are very near to the bright ionized shell. \citet{Terebey:2003} and \citet{Gray:1999} discuss a halo of ionized gas that surrounds IC 1805, which may be causing the high RM values. \citet{Gray:1999} speculate that the diffuse extended structure is an extended \HII envelope as suggested by \citet{Anantharamaiah:1985}. Another possibility is that these high RMs arise in the PDR surrounding the IC 1805 \HII region. 

PDRs are the regions between ionized gas, which is fully ionized by photons with \textit{h$\nu$} $>$ 13.6 eV, and neutral or molecular material. PDRs can be partially ionized and heated by far-ultraviolet photons (6 eV $<$ \textit{h$\nu$} $<$ 13.6 eV) \citep{Tielens:1985,Hollenbach:1999}. Typically, the PDR consists of neutral hydrogen, ionized carbon, and neutral oxygen nearest to the ionization front, and with increasing distance, molecular species (e.g., CO, H$_2$, and O$_2$) dominate the chemical composition of a PDR \citep{Hollenbach:1999}. One tracer of PDRs is polycyclic aromatic hydrocarbon (PAH) emission at infrared (IR) wavelengths. \citet{Churchwell:2006} identify more than 300 bubbles at IR wavelengths in the Galactic Legacy Infrared Mid-Plane Survey Extraordinaire (GLIMPSE), and 25$\%$ of these bubbles coincide with known \HII regions. \citet{Watson:2008} examine three bubbles from the \citet{Churchwell:2006} catalog with the \textit{Spitzer} Infrared Array Camera (IRAC) bands 4.5, 5.8, and 8.0 \micron~and the 24 $\mu$m band from the \textit{Spitzer} Multiband Imaging Photometer (MIPS) to determine the extent of the PDR around three young \HII regions. One of their main results is that the 8 \micron~emission, which is due to PAHs, encloses the 24 \micron~emission, which traces hot dust. \citet{Kerton:2013} discuss similar observations near the W 39 \HII region. \citet{Watson:2008} use ratios between the 4.5, 5.8, and 8.0 \micron~bands to determine the extent of the PDRs, as the 4.5 \micron~emission does not include PAHs but the 5.8 and 8.0 \micron~bands do (see their Section 1 for details).

To determine the presence and extend of a potential PDR around IC 1805, we analyze Wide-field Infrared Survey Explorer (WISE) data from the IPAC All-Sky Data Release\footnote{http://wise2.ipac.caltech.edu/docs/release/allsky/} at 3.6, 4.6, 12, and 22 $\mu$m. The 4.6 \mircon~WISE bands is similar in bandwidth and center frequency to the IRAC 4.5 \micron~band, and the WISE 22 \micron~band is also similar to the MIPS 24 \micron~band \citep{Anderson:2014}. The 12 \micron~WISE band does not overlap with the 8.0 \micron~band of IRAC, but the WISE band traces PAH emission at 11.2  and 12.7 \micron. \citet{Anderson:2012} note, however, that the 12 \micron~flux is on average lower than the 8.0 \micron~IRAC band, which is most likely due to the WISE band sampling different wavelengths of PAH emission instead of the 7.7 and 8.6 \micron~PAH emission in the IRAC band.

\begin{figure}[htbp!]
\centering

\includegraphics[width=0.95\textwidth]{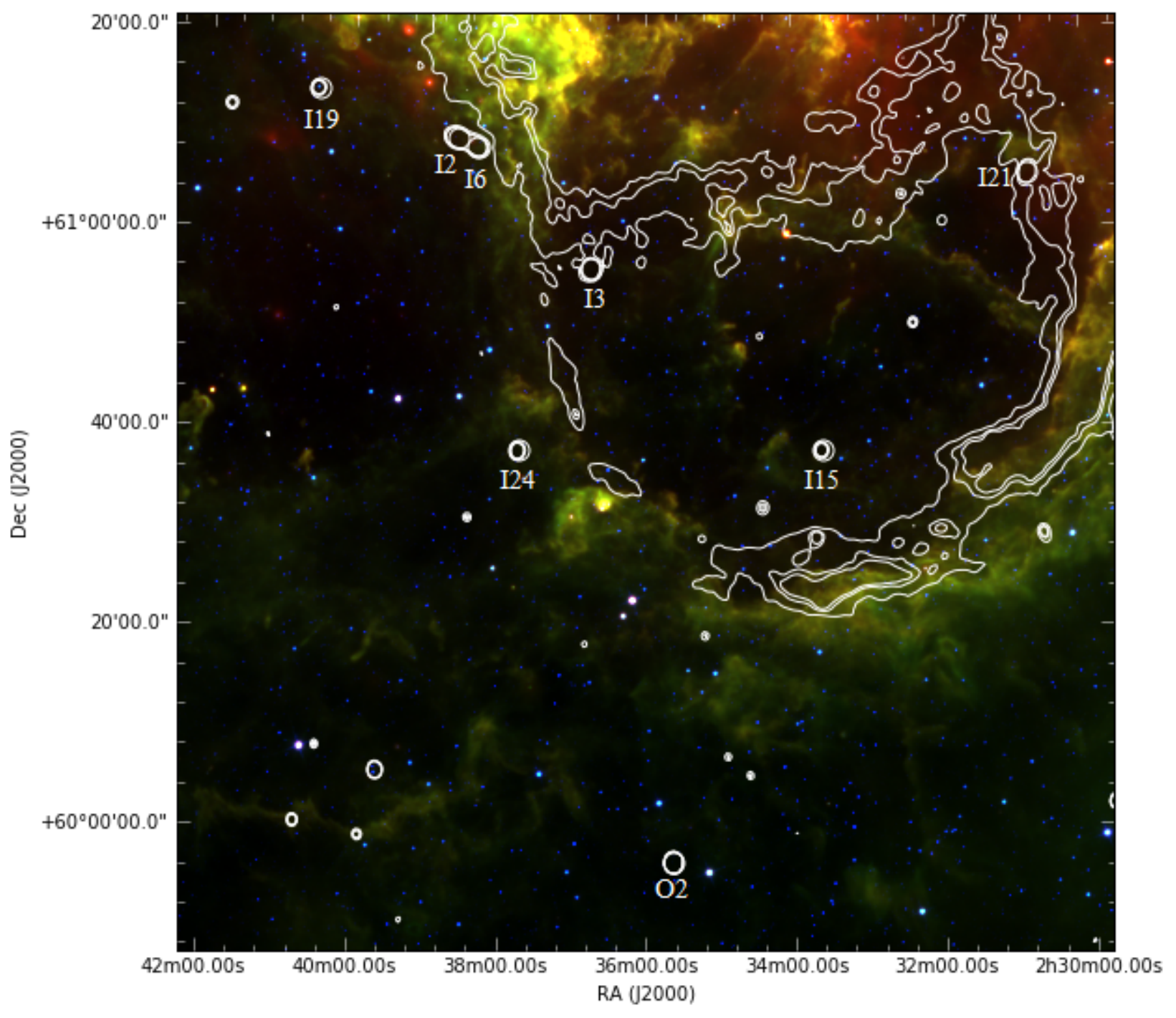}

\caption[WISE image of W4]{Inset from Figure \ref{fig:w4}. A RGB image of archive WISE data at 4.6 \micron~(blue), 12 \micron~(green), and 22 \micron~(red) with CGPS contours at 8.5, 9.5, and 10 K in white. The lines of sight from the present study are shown with circles and are labeled according to Table \ref{tab:sources}.}
\label{fig:wise}
\end{figure}

Figure \ref{fig:wise} is a RGB image of the southern loop of IC 1805 at 4.6 \micron~(blue), 12 \micron~(green), and 22 $\mu$m (red). The 1.42 GHz radio continuum emission is shown in the white contours at 8.5, 9.5, and 10 K, and the lines of sight that intersect this region are labeled as well. Similar to the results of \citet{Watson:2008}, the majority of the 22 \micron~emission is located inside the bubble. The radio contours trace the ionized shell of the \HII region, which show a patchy ionized shell. Outside of the radio contours, there is a shell of 12 \micron~(green) PAH emission that encloses the 22 \micron~emission as well.  In the northeastern portion of the image, there is extended 22 \micron~(hot dust) emission, which is spatially coincident with a CO clump \citep{Lagrois:2009}.

The PDR model predicts the presence of neutral hydrogen and molecular CO (see Figure 3 of \citealt{Hollenbach:1999}) at increasing distance from the exciting star cluster. Figure 1 of \citet{Sato:1990} and Figure 2 of \citet{Hasegawa:1983} show \hi contours in the vicinity of IC 1805, and the \hi emission appears to completely enclose the southern loop except 
near 135.5\ddeg{} $\leq$ $\ell$ $\leq$ 136\ddeg, 0.2\ddeg $\leq$ $b$ $\leq$ 0.9\ddeg.
\citet{Braunsfurth:1983} report \hi emission near IC 1805, and he notes that the hole could be due to cold \hi gas or the lack of gas if the winds have sufficiently swept the material away or ionized it. Figure 6 of \citet{Digel:1996} shows the CO emission, with the W3 molecular cloud on the western side of IC 1805, CO emission along the southern loop of IC 1805, and the molecular material associated with the W5 ($\ell$ = 137.1, $b$ = +0.89) \HII region on the eastern side of IC 1805.  We interpret the WISE data, the radio contours, and the CO and \hi maps as a patchy ionized shell surrounded by a PDR.

If there is a PDR surrounding IC 1805, then the highest RM values from our data set, RM = --954 \radm{} and --961 \radm{} for W4-I2 and -I6, respectively, lie outside the ionized shell of the \HII region and in the PDR. Similarly, the sources W4-I19 and -I24 are also outside the radio continuum contours but appear to be within the 12 \micron~(green) emission. This is a surprising result compared with our results from the Rosette Nebula, where we found the highest RM values for lines of sight that pass through the ionized shell.  \citet{Gray:1999} note zones of depolarization near the southern portion of IC 1805, which require RMs on order 10$^3$ \radm, and spatial RM gradients. The RMs for W4-I2 and -I6 are on this order, but those for W4-I24 and -I19 are not, and we do not find that these lines of sight are affect by depolarization. W4-I24 has two components for which we measure RMs, and the components are separated by \about~18 arcseconds. The $\Delta$RM, which is the difference in RM between the two components is 38 \radm, which is not a large change in the RM and is consistent within the errors.

The presence of the PDR is complicated, however, by the extended diffusion ionized emission reported by \citet{Terebey:2003} and \citet{Gray:1999}.  At lower contours, the high RM sources do lie within the radio continuum emission. To fully understand the presence and extent of a PDR or an extended \HII envelope, observations of radio recombination lines on the eastern side of IC 1805 would clarify the structure as well as observations of other tracers of PDRs (e.g., fine structure lines of C and C$^+$, H$_2$, and CO). It may be the case that the ionized shell is patchy along the shell wall, which allows photons $>$ 13.6 eV to escape the shell at places, but the shell is sufficiently ionization-bounded at other places such that a PDR can form.


\subsection{Faraday Rotation Through the Cavity and Shell of the Stellar Bubble}

There are four lines of sight through the cavity of the stellar bubble, assuming an inner radius from the \citet{Terebey:2003} model. The sources W4-I3, -I15, -I18, and -I21 are through the cavity, and including multiple components, we find 6 RM values. W4-I3 has a high RM (--878 $\pm$ 18 \radm), and W4-I15 and -I21 have comparatively low RM values ( --79 to --232 \radm). Examination of Figures \ref{fig:w4} and \ref{fig:wise} does not reveal enhanced emission near W4-I3 in comparison to W4-I21. W4-I15, however, is in a region of relatively low emission, which may explain why W4-I15 has a RM value at least 4 times smaller than W4-I3.

W4-I13 is outside the shell, assuming a shell radius from either \citet{Terebey:2003} model. From Figure \ref{fig:w4}, it does appear to be outside the ionized shell. However W4-I13 is within a 8.5 K contour on the 1.42 GHz radio continuum map, which may indicate that it is probing the ionized shell. We find a high RM for both components of this source, which is similar to the RM values for W4-I2 and -I6.


Across IC 1805, we observe negative RM values for all lines of sight except one: W4-I18, which is 5.6 arcmin (4 pc) from the center of the star cluster. The absolute value of the RM for W4-I18 is also large (+501 $\pm$ 33 \radm), indicating a large change in RM along this line of sight relative to other lines of sight in this part of the sky. This line of sight is probing the space close to the massive O and B stars responsible for IC 1805. In the \citet{Weaver:1977} model for a stellar bubble, the hypersonic stellar wind dominates the region between the star responsible for the bubble and the inner termination shock. Equation (12) of \citet{Weaver:1977} states that the distance of the inner shock, $R_t$ is 

\begin{equation}
R_{\textrm{t}} = 0.90\; \alpha^{3/2} \left(\frac{1}{\rho_0}\frac{\textrm{d}M_w}{\textrm{d}t}\right)^{3/10} \ V_w^{1/10} \ t^{2/5},
\label{eq:weaver}
\end{equation}
where $\alpha$ is a constant equal to 0.88, $\rho_0$ is the mass density in the external ISM, d$M_w$/d$t$ is the mass loss rate, $V_w$ is the terminal wind speed, and $t$ is time. For a rough estimate of the inner shock distance, we utilize general stellar parameters for OCl 352 of d$M_w/dt$ = 10$^{-5}$ \MLu, $t$ = 10$^6$ yr,  and V$_w$ = 2200 \kms (see Section \ref{sec:structure} or Table \ref{tab:stellarpar}). From the discussion in Section \ref{sec:ferriere}, we adopt $n_0$ = 4.5 \cm~for $\rho_0$ = $n_0 m_p$, where $m_p$ is the mass of a proton. With these values in the appropriate SI units, $R_t$ $\sim$ 6 pc. It is possible that the line of sight to W4-I18 passes inside the inner shock, and the large, positive RM is due to material modified by the hypersonic stellar wind and not the shocked interstellar material. 
Because the inner shock is interior to the contact discontinuity between the stellar wind and the ambient ISM, the magnetic field close to the star cluster may be oriented in any direction relative to the exterior (upstream) field. With a positive value of the RM for W4-I18, the line of sight component of the field points toward us while the remaining lines of sight in the cavity are negative, meaning \blos~points away.

\subsection{Low Rotation Measure Values Through the W4 Superbubble}
North of IC 1805 is the W4 Superbubble, which is an extended ``egg-shaped'' structure closed at $b$ $\sim$ 7\ddeg{}  \citep{West:2007}. \citet{Basu:1999} utilize an H$\alpha$ map to define the shape, which would include the southern loop (134\ddeg $<$  $\ell$ $<$ 136\ddeg, \textit{b} $<$ 0.5\ddeg); \citet{Normandeau:1996} examine the \hi distribution, however, and place the base of the structure at OCl 352. Similarly, \citet{West:2007} place an offset bottom of the ``egg'' at OCl 352. The southern loop of IC 1805 is seemingly sufficiently different from the northern latitudes, as it is often not included in the discussion of the W4 Superbubble in spite of the fact that OCl 352 is thought to be responsible for the formation of both structures \citep{Terebey:2003,West:2007}.

Nine lines of sight in the present study are north of OCl 352 in the W4 Superbubble. These sources are W4-I1, -I4, -I8, -I12, -I14, -I17, -O4, -O6, and -O7, and they have a mean RM of --293 \radm~and a standard deviation of 178 \radm. Of these sources, W4-I14 and -I17 have the largest RM values, --666 \radm~and --460 \radm, respectively, and they are close to OCl 352, with distances of 31 arcminutes (20 pc) and 17 arcminutes (11 pc), respectively. As discussed in Section \ref{sec:structure}, \citet{Lagrois:2009} argue that the ionized ``v'' structure north of OCl 352 is part of the bubble wall and not a cap to southern loop structure, but examination of Figure \ref{fig:w4} suggests that the bubble walls are denser, or thicker, at latitudes $<$ 1.5\ddeg~than higher latitudes, which may explain the larger RM associated with W4-I14 and -I17. The remaining lines of sight, however, in the W4 Superbubble have some of the lowest RM values in the data set and are consistently lower RM values than the lines of sight through the PDR.

At higher latitudes, \citet{Gao:2015} modeled the polarized emission and applied a Faraday screen model to the W4 Superbubble. They report RMs on the western side of W4 ($\ell$ $\sim$ 132\ddeg, $b$ $\sim$ 4.8\ddeg) between --70 and --300 \radm~and $\sim$ +55 \radm~for the eastern shell ($\ell$ $\sim$ 136\ddeg, $b$ $\sim$ 7\ddeg). \citet{Gao:2015} argue that since W4 is tilted at an angle towards the observer \citep{Normandeau:1997}, a change in the sign of the RM is consistent with a scenario in which the superbubble lifts up a clockwise running Galactic magnetic field \citep{Han:2006} out of the Galactic plane. The magnetic field would go up the eastern side of the superbubble and then down the western side, resulting in the field being pointed toward the observer in the east and away from the observer in the west. While the lines of sight reported in this paper are at $b$ $<$ 2\ddeg, we find a similar range of RM values as reported by \citet{Gao:2015} for the western side. However, we measure RM values 3 -- 4.5 times higher on the eastern side, and we do not observe a sign reversal on the eastern side as suggested by \citet{Gao:2015}.  

\citet{West:2007} report positive values of the magnetic field for the western side from a change in polarization position angle of $\sim$ 60\ddeg~at 21 cm, which gives a RM value on order of 20 \radm. We do not observe RM values this low for any of our lines of sight through the northern latitudes. Our lines of sight, however, do not probe the same regions as the \citet{West:2007} and \citet{Gao:2015} studies.

The line of sight W4-I4 is arguably within the W4 Superbubble; however, it is also \about~8 arcmin (5 pc) on the sky from W3-North (G133.8 +1.4), which is a star forming region within W3. W4-I4 has two components, separated by 15 arcsec (0.2 pc), and a difference in RM values between the two components of $\Delta$RM = 85 \radm. The RM values for both components are low (--153 \radm~and --68 \radm) despite being in the superbubble and near to W3, which may have variable but potentially large magnetic fields \citep{vanderWerf:1990,Roberts:1993} (see Section \ref{sec:structure}).

\section{Models for the Structure of the \HII region and Stellar Bubble\label{sec:models}}

\subsection{\citet{Whiting:2009} Model of the Rotation Measure in the Shell of a Magnetized Bubble\label{sec:whitingmod}}

\citet{Whiting:2009} developed a simple analytical shell model intended to represent the Faraday rotation due to a \citet{Weaver:1977} solution for a wind-blown bubble. We employed this model in \citet{Savage:2013} and \citet{Costa:2016} to model the magnitude of the RM in the shell of the Rosette Nebula as a function of distance from the exciting star cluster. Figure 6 of \citet{Whiting:2009} and their Section 5.1 give the details of the model, and Sections 4.1 of \citet{Savage:2013} and 5 of \citet{Costa:2016} describe the application of the model to the Rosette Nebula. This model takes as inputs the general interstellar magnetic field ($\textbf{B}$) in $\mu$G, the inner ($R_1$) and outer ($R_0$) radii of the shell in parsecs, and the electron density in the shell, $n_e$ (\cm). $R_0$ represents the shock between the ambient ISM and the shocked, compressed ISM, and $R_1$ separates the shocked ISM from the hot, diffuse stellar wind in the cavity. Only the component of the ambient interstellar magnetic field that is perpendicular to the shock normal is amplified by the density compression ratio, X. The resulting expression for the RM through the shell is
\begin{equation}
\textrm{RM}=C\, n_{e}\, L(\xi)\, B_{0z} \left(1+(X-1)\left(\frac{\xi}{R_{0}}\right)^{2}\right),
\label{eq:rmmodelW}
\end{equation}
where L($\xi$) is the cord length through the shell in parsecs (see Equation 10 in \citealt{Whiting:2009} or Equation 6 in \citealt{Costa:2016}), and $B_{0z}$ is the z-component of \textbf{B$_0$}, the magnetic field in the ISM. If $n_e$ has units of cm$^{-3}$, $B_{0z}$ is in $\mu$G, and $L$ is in parsecs, $C=0.81$ (see Equation \ref{eq:rmprat}). $B_{0z}$ is at an angle $\Theta$ with respect to the LOS and is written as

\begin{equation}
B_{0z}=B_{0}\cos{\Theta}.
\end{equation}
In our previous work, we presented two cases for the behavior of the magnetic field in the shell. The first is that the magnetic field is amplified by a factor of 4 in the shell. The second case, in which there is not an amplification of the magnetic field in the shell, sets X = 1. Equation \ref{eq:rmmodelW} then simplifies to

\begin{equation}
\textrm{RM}(\xi)=0.81\, n_{e} \, L(\xi) \, B_{0z}.
\label{eq:rmmodelH}
\end{equation}
In \citet{Costa:2016}, we employed a Bayesian analysis to determine which of the two models better reproduces the observed dependence of the RM as a function of distance. We found that neither model was strongly favored in the case of the Rosette. The model given in Equation (\ref{eq:rmmodelW}) is subject to the criticism that it applies shock jump conditions for \textbf{B} over a large volume of a shell, and that the outer radius of an observed \HII region need not be the outer shock of a Weaver bubble (see remarks in Section 5.1.1 of \citealt{Costa:2016}). It is worth including this model, however, in our analysis of IC 1805 for completeness and in order to compare our results to those of the Rosette Nebula.

In Section \ref{sec:structure}, we discussed the the structure of IC 1805, and we present evidence from the literature that north of OCl 352 is part of the W4 Superbubble. Thus, lines of sight north of OCl 352 may have different model parameters for the shell radii and electron density than the southern loop.  For the southern latitudes, we utilize the \citet{Terebey:2003} thin and thick shell values for the shell radii and electron density. The remaining parameters in Equation (\ref{eq:rmmodelH}) are $B_0$ and $\Theta$.  As in \citet{Savage:2013} and \citet{Costa:2016}, we adopt $B_0$ = 4 $\mu$G for the general Galactic field in front of the \HII region. The angle $\Theta$ is calculated as follows. 
Assuming a distance of 8.5 kpc to the Galactic center, a distance to OCl 352 of 2.2 kpc, and given a Galactic longitude of 135\ddeg, the angle between the line of sight and an azimuthal magnetic field is $\Theta$ = 55\ddeg. We discuss our comparison of this model with the data in Section \ref{sec:results}.

\begin{figure}[htb!]
\centering

\subfloat[\label{fig:southshell-thick}]{
\includegraphics[width=0.48\textwidth]{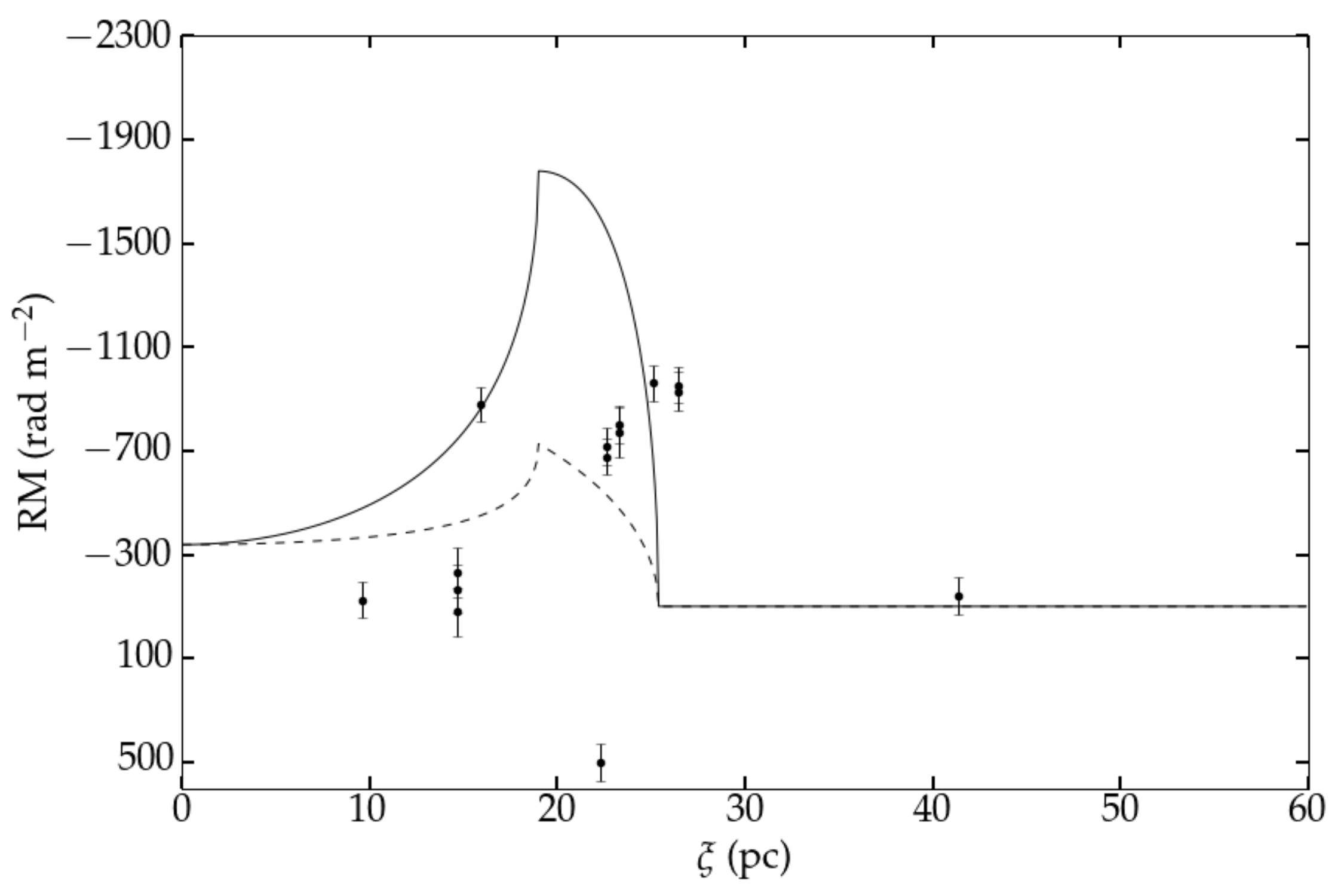}}
\quad
\subfloat[\label{fig:southshell-thin}]{
\includegraphics[width=0.48\textwidth]{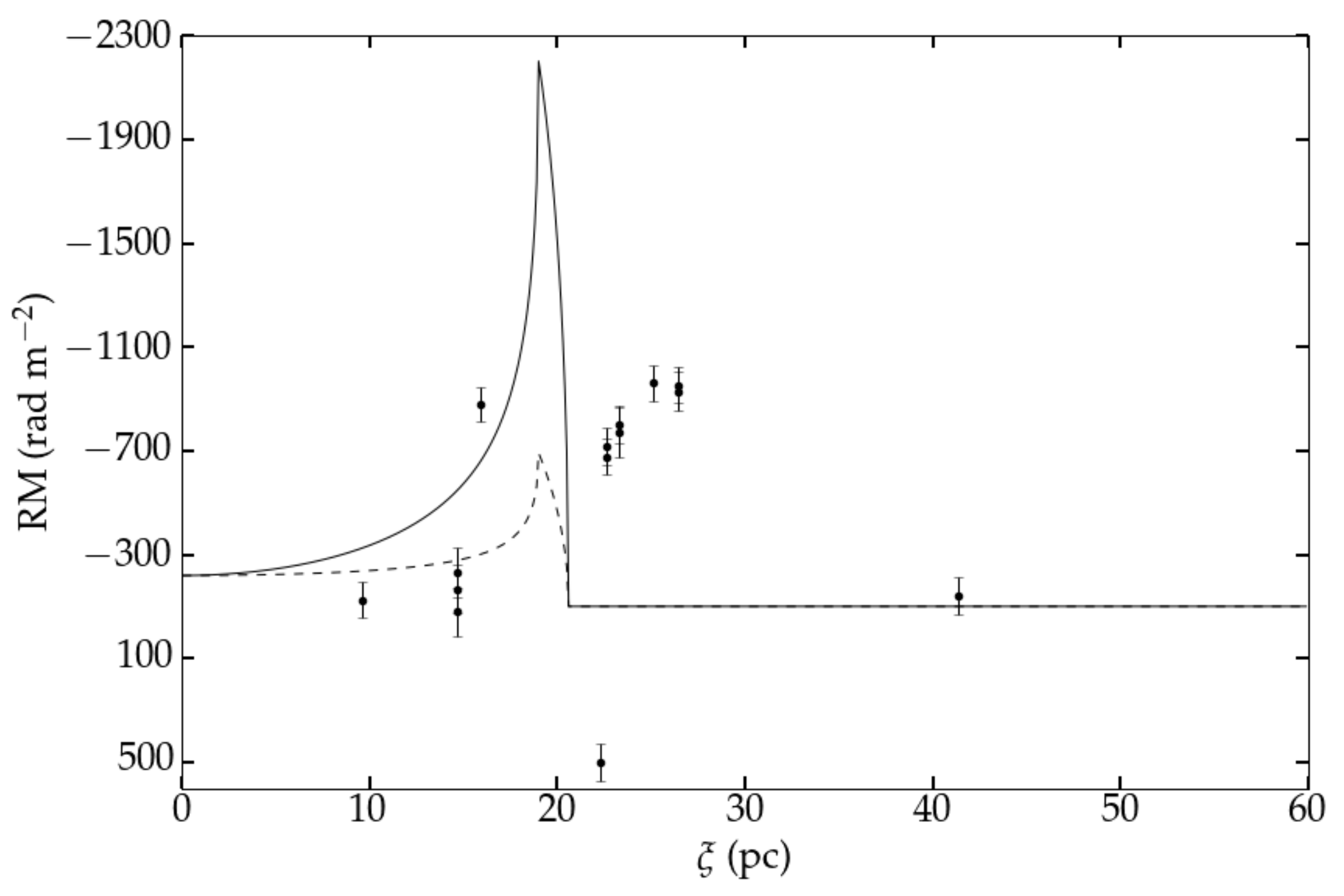}}
\caption[]{ Plots of LOS versus distance with the \citet{Whiting:2009} model for X = 4 (solid) and X = 1 (dashed) for $b$ $<$ +0.9\ddeg{} using (a) the thick shell and (b) the thin shell parameters from \citet{Terebey:2003}. The background RM = --100 \radm~from the \citet{vanEck:2011} model. The errors on the RM values include the measurement errors and an expected deviation of \about{} 67 \radm{} from the \citet{vanEck:2011} model.  See Table \ref{tab:model} for model parameters.}
\label{fig:shell}
\end{figure}

%
%
%
%
%
%
\begin{table}[hbtp]
\centering
\begin{threeparttable}
\caption{Model Parameters \label{tab:model}}
\begin{tabular}{ccccccc}
\hline
\multicolumn{7}{c}{\citet{Whiting:2009} Model} \\
\hline
Center\tnote{a}\phantom{2} & R$_i$  & $R_o$  & $n_e$  & X\tnote{b} & $\Theta$ & Figure \\
(\ddeg) & (pc) & (pc) & (\cm) & & (\ddeg) & \\



(135, +0.42) & 19 & 25 & 10  & 1, 4 & 55 & \ref{fig:southshell-thick} \\

(135, +0.42) & 19 & 21 & 20  & 1, 4 & 55 & \ref{fig:southshell-thin} \\

 \hline \hline
\multicolumn{7}{c}{\citet{Ferriere:1991} Model} \\

\hline
 Center & $\Delta R$ & $R_s$  & $n_s$ & $\epsilon$ &     $\Theta$ &Figure \\
(\ddeg) & (pc) & (pc) & (\cm) & & (\ddeg) & \\
 (135, +0.42) & 6 & 25 & 10 & 0.25 & 55 & \ref{fig:steve2}\\
\hline

\end{tabular}
\begin{tablenotes}
\item[a] Position of model center in Galactic coordinates in the format of ($\ell$, $b$).
\item[b] The model uses either X = 1 or X = 4.

\end{tablenotes}
\end{threeparttable}
\end{table}

\subsection{Analytical Approximation to Magnetized Bubbles of \citet{Ferriere:1991}\label{sec:ferriere}} 

\begin{figure}[htb!]
\centering
\includegraphics[width=0.6\textwidth]{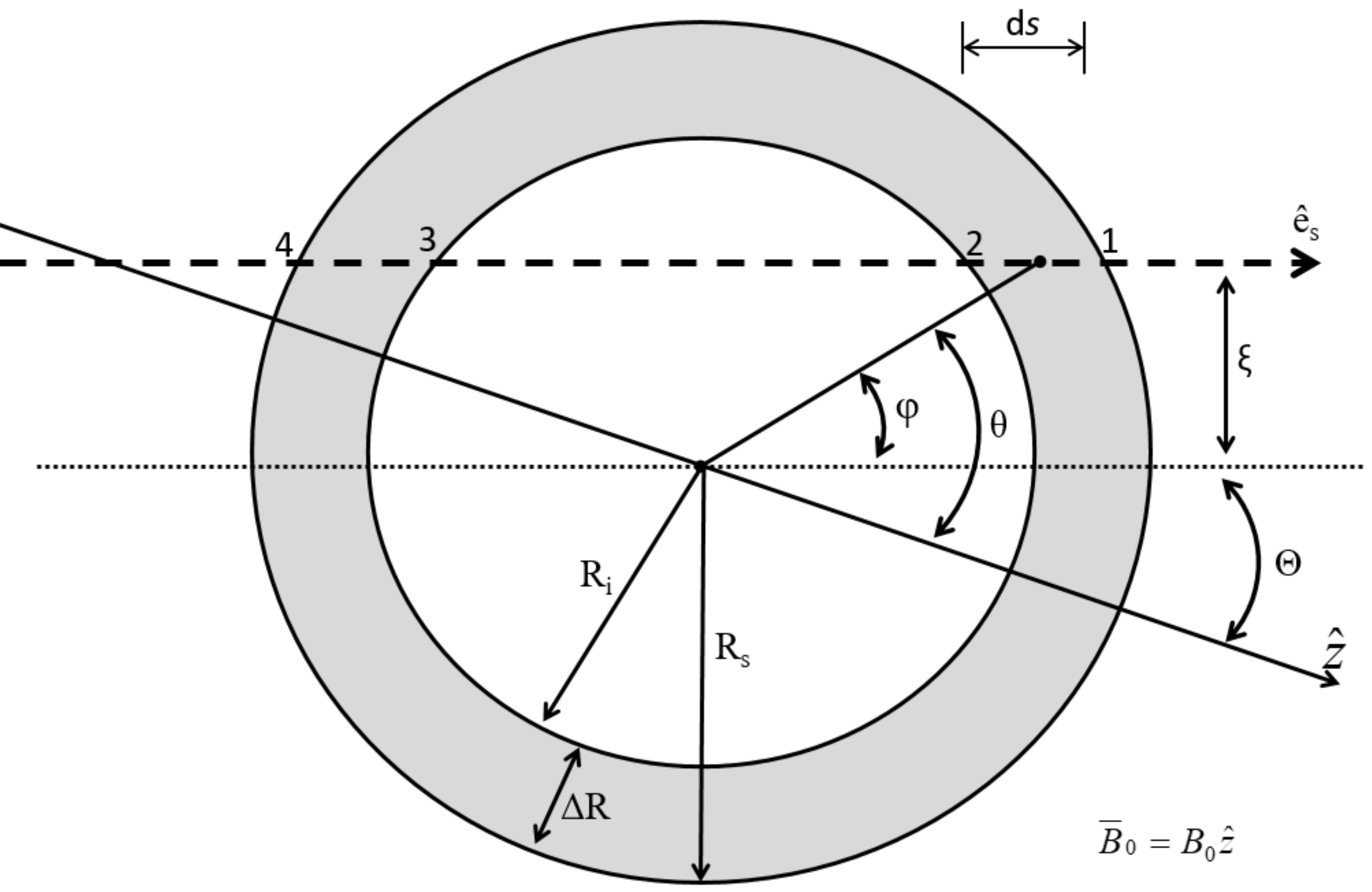}
\caption{Illustration of a simplified version of the shell and cavity produced by a stellar wind, as discussed by \citet{Ferriere:1991}.  The z direction is that of the interstellar magnetic field, and $\Theta$ is the angle between the magnetic field and the line of sight. A line of sight passes at a closest distance $\xi$ from the center of the cavity (the ``impact parameter''). Other parameters in the figure are defined in the text. A Faraday rotation measurement is along a line of sight offset a linear distance $\xi$ from the center of the bubble.  The quantity d$s$ represents an incremental spatial interval along the LOS. }
\label{fig:steve1}
\end{figure} 

\citet{Ferriere:1991} presented a semi-analytic discussion of the evolution of a stellar bubble in a magnetized interstellar medium.  The theoretical object discussed by \citet{Ferriere:1991} could describe a shock wave produced by a supernova explosion or energy input due to a stellar wind.  The main features of the model were an outer boundary (e.g. outer shock) which was the first interface between the undisturbed ISM and the bubble, and an inner contact discontinuity between ISM material, albeit modified by the bubble, and matter that originated from the central star or star cluster.  

The main feature of the model is that plasma passing through the outer boundary is concentrated in a region between the outer boundary and the contact discontinuity. In what follows, we will refer to this region as the shell of the bubble.  The equation of continuity then indicates that there will be higher plasma density in the shell, and the law of magnetic flux conservation indicates that there will be an increase in the strength of the magnetic field in the shell relative to the general ISM field. \citet{Ferriere:1991} were interested in the structure of the bubble, and their results have been corroborated by the fully numerical studies of \citet{Stil:2009}.  However, \citet{Ferriere:1991} did not calculate the Faraday rotation measure through their model for diagnostic purposes.  \citet{Stil:2009} explicitly considered the model RMs from their calculations, but only for a couple of cases and for two values of bubble orientation.  It is our goal in this section to use a simplified, fully analytic approximation of the results of \citet{Ferriere:1991}, that permits RM profiles RM($\xi$) for a wide range of bubble parameters and orientation with respect to the LOS.    

The geometry of the bubble is shown in Figure \ref{fig:steve1}, which is an adaptation of, and approximation to Figure 1 and Figure 4 from \citet{Ferriere:1991}.  An important feature of Figure \ref{fig:steve1}, not present in \citet{Ferriere:1991}, is the orientation of the line of sight at an angle $\Theta$ with respect to the ISM magnetic field at the position of the bubble, and the impact parameter $\xi$ indicating the separation of the LOS from the center of the bubble. The region interior to the contact discontinuity is referred to as the cavity, and for the purposes of our discussion will be considered a vacuum. Another important shell parameter is the thickness $\Delta R$ $\equiv$ $R_s$ - $R_i$, where $R_s$ and  $R_i$ are the outer radius of the bubble and the radius of the contact discontinuity, respectively (see Figure \ref{fig:steve1}).  We also define and use the dimensionless shell thickness
\begin{equation}
  \epsilon \equiv \frac{\Delta R}{R_s}
\label{eq:steve8}
\end{equation}
A major simplification that we adopt, based on an approximation of the results of \citet{Ferriere:1991}, is that the magnetic field in the shell ($\textbf{B}_s$) is entirely in the azimuthal direction, and that we ignore radial variations within the shell, i.e. 
\begin{equation}
\textbf{B}_s(r, \theta) \equiv \pm B_s(\theta) \hat{e}_{\theta}
\label{eq:steve9}
\end{equation}
where $\hat{e}_{\theta}$ is a unit vector in the azimuthal direction, and the $\pm$ is selected by the polarity of the interstellar field at the bubble. 

We need expressions for the electron density and vector magnetic field within the shell, as well as the geometry of the line of sight. The most important aspect of the \citet{Ferriere:1991} theory is the conservation of magnetic flux as the magnetic field in the external medium is swept up and accumulated in the shell.  This results in the azimuthal component of the magnetic field increasing as $\theta$ increases from $0$ to $\frac{\pi}{2}$, as given by Equation (40) of  \citet{Ferriere:1991}.
In  \citet{Ferriere:1991} the shell thickness also depends on $\theta$ (Equation 46 of \citealt{Ferriere:1991}), and as a consequence, so does the plasma density in the shell $n_s$ (Equation 38 of that paper).

In the initial version of this paper, we calculated the RM through model bubbles in which $B_s$, $n_s$, and $\Delta$ R all varied with $\theta$ as prescribed by \citet{Ferriere:1991}.  These calculations utilized an approximate form for lines of sight that intersected the bubble in two segments (passing through the central cavity between), and a form that contained a numerically-evaluated expression for lines of sight that remained within the shell from ingress to egress.  The algebraic distinction between these 2 cases is discussed below (Sections \ref{sec:walls} and \ref{sec:allshell}).  These expressions for RM($\xi$), including a comparison with our RM measurements, are given in \citet{Costa:2018phd}. 
After examining the results of these calculations, it was decided to simplify our bubble model to that of a spherical shell with constant $\epsilon$. The motivation for this suggestion was the very limited success of the more general model in representing our data, which did not justify the extensive algebraic presentation and non-compact expressions that resulted. The calculations with the approximation of constant $\epsilon$ are presented below. 

Due to magnetic flux conservation, the expanding shell (now approximated as spherical) will have a magnetic field that is larger than in the external medium, and increases with $\theta$, as in the original discussion of \citet{Ferriere:1991}.  For our spherical case, it may be shown that the magnetic field in the shell is
\begin{equation}
  B_s(\theta) = \frac{B_0}{2 \epsilon} \sin \theta
  \label{eq:steve10}
\end{equation}
where $B_0$ is the magnitude of the magnetic field in the external medium.  The dimensionless shell thickness $\epsilon$ remains a free parameter of the model, or one that can be determined by observations.  Finally, the plasma density in the shell, determined by mass conservation, is
\begin{equation}
  n_s = \frac{n_0}{3 \epsilon}
  \label{eq:steve11}
\end{equation}
where $n_0$ is the plasma density in the external medium.  \equa{steve11} is a valid approximation for $\epsilon \ll 1$.  

\subsubsection{RM Calculation for Lines of Sight Through the Walls of the Shell\label{sec:walls}}
In evaluating the integral Equation (\ref{eq:rmorg}) or (\ref{eq:rmprat}) through the model shell shown in Figure \ref{fig:steve1}, we consider two cases.  The first calculation is for lines of sight that pass through a portion of the shell, emerge into the cavity, and then reenter the shell on the opposite side before exiting the shell entirely.  This is the case illustrated in Figure \ref{fig:steve1}. The incremental RM for a spatial interval d$s$ along the line of sight is 
\begin{equation}
\textrm{d(RM)} = \pm \, C \, n_s \, B_s(\theta) \, (\hat{e}_s \cdot \hat{e}_{\theta})\, \textrm{d}s
\label{eq:steve12}
\end{equation}

The $\pm$ in front of the RHS indicates that the polarity of the field in the external medium determines the sign of the measured RM. We introduce the variable $s$ as a coordinate along the line of sight; d$s$ is an incremental vector along the line of sight from the source to the observer, and $\hat{e}_s$ is the corresponding unit vector. The constant $C$ is the same as introduced in Equation (\ref{eq:rmmodelW}).

It is convenient to change the variable of integration over the LOS from $s$ to $\phi$, an angle defined in Figure \ref{fig:steve1}.  With the introduction of this variable, the term $(\hat{e}_s \cdot \hat{e}_{\theta}) = -\sin \phi$.  Integration through the shell segments along the line of sight then corresponds to an appropriate integration over $\phi$.  The shell segment closest to the observer corresponds to an integration from  $\phi_1$ to  $\phi_2$, and the segment furthest from the observer is given by an integration from  $\phi_3$ to  $\phi_4$.

Substitution of Equations (\ref{eq:steve10}) and (\ref{eq:steve11}) into (\ref{eq:steve12}), followed by integration over $\phi$ and straightforward algebraic manipulation yields the following expression for the RM
\begin{equation}
  \textrm{RM}(x) = \pm \left( \frac{C \, n_0\, B_0\, R_s}{3 \epsilon^2} \right) x \left[ \arcsin \left(\frac{x}{1 - \epsilon}\right) - \arcsin (x)  \right] \cos \Theta
  \label{eq:steve13}
  \end{equation}

where the new dependent variable is the normalized impact parameter $x \equiv \frac{\xi}{R_s}$. The identity (\ref{eq:steve11}) may be used to convert Equation (\ref{eq:steve13}) into a form in which the observed plasma density in the shell ($n_s$) is the density parameter rather than that in the external medium ($n_0$).  This substitution makes Equation (\ref{eq:steve13}) more directly comparable to Equation (\ref{eq:rmmodelW}).  

\subsubsection{RM for Lines of Sight Entirely Within the Shell \label{sec:allshell}}
\begin{figure}[htb!]
 \centering
\includegraphics[width=0.6\textwidth]{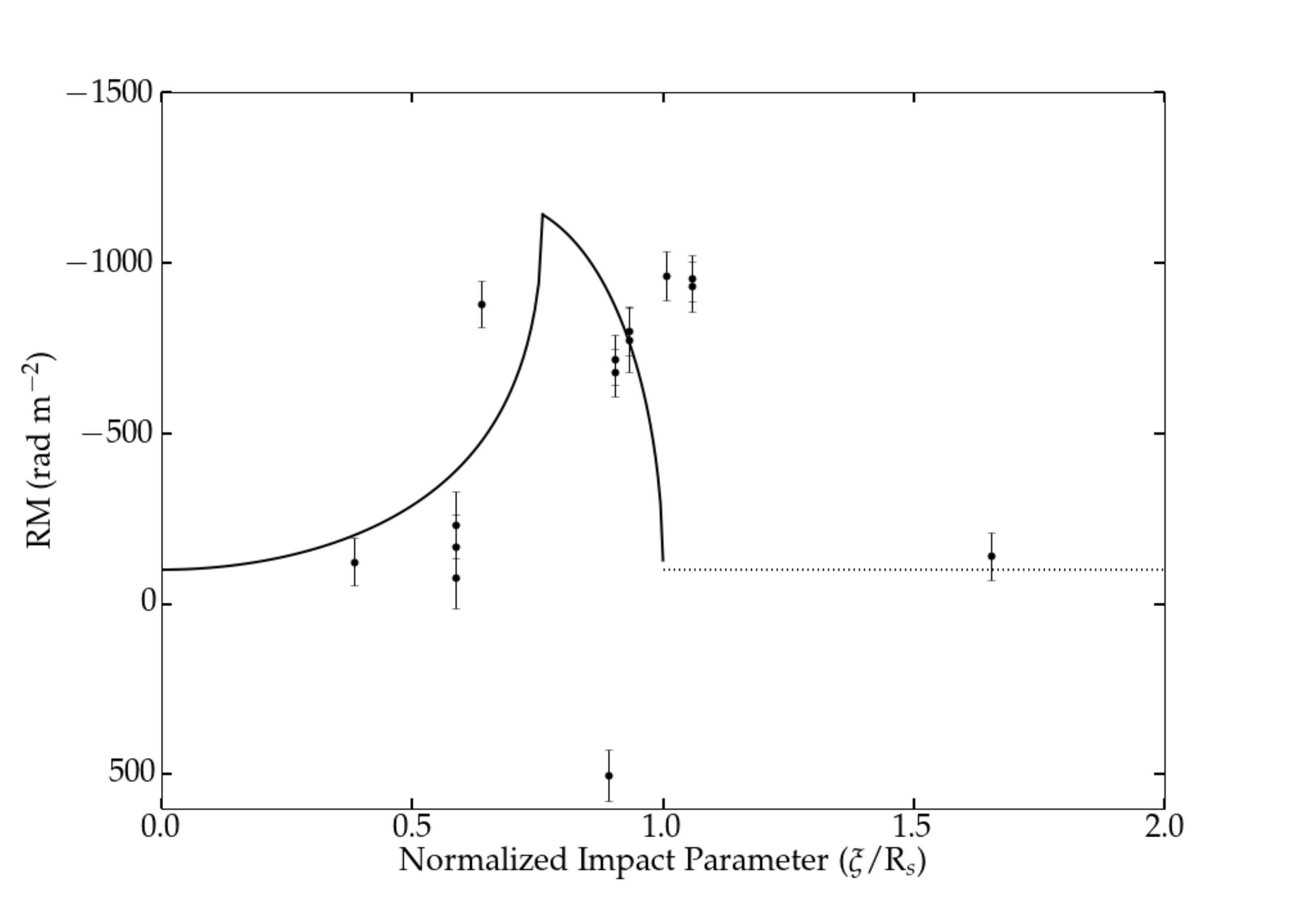}
\caption{Model for the analytic approximation to the bubble model of \citet{Ferriere:1991}, Equations (\ref{eq:steve13}) and (\ref{eq:steve14}).  The model RM is function of the normalized impact parameter $x = \frac{\xi}{R_s}$.  The plotted points represent measured RMs presented in this paper. }
\label{fig:steve2}
\end{figure}

If the ``impact parameter'' $\xi$ is sufficiently large, the entire line of sight is within the shell from the point of ingress to that of egress. From Figure \ref{fig:steve1}, it can be seen that this occurs if
\begin{equation}
  x \equiv \frac{\xi}{R_s}\, \geq\, x_{min} = 1 -\epsilon
  \label{eq:steve14}
  \end{equation}
 The RM in this case is a simple generalization of the algebra involved in obtaining Equation (\ref{eq:steve13}) via an integration over the angular variable $\phi$; the upper limit of integration in the segment closest to the observer $\phi_2 \rightarrow \frac{\pi}{2}$, and the lower limit of integration for the shell segment further from the observer  $\phi_3 \rightarrow \frac{\pi}{2}$.
\begin{equation}
 \textrm{RM}(x) = \pm \left( \frac{C\, n_0\, B_0\, R_s}{3 \epsilon^2} \right)\, x\, \left[ \frac{\pi}{2} - \arcsin (x)  \right] \cos \Theta \mbox{ , if: } x_{min} \leq x \leq 1
 \label{eq:steve15}
\end{equation}
A plot of the expression RM($x$) given by (\ref{eq:steve13}) and (\ref{eq:steve15}) is shown in Figure \ref{fig:steve2} for a set of parameters that are representative for the IC 1805 \HII region (see Table \ref{tab:model}).  The curve is very similar in form to the Whiting model, for the case of no magnetic compression, Equation (\ref{eq:rmmodelW}) with $X = 1$ or Equation (\ref{eq:rmmodelH}).  The model expression for RM($x$) is dependent on $n_0$ (or the shell density $n_s$), $B_0$, $R_s$, $\Theta$, and $\epsilon$, the shell thickness parameter.  For comparison with observations, we also need to specify the background Galactic rotation measure, RM$_{off}$.

Our simple model contained in Equations (\ref{eq:steve13}) and (\ref{eq:steve15}) immediately accounts for one of the main results emergent from the numerical simulations of \citet{Stil:2009}.  The RM through a bubble is maximized when the LOS is parallel to \textbf{B}$_0$ ($\cos \Theta = 1$) and small or zero when the LOS is  $\perp \mbox{ to } \textbf{B}_0$ ($\cos \Theta = 0$).   

\section{Discussion of Observational Results\label{sec:results}}
\subsection{Comparison of Models with Observations in the \HII Region}

In this section we discuss the results of the two models presented in Sections \ref{sec:whitingmod} and \ref{sec:ferriere}. In both cases, we adopt the \citet{Terebey:2003} center for geometric ease and spherical symmetry as well as the parameters given in their Table 3 for a thick shell. 

Figures \ref{fig:southshell-thick} and \ref{fig:southshell-thin} show model RM values for lines of sight south of IC 1805 (\textit{b} $<$ 0.9\ddeg) with the \citet{Whiting:2009} model for the RM as a function of distance and the shell parameters from \citet{Terebey:2003}. Table \ref{tab:model} gives the values of the center of the bubble, the shell radii, the electron density, X, and $\Theta$ for Figure \ref{fig:shell}. Neither model reliably reproduces the observed RM as a function of distance, and as in \citet{Costa:2016}, the model can not account for the dispersion of RM values at similar distances. Generally, the lines of sight in the cavity are low and are more consistent with the background RM. In the thin shell approximation, the largest RM values are associated with lines of sight outside the shell. While the model without amplification of the magnetic field in the shell can marginally account for the magnitude of the RM, the model with amplification (Equation \ref{eq:rmmodelW}) predicts far too high values for the RM for $\Theta$ = 55\ddeg. The analysis contained here mildly supports a result from \citet{Costa:2016} for the Rosette Nebula; Faraday rotation values through these \HII regions do not permit a substantial increase in $|$B$|$ over the general Galactic field.



To reproduce the observed RM in the shell at $\xi$ \about{} 20 pc, the angle between the magnetic field and the observer would need to be tilted more into the plane of the sky for the X = 4 case or into the line of sight for the X = 1 case. For the former case, an angle of \about{} 75\ddeg{} would reproduce the magnitude of the RM in the shell; such an angle is greater than that expected from a geometric argument, even accounting for a magnetic field pitch angle of \about{} 8\ddeg. Also, no one angle can account for the range of the RM values in the cavity.

With our analytic solution for the RM due to a magnetized bubble as described by \citet{Ferriere:1991}, we can examine the dependence of RM on $\Theta$ as well as $\xi$.  The most obvious choice for the latter parameter is $\Theta = 55^{\circ}$, based on the geometry as described in Section \ref{sec:whitingmod}.  Figure \ref{fig:steve2} shows our model RM(x) for $\Theta = 55^{\circ}$, with other parameters given in Table 6. Data for sources south of IC 1805 are superposed on the model.  Although the model obviously does not reproduce the measurements in detail, it can describe the overall scale of the ``rotation measure anomaly'' associated with W4, as well as the approximate magnitude of the largest measured RMs ($|$RM$|$ $\sim$ 1000 \radm).  The peak model RM values shown in Figure 12 do not significantly exceed the measured values, unlike the case for the Whiting model with X = 4 (see Figure \ref{fig:shell}). It should be kept in mind that the shell modeled in Figure \ref{fig:steve2} is the ``thick shell model'' of \citet{Terebey:2003}; the center of that shell is not the star cluster OCl 352, as might be expected.

\citet{Stil:2009} carried out numerical MHD simulations of the Ferri\`ere bubbles, which are obviously more accurate than our analytic approximations. Furthermore, they specifically consider and calculate the Faraday rotation through their models. However, \citet{Stil:2009} only consider $\Theta$ = 0\ddeg{} and $\Theta$ = 90\ddeg, so the calculations reported in that paper can not explore the changes in RM structure with $\Theta$. Furthermore, the Faraday rotation calculation of \citet{Stil:2009} is done when the outer radius $R_s$ \about{} 200 pc (see Figure 14 of \citealt{Stil:2009}), which is much larger than the structure we are modeling in Section \ref{sec:ferriere} of this paper. In what follows, we compare our observations with the results presented in Section 6 of \citet{Stil:2009}.

If LOS $||$ \bext, then the highest values of RM will be through the shell closest to the Galactic plane, but the mean RM across the region will be similar to the mean RM exterior to the bubble (see Figure 14 of \citealt{Stil:2009}). Out of the Galactic plane, the RM is 20 -- 30$\%$ of the mean RM exterior to the bubble. Effectively, the largest RMs will always be found in the Galactic plane, and different lines of sight through the bubble will have varying RM values.

In comparing the simulations of \citet{Stil:2009} to our observational results, we find low RM measures for lines of sight through the cavity, though not always low (e.g., W4-I14 and -I17 vs -I15 and -I21). Lines of sight through the shell have generally large RMs, which is inconsistent with a \bext{} perpendicular to the LOS. The case of LOS $||$ \textbf{B$_{\textrm{ext}}$} is inconsistent as well because far from the bubble, the RM is low (e.g., W4-O26 vs -I24) even at similar latitudes, and the lines of sight at $b$ $>$ 1\ddeg{} are consistent with the background RM instead of being reduced by 70 -- 80$\%$. Unsurprisingly, our results indicate a case somewhere between these two predictions. As a reminder, we note that the largest values of the RM are for lines of sight exterior to the shell, which is not a prediction from \citet{Stil:2009}, most likely due to their simulations modeling the ionized bubble and not a PDR structure.



\subsection{Magnetic Fields in the PDR}
In Section \ref{sec:pdr}, we examine evidence for a PDR outside the southern loop of IC 1805. \citet{Brogan:1999}, \citet{Troland:2016}, and \citet{Pellegrini:2007} report large (\about{} 150 $\mu$G) magnetic fields in PDRs associated with the Orion Veil and M17. In the analysis that follows, we attempt to understand the large RM values for lines of sight through the IC 1805 PDR.

If we consider the PDR and the \HII region to be in pressure equilibrium and include magnetic pressure in the PDR, then 
\begin{equation}
\mPhii = \mPpdr \; + P^{\textrm{PDR}}_{\textrm{mag}},
\label{eq:pbal}
\end{equation}
where \Phii{} and \Ppdr{} are the thermal pressures in the \HII region and PDR, respectively, and \(P^{\textrm{PDR}}_{\textrm{mag}}=\frac{B^2}{8\pi} \) is the magnetic  pressure in the PDR. In the \HII region, \(\mPhii = 2n_e^{\mhii}\, k \,T_{\mhii} \), where $n_e^{\mhii}$ and T$_{\mhii}$ are the electron density and temperature, $k$ is the Boltzmann constant, and the factor of 2 accounts for the contribution from both ions and electrons. For P$^{\textrm{PDR}}_{\textrm{th}}$ = $N_{\textrm{PDR}}$ $k$ $T_{\textrm{PDR}}$, $N_{\textrm{PDR}}$ and $T_{\textrm{PDR}}$ are the neutral hydrogen density and the temperature in the PDR.  

Near the interface of the PDR and the \HII region, the electron density in the PDR is governed by photoioniziation of carbon \citep{Tielens:1985}, so we estimate $n_e^{\textrm{PDR}}$ by \[n_e^{\textrm{PDR}} = N_{\textrm{PDR}}X_C,\] where $X_C$ is the cosmic abundance of carbon given in Table 1.4 of \citet{Draine:2011} ($X_C$ \about{} 2.95 $\times$ 10$^{-4}$). Solving for $B$ in \equa{pbal} gives
\begin{equation}
B = \sqrt{8\pi \; k(2\; n_e^{\mhii} \;T_{\mhii} - N_{\textrm{PDR}}\;T_{\textrm{PDR}})},
\label{eq:bpbal}
\end{equation}
and inserting it into \equa{rmprat}, we express the RM in the PDR as
\begin{equation}
RM = 0.81\; L \; X_C  N_{\textrm{PDR}} \sqrt{8\pi \; k \; (2\;n_e^{\mhii} T_{\mhii} - N_{\textrm{PDR}}\; T_{\textrm{PDR}})}.
\label{eq:rmpdr}
\end{equation}
It should be emphasized that this RM estimate is in the nature of an upper limit to the rotation measure through the PDR.  The reason is that it is obtained from a value of $B$, given by Equation (\ref{eq:rmpdr}), which is based on the magnetic pressure $\frac{B^2}{8 \pi}$. The magnetic pressure includes contributions from turbulent fluctuations on all scales, as well as that from a mean or large scale field that produces the net Faraday rotation.  In general then, a magnetic field value obtained from an estimate of the magnetic pressure will exceed that obtained from a Faraday rotation measurement.

We differentiate \equa{rmpdr} with respect to $N_{\textrm{PDR}}$ to find the value of $N_{\textrm{PDR}}$ that maximizes the RM, which is
\begin{equation}
N_{\textrm{PDR}} = \frac{4}{3}\frac{n_e^{\mhii}\;T_{\mhii}}{T_{\textrm{PDR}}}.
\label{eq:maxN}
\end{equation}
Inserting values of $T_{\mhii}$ = 8000 K, $n_e^{\mhii}$ = 10 \cm{} \citep{Terebey:2003}, and $T_{\textrm{PDR}}$ = 100 K \citep{Tielens:1985}, gives $N_{\textrm{PDR}}$ \about{} 1000 \cm{}, $B$ \about{} 14 $\mu$G (Eq \ref{eq:bpbal}), and RM \about{} 100 \radm. The electron density in the \HII region is governing the maximum $B$ expected in the PDR given pressure balance. For the IC 1805 \HII region, $n_e$ is low compared to M17 ($n_e$ \about{} 560 \cm) \citep{Pellegrini:2007}, which suggests that a high density (pressure) \HII region is needed to explain large magnetic fields in the PDR. 


Our analysis suggests that a simple pressure balance analysis predicts low RM values from the PDR that are inconsistent with our observations. It appears that a different mechanism is required to achieve the magnetic fields strengths observed in \citet{Brogan:1999}.

\citet{Terebey:2003} discuss an extended halo of ionized emission around the southern loop, which may indicate that there are more free electrons present outside the obvious ionized shell as seen in Figure \ref{fig:w4}. This may account for the larger values of the RM we observe. It is clear that knowing the electron density in this region and determining the presence of a PDR through observations, such as carbon radio recombination lines, is necessary to understand how the magnetic field is modified in this complex region.

\section{A Comparison of IC 1805 and the Rosette Nebula as ``Rotation Measure Anomalies'' \label{sec:rosette}}

\begin{table}[htb!]
\centering
\caption{Stellar Parameters \label{tab:stellarpar}}
\begin{threeparttable}
\centering
\begin{tabular}{llllll}
\hline
\multicolumn{1}{c}{Star Cluster} & \multicolumn{1}{c}{Star} & \multicolumn{1}{c}{Type} & \multicolumn{1}{c}{\ML} & \multicolumn{1}{c}{V$_{\infinity}$} & \multicolumn{1}{c}{ L$_{W}$=$\frac{1}{2}\dot{M}$v$_{\infinity}^{2}$}\tnote{a} \\
\multicolumn{1}{c}{ } &\multicolumn{1}{c}{ } &\multicolumn{1}{c}{ } & \multicolumn{1}{c}{ (\Msun yr$^{-1}$) } & \multicolumn{1}{c}{(km/s)} & \multicolumn{1}{c}{ (erg s$^{-1}$)} \\
\hline
{\multirow{4}{*}{NGC 2244}} & HD 46223 & O4V(f)\tnote{b} & 1.6$\times$10$^{-6}$ \tnote{c} & 3100\tnote{d} & 4.8$\times$10$^{36}$ \\
& HD 46150 & O5.5V\tnote{e} & 2.0$\times$10$^{-6}$ \tnote{c} & 3100\tnote{d} & 6.0$\times$10$^{36}$ \\
& HD 46202 & O9V(f)\tnote{b} & 6.3$\times$10$^{-8}$ \tnote{c} & 1150\tnote{d} & 2.6$\times$10$^{34}$ \\
& HD 46149 & O8.5V(f)\tnote{b} & 2.0$\times$10$^{-7}$ \tnote{c} & 1700\tnote{f} & 1.8$\times$10$^{35}$\\
\hline
{\multirow{3}{*}{OCl 352}} & HD15570 & O4I\tnote{b} & 1.0$\times$10$^{-5}$ \tnote{c} & 2200\tnote{g} & 1.5$\times$10$^{37} $ \\
& HD15558 & O4III\tnote{b} & 6.3$\times$10$^{-6}$ \tnote{c} & 3000\tnote{f} & 1.8$\times$10$^{37} $ \\
& HD 15629 & O5V\tnote{b} & 2.0$\times$10$^{-6}$ \tnote{c} & 2900\tnote{h} & 5.3$\times$10$^{36}$ \\
\hline
\end{tabular}
\begin{tablenotes}
\item[a] Calculated mechanical wind luminosity based on cited mass loss rates and terminal velocities.
\item[b] \citet{Massey:1995}
\item[c] \citet{Howarth:1989}
\item[d] \citet{Chlebowski:1991}
\item[e] \citet{Roman:2008}
\item[f] \citet{Garmany:1988}
\item[g] \citet{Bouret:2012}
\item[h] \citet{Groenewegen:1989}

\end{tablenotes}
\end{threeparttable}
\end{table}

We are interested in how the Galactic magnetic field is modified by OB associations via their stellar winds and ionizing photons, and we started our study with the Rosette Nebula, where we found large (\about~10$^{3}$ \radm) RM measurements through the ionized shell of the \HII region \citep{Costa:2016}.

In the case of the Rosette, we find positive RM across the region, and for IC 1805, we find negative values. If the Galactic magnetic field follows the spiral arms in a clockwise direction, then we would expect the LOS magnetic field component to be pointed towards us (positive B) for $\ell$ $>$ 180\ddeg, and pointed away from us (negative B) for $\ell$ $<$ 180\ddeg. Except for one line of sight in each nebula, we find that the polarity of the Galactic magnetic field is preserved across each nebula and is consistent with the large scale field through the arm.

In our study of the Rosette, we investigated whether the magnetic field is amplified in the shell of the nebula. We found that the model without amplification was weakly favored over the case when the magnetic field is amplified in the shell. When we applied the same model to IC 1805, however, it is difficult to conclude in favor of either model, but in both cases, the model with an enhanced magnetic field overpredicts the RM. From inspection of Figures \ref{fig:southshell-thick} and \ref{fig:southshell-thin}, it seems that the model without amplification better accounts for the magnitude of the observed RMs, but the observations do not conform to the model prediction of RM($\xi$), and the model can not account for the wide range in observed values of RM at a given $\xi$.

In the present study, we find the highest RMs for lines of sight outside the obvious shell structure, though one line of sight (W4-I13) does appear to intersect the ionized shell and it has a large RM. These lines of sight may be probing the magnetic field within the PDR. In the case of the Rosette, we found that the highest RM values were for lines of sight through the bright ionized shell. However with our work on IC 1805 and the PDR associated with it, we have briefly revisited our results in the Rosette, particularly Figure 1 from \citet{Costa:2016}. There are a few lines of sight with RM of order a few 10$^2$ \radm{} that appear to be outside the ionized shell. These lines of sight were included in the background estimate for the Rosette, but if the Rosette also has a PDR, then these lines of sight may actually be probing that material.

Table \ref{tab:stellarpar} lists spectral type, mass loss rate, terminal wind velocity, and calculated wind luminosity from the literature for O stars with the largest wind luminosities in both NGC 2244, which is associated with the Rosette Nebula, and OCl 352. The sum of the wind luminosities of the three main stars in OCl 352 is 3.8 $\times$ 10$^{37}$ ergs s$^{-1}$, while the corresponding number for NGC 2244 (4 stars) is 1.1 $\times$ 10$^{37}$ ergs s$^{-1}$. In addition, OCl 352 appears to have more luminous stars. As such, OCl 352 might be expected to produce a more energetic stellar bubble than NGC 2244. Our Faraday rotation measurements show no indication of this, in that the largest RMs observed are similar for the two objects. In fact, higher RMs were measured for the Rosette than for any line of sight through IC 1805. A number of factors can control the impact a star cluster has on the ISM. If some relationship exists between the total wind luminosity of a star cluster and properties of an interstellar bubble that can be measure with Faraday rotation, it will apparently require a large sample of clusters/ \HII regions to reveal it.
%
%

\section{Future Research\label{sec:fut}}
In the future, we will continue our investigation of \HII regions and how they modify their surroundings and the Galactic magnetic field. An immediate investigation will be centered on observations of the \HII region IC 1396. This will provide a third \HII region with different age, stellar content, and Galactic location. The observations are similar to those we have made of the Rosette Nebula and IC 1805.  The observations of IC 1396 have been made with the VLA and are awaiting analysis. By adding more \HII regions to our study, we can begin to address questions such as 
\begin{enumerate}
\item Since the electron density distributions in \HII regions are known from radio continuum observations, we can inquire what conditions would result in an RM $>$$>$ 10$^3$ \radm~through the shell of an \HII region.
\item Is it a general property of \HII regions and stellar bubbles that the polarity of the Galactic magnetic field is preserved within the region? The answer to this question has implications for the amplitude of MHD turbulence in the ISM on scales of the order of the \HII regions, $\sim 10 - 30$ pc.
\item Do PDRs around other nebulae produce high RMs? What is the magnitude of the RM due to the PDR relative to that of the shell of an \HII region?
\end{enumerate}
In addition to increasing the number of \HII regions, understanding Faraday complexity and how to interpret the associated RM measurements is important to studies of Galactic magnetic fields, particularly with large polarization surveys like the \VLASS{} and \POSSUM{} with the \ASKAP{} in the near future.

\section{Summary and Conclusions\label{sec:sum}}
\begin{enumerate}
\item We performed polarimetric observations using the VLA for 27 lines of sight through or near the shell of the \HII region and stellar bubble associated with the OB association OCl 352. 
\item We obtain RM measurements for 20 sources using two methods. The first is through the traditional least-squares fit to \chilam, and the second is using RM Synthesis. Including components that are resolved, we report 27 RM values, and we find good agreement between the two methods. We find the same sign of the RM across the entire region with the exception of one source, W4-I18. We estimate a background RM due to the general ISM of --145 \radm{} in this part of the Galactic plane. 
We measure an excess of RM of \about~+600 to --800 \radm~due to W4. 
\item Only one line of sight has a positive RM value, W4-I18. It has a RM of +501 $\pm$ 33 \radm, and it is located 5.6 arcminutes from the center of OCl 352. This line of sight may be probing the material close to the massive stars. 
The orientation of the line of sight component of the magnetic field is directed towards the observer, whereas in the rest of the region, the magnetic field is directed away. 
\item We find that some of the lines of sight with the largest RM values occur just outside the obvious ionized shell of IC 1805 and are potentially probing the magnetic field in the PDR. The lines of sight through the cavity of the bubble have lower RM values than those through the shell. In the W4 Superbubble, which is north of OCl 352, we find RM values consistent with the background RM. 
\item We discuss two shell models to reproduce the magnitude of the RM and its dependency on distance from the center of the star cluster. We employed the first of these models in \citet{Savage:2013} and \citet{Costa:2016}, and it is based on the \citet{Weaver:1977} solution for a stellar bubble, which includes a shock expanding into an ambient medium. The second model uses magnetic flux conservation to describe how the magnetic field is modified in the shell and consists of a simplified analytic approximation to the results presented by \citet{Ferriere:1991}. Neither of these simplified models satisfactorily accounts for the dependence of RM on spatial location within the shell, although the Whiting model without field amplification (X = 1) and the simplified Ferri\`ere model approximately reproduce the magnitude of the largest  RMs.  However, both models predict a single-valued dependence of RM on $\xi$, the separation of the line of sight from the center of the nebula, whereas the observations show a large range of RM for sources with similar values of $\xi$. 
\item Because we have independent information on the electron density from radio continuum observations of both IC 1805 and the Rosette Nebula, our observations can limit the magnitude of the magnetic field in the \HII regions. Our RM measurements indicate that the field does not greatly exceed the value in the general ISM.
\item We compare our results from the current study of IC 1805 and our previous study of the Rosette Nebula. Notably, we find the same order of magnitude for the RM for the two nebulae, but the sign of the RM in each region is opposite. Since IC 1805 and the Rosette are at different Galactic longitudes and on either side of $b$ = 180\ddeg, the sign difference between the two nebula is consistent with a Galactic magnetic field that follows the spiral arm structure in a clock-wise direction, as suggested in models \citep{vanEck:2011}.
\end{enumerate}

\acknowledgments
This research was partially supported at the University of Iowa by grants AST09-07911 and ATM09-56901 from the National Science Foundation. This publication makes use of data products from the Wide-field Infrared Survey Explorer \citep{2010AJ....140.1868W}, which is a joint project of the University of California, Los Angeles, and the Jet Propulsion Laboratory/California Institute of Technology, funded by the National Aeronautics and Space Administration. Additionally, the research presented in this paper uses data from the Canadian Galactic Plane Survey, a Canadian project with international partners, supported by the Natural Sciences and Engineering Research Council. This research also uses the Python packages Astropy, a community-developed core Python package for Astronomy \citep{2013A&A...558A..33A} and NumPy \citep{van2011numpy}. Finally, we thank the referee of this paper for a helpful and collegial review.

\clearpage

\newpage
\bibliography{FRbib,LBVbib,compbib,FollowupBib}
\bibliographystyle{apj}

\end{document}

%% file: defs.tex
\newcommand{\bext}{\textbf{B}$_{\textrm{ext}}$}
\newcommand{\mhii}{\textrm{H}\,\textsc{ii}}
\newcommand{\mPhii}{P^{\textrm{H}\,\textsc{ii}}_{\textrm{th}}}
\newcommand{\Phii}{$P^{\textrm{H}\,\textsc{ii}}_{\textrm{th}}$}
\newcommand{\mPpdr}{P^{\textrm{PDR}}_{\textrm{th}}}
\newcommand{\Ppdr}{$P^{\textrm{PDR}}_{\textrm{th}}$}

\newcommand{\VLASS}{VLA Sky Survey (VLASS)}

\newcommand{\Q}{\textit{Q}}
\newcommand{\U}{\textit{U}}

\newcommand{\ASKAP}{Australian Square Kilometre Array Pathfinder (ASKAP)}
\newcommand{\POSSUM}{Polarisation Sky Survey of the Universe's Magnetism (POSSUM)}

\newcommand{\chilam}{$\chi(\lambda^{2})$}

\newcommand{\hi}{H\,{\sc i} } 
\newcommand{\HII}{H\,{\sc ii} } 
\newcommand{\ddeg}{\ensuremath{^{\circ}}} 
\newcommand{\lsi}{LSI +61\ddeg303}
\newcommand{\Msun}{\ensuremath{\textrm{M}_{\odot} }} 
\newcommand{\MLu}{\textrm{M}$_{\odot}$ yr$^{-1}$}
\newcommand{\kms}{km s$^{-1}$}
\newcommand{\infinity}{\ensuremath{\infty}}
\newcommand{\ML}{\ensuremath{\dot{M}}} 
\newcommand{\radm}{rad m$^{-2}$}

\newcommand{\blos}{B$_{\textrm{LOS}}$}
\newcommand{\cm}{cm$^{-3}$}
\newcommand{\plam}{\textit{P($\lambda^2$)}}
\newcommand{\fsimp}{\textit{Faraday simple}}
\newcommand{\fcomp}{\textit{Faraday complex}}
\newcommand{\fphi}{\textit{F($\phi$)}}
\newcommand{\lamsq}{$\lambda^2$}
\newcommand{\rmclean}{\textsc{rmclean}}
\newcommand{\clean}{\textsc{clean}}
\newcommand{\immath}{\textsc{immath}}
\newcommand{\polcal}{\textsc{polcal}}
\newcommand{\equa}[1]{Equation (\ref{eq:#1})}

\newcommand{\about}{$\sim$}
\newcommand{\mircon}{$\mu$m}

%% file: costa2018.bbl
\begin{thebibliography}{}
\expandafter\ifx\csname natexlab\endcsname\relax\def\natexlab#1{#1}\fi

\bibitem[{{Albert} {et~al.}(2008){Albert}, {Aliu}, {Anderhub}, {Antoranz},
  {Backes}, {Baixeras}, {Barrio}, {Bartko}, {Bastieri}, {Becker}, {Bednarek},
  {Berger}, {Bigongiari}, {Biland}, {Bock}, {Bonnoli}, {Bordas}, {Bosch-Ramon},
  {Bretz}, {Britvitch}, {Camara}, {Carmona}, {Chilingarian}, {Commichau},
  {Contreras}, {Cortina}, {Costado}, {Curtef}, {Dazzi}, {De Angelis}, {de los
  Reyes}, {De Lotto}, {De Maria}, {De Sabata}, {Delgado Mendez}, {Dorner},
  {Doro}, {Errando}, {Fagiolini}, {Ferenc}, {Fern{\'a}ndez}, {Firpo},
  {Fonseca}, {Font}, {Galante}, {Garc{\'{\i}}a L{\'o}pez}, {Garczarczyk},
  {Gaug}, {Goebel}, {Hayashida}, {Herrero}, {H{\"o}hne}, {Hose}, {Hsu},
  {Huber}, {Jogler}, {Kosyra}, {Kranich}, {Laille}, {Leonardo}, {Lindfors},
  {Lombardi}, {Longo}, {L{\'o}pez}, {Lorenz}, {Majumdar}, {Maneva},
  {Mankuzhiyil}, {Mannheim}, {Mariotti}, {Mart{\'{\i}}nez}, {Mazin}, {Merck},
  {Meucci}, {Meyer}, {Miranda}, {Mirzoyan}, {Mizobuchi}, {Moles}, {Moralejo},
  {Nieto}, {Nilsson}, {Ninkovic}, {O{\~n}a-Wilhelmi}, {Otte}, {Oya},
  {Panniello}, {Paoletti}, {Paredes}, {Pasanen}, {Pascoli}, {Pauss}, {Pegna},
  {P{\'e}rez-Torres}, {Persic}, {Peruzzo}, {Piccioli}, {Prada}, {Prandini},
  {Puchades}, {Raymers}, {Rhode}, {Rib{\'o}}, {Rico}, {Rissi}, {Robert},
  {R{\"u}gamer}, {Saggion}, {Saito}, {S{\'a}nchez}, {S{\'a}nchez-Conde},
  {Sartori}, {Scalzotto}, {Scapin}, {Schmitt}, {Schweizer}, {Shayduk},
  {Shinozaki}, {Shore}, {Sidro}, {Sillanp{\"a}{\"a}}, {Sobczynska}, {Spanier},
  {Stamerra}, {Stark}, {Takalo}, {Temnikov}, {Tescaro}, {Teshima}, {Torres},
  {Turini}, {Vankov}, {Venturini}, {Vitale}, {Wagner}, {Wittek}, {Zandanel},
  {Zanin}, {Zapatero}, {MAGIC Collaboration}, {Guerrero}, {Alberdi}, {Paragi},
  {Muxlow}, \& {Diamond}}]{Albert:2008}
{Albert}, J., {Aliu}, E., {Anderhub}, H., {et~al.} 2008, \apj, 684, 1351

\bibitem[{{Anantharamaiah}(1985)}]{Anantharamaiah:1985}
{Anantharamaiah}, K.~R. 1985, Journal of Astrophysics and Astronomy, 6, 203

\bibitem[{{Anderson} {et~al.}(2015){Anderson}, {Gaensler}, {Feain}, \&
  {Franzen}}]{Anderson:2015}
{Anderson}, C.~S., {Gaensler}, B.~M., {Feain}, I.~J., \& {Franzen}, T.~M.~O.
  2015, \apj, 815, 49

\bibitem[{{Anderson} {et~al.}(2014){Anderson}, {Bania}, {Balser}, {Cunningham},
  {Wenger}, {Johnstone}, \& {Armentrout}}]{Anderson:2014}
{Anderson}, L.~D., {Bania}, T.~M., {Balser}, D.~S., {et~al.} 2014, \apjs, 212,
  1

\bibitem[{{Anderson} {et~al.}(2012){Anderson}, {Zavagno}, {Barlow},
  {Garc{\'{\i}}a-Lario}, \& {Noriega-Crespo}}]{Anderson:2012}
{Anderson}, L.~D., {Zavagno}, A., {Barlow}, M.~J., {Garc{\'{\i}}a-Lario}, P.,
  \& {Noriega-Crespo}, A. 2012, \aap, 537, A1

\bibitem[{{Aragona} {et~al.}(2009){Aragona}, {McSwain}, {Grundstrom}, {Marsh},
  {Roettenbacher}, {Hessler}, {Boyajian}, \& {Ray}}]{Aragona:2009}
{Aragona}, C., {McSwain}, M.~V., {Grundstrom}, E.~D., {et~al.} 2009, \apj, 698,
  514

\bibitem[{{Astropy Collaboration} {et~al.}(2013){Astropy Collaboration},
  {Robitaille}, {Tollerud}, {Greenfield}, {Droettboom}, {Bray}, {Aldcroft},
  {Davis}, {Ginsburg}, {Price-Whelan}, {Kerzendorf}, {Conley}, {Crighton},
  {Barbary}, {Muna}, {Ferguson}, {Grollier}, {Parikh}, {Nair}, {Unther},
  {Deil}, {Woillez}, {Conseil}, {Kramer}, {Turner}, {Singer}, {Fox}, {Weaver},
  {Zabalza}, {Edwards}, {Azalee Bostroem}, {Burke}, {Casey}, {Crawford},
  {Dencheva}, {Ely}, {Jenness}, {Labrie}, {Lim}, {Pierfederici}, {Pontzen},
  {Ptak}, {Refsdal}, {Servillat}, \& {Streicher}}]{2013A&A...558A..33A}
{Astropy Collaboration}, {Robitaille}, T.~P., {Tollerud}, E.~J., {et~al.} 2013,
  \aap, 558, A33

\bibitem[{{Balser} {et~al.}(2016){Balser}, {Anish Roshi}, {Jeyakumar}, {Bania},
  {Montet}, \& {Shitanishi}}]{Balser:2016}
{Balser}, D.~S., {Anish Roshi}, D., {Jeyakumar}, S., {et~al.} 2016, \apj, 816,
  22

\bibitem[{{Basu} {et~al.}(1999){Basu}, {Johnstone}, \& {Martin}}]{Basu:1999}
{Basu}, S., {Johnstone}, D., \& {Martin}, P.~G. 1999, \apj, 516, 843

\bibitem[{{Bell} \& {En{\ss}lin}(2012)}]{Bell:2012}
{Bell}, M.~R., \& {En{\ss}lin}, T.~A. 2012, \aap, 540, A80

\bibitem[{{Bignami} {et~al.}(1981){Bignami}, {Caraveo}, {Lamb}, {Markert}, \&
  {Paul}}]{Bignami:1981}
{Bignami}, G.~F., {Caraveo}, P.~A., {Lamb}, R.~C., {Markert}, T.~H., \& {Paul},
  J.~A. 1981, \apjl, 247, L85

\bibitem[{{Bignell}(1982)}]{Bignell:1982}
{Bignell}, C. 1982, in Synthesis Mapping, ed. A.~R. {Thompson} \& L.~R.
  {D'Addario}, 6

\bibitem[{{Bik} {et~al.}(2012){Bik}, {Henning}, {Stolte}, {Brandner},
  {Gouliermis}, {Gennaro}, {Pasquali}, {Rochau}, {Beuther}, {Ageorges},
  {Seifert}, {Wang}, \& {Kudryavtseva}}]{Bik:2012}
{Bik}, A., {Henning}, T., {Stolte}, A., {et~al.} 2012, \apj, 744, 87

\bibitem[{{Bouret} {et~al.}(2012){Bouret}, {Hillier}, {Lanz}, \&
  {Fullerton}}]{Bouret:2012}
{Bouret}, J.-C., {Hillier}, D.~J., {Lanz}, T., \& {Fullerton}, A.~W. 2012,
  \aap, 544, A67

\bibitem[{{Braunsfurth}(1983)}]{Braunsfurth:1983}
{Braunsfurth}, E. 1983, \aap, 117, 297

\bibitem[{{Brentjens} \& {de Bruyn}(2005)}]{Brentjens:2005}
{Brentjens}, M.~A., \& {de Bruyn}, A.~G. 2005, \aap, 441, 1217

\bibitem[{{Brogan} {et~al.}(1999){Brogan}, {Troland}, {Roberts}, \&
  {Crutcher}}]{Brogan:1999}
{Brogan}, C.~L., {Troland}, T.~H., {Roberts}, D.~A., \& {Crutcher}, R.~M. 1999,
  \apj, 515, 304

\bibitem[{{Brown} {et~al.}(2003){Brown}, {Taylor}, \& {Jackel}}]{Brown:2003}
{Brown}, J.~C., {Taylor}, A.~R., \& {Jackel}, B.~J. 2003, \apjs, 145, 213

\bibitem[{{Carpenter} {et~al.}(2000){Carpenter}, {Heyer}, \&
  {Snell}}]{Carpenter:2000}
{Carpenter}, J.~M., {Heyer}, M.~H., \& {Snell}, R.~L. 2000, \apjs, 130, 381

\bibitem[{{Casares} {et~al.}(2005){Casares}, {Ribas}, {Paredes},
  {Mart{\'{\i}}}, \& {Allende Prieto}}]{Casares:2005}
{Casares}, J., {Ribas}, I., {Paredes}, J.~M., {Mart{\'{\i}}}, J., \& {Allende
  Prieto}, C. 2005, \mnras, 360, 1105

\bibitem[{{Chlebowski} \& {Garmany}(1991)}]{Chlebowski:1991}
{Chlebowski}, T., \& {Garmany}, C.~D. 1991, \apj, 368, 241

\bibitem[{{Churchwell} {et~al.}(2006){Churchwell}, {Povich}, {Allen}, {Taylor},
  {Meade}, {Babler}, {Indebetouw}, {Watson}, {Whitney}, {Wolfire}, {Bania},
  {Benjamin}, {Clemens}, {Cohen}, {Cyganowski}, {Jackson}, {Kobulnicky},
  {Mathis}, {Mercer}, {Stolovy}, {Uzpen}, {Watson}, \&
  {Wolff}}]{Churchwell:2006}
{Churchwell}, E., {Povich}, M.~S., {Allen}, D., {et~al.} 2006, \apj, 649, 759

\bibitem[{{Cioffi} \& {Jones}(1980)}]{Cioffi:1980}
{Cioffi}, D.~F., \& {Jones}, T.~W. 1980, \aj, 85, 368

\bibitem[{{Condon} {et~al.}(1998){Condon}, {Cotton}, {Greisen}, {Yin},
  {Perley}, {Taylor}, \& {Broderick}}]{Condon:1998}
{Condon}, J.~J., {Cotton}, W.~D., {Greisen}, E.~W., {et~al.} 1998, \aj, 115,
  1693

\bibitem[{{Costa}(2018)}]{Costa:2018phd}
{Costa}, A.~H. 2018, PhD thesis, The University of Iowa

\bibitem[{{Costa} {et~al.}(2016){Costa}, {Spangler}, {Sink}, {Brown}, \&
  {Mao}}]{Costa:2016}
{Costa}, A.~H., {Spangler}, S.~R., {Sink}, J.~R., {Brown}, S., \& {Mao}, S.~A.
  2016, \apj, 821, 92

\bibitem[{{Dambis} {et~al.}(2001){Dambis}, {Mel'Nik}, \&
  {Rastorguev}}]{Dambis:2001}
{Dambis}, A.~K., {Mel'Nik}, A.~M., \& {Rastorguev}, A.~S. 2001, Astronomy
  Letters, 27, 58

\bibitem[{{de Avillez} \& {Breitschwerdt}(2005)}]{deAvillez:2005}
{de Avillez}, M.~A., \& {Breitschwerdt}, D. 2005, \aap, 436, 585

\bibitem[{{Dennison} {et~al.}(1997){Dennison}, {Topasna}, \&
  {Simonetti}}]{Dennison:1997}
{Dennison}, B., {Topasna}, G.~A., \& {Simonetti}, J.~H. 1997, \apjl, 474, L31

\bibitem[{{Dhawan} {et~al.}(2006){Dhawan}, {Mioduszewski}, \&
  {Rupen}}]{Dhawan:2006}
{Dhawan}, V., {Mioduszewski}, A., \& {Rupen}, M. 2006, in VI Microquasar
  Workshop: Microquasars and Beyond, 52.1

\bibitem[{{Dickel}(1980)}]{Dickel:1980}
{Dickel}, H.~R. 1980, \apj, 238, 829

\bibitem[{{Dickel} {et~al.}(1980){Dickel}, {Dickel}, {Wilson}, \&
  {Werner}}]{Dickel:1980b}
{Dickel}, H.~R., {Dickel}, J.~R., {Wilson}, W.~J., \& {Werner}, M.~W. 1980,
  \apj, 237, 711

\bibitem[{{Digel} {et~al.}(1996){Digel}, {Lyder}, {Philbrick}, {Puche}, \&
  {Thaddeus}}]{Digel:1996}
{Digel}, S.~W., {Lyder}, D.~A., {Philbrick}, A.~J., {Puche}, D., \& {Thaddeus},
  P. 1996, \apj, 458, 561

\bibitem[{{Draine}(2011)}]{Draine:2011}
{Draine}, B.~T. 2011, {Physics of the Interstellar and Intergalactic Medium}

\bibitem[{{Dubus}(2006)}]{Dubus:2006}
{Dubus}, G. 2006, \aap, 456, 801

\bibitem[{{Everett} \& {Weisberg}(2001)}]{Everett:2001}
{Everett}, J.~E., \& {Weisberg}, J.~M. 2001, \apj, 553, 341

\bibitem[{{Farnsworth} {et~al.}(2011){Farnsworth}, {Rudnick}, \&
  {Brown}}]{Farnsworth:2011}
{Farnsworth}, D., {Rudnick}, L., \& {Brown}, S. 2011, \aj, 141, 191

\bibitem[{{Feigelson} \& {Townsley}(2008)}]{Feigelson:2008}
{Feigelson}, E.~D., \& {Townsley}, L.~K. 2008, \apj, 673, 354

\bibitem[{{Feigelson} {et~al.}(2013){Feigelson}, {Townsley}, {Broos}, {Busk},
  {Getman}, {King}, {Kuhn}, {Naylor}, {Povich}, {Baddeley}, {Bate},
  {Indebetouw}, {Luhman}, {McCaughrean}, {Pittard}, {Pudritz}, {Sills}, {Song},
  \& {Wadsley}}]{Feigelson:2013}
{Feigelson}, E.~D., {Townsley}, L.~K., {Broos}, P.~S., {et~al.} 2013, \apjs,
  209, 26

\bibitem[{Ferri{\`e}re {et~al.}(1991)Ferri{\`e}re, {Mac Low}, \&
  {Zweibel}}]{Ferriere:1991}
Ferri{\`e}re, K.~M., {Mac Low}, M.-M., \& {Zweibel}, E.~G. 1991, \apj, 375, 239

\bibitem[{{Frail} \& {Hjellming}(1991)}]{Frail:1991}
{Frail}, D.~A., \& {Hjellming}, R.~M. 1991, \aj, 101, 2126

\bibitem[{{Gao} {et~al.}(2015){Gao}, {Reich}, {Reich}, {Han}, \&
  {Kothes}}]{Gao:2015}
{Gao}, X.~Y., {Reich}, W., {Reich}, P., {Han}, J.~L., \& {Kothes}, R. 2015,
  \aap, 578, A24

\bibitem[{{Garmany}(1988)}]{Garmany:1988}
{Garmany}, C.~D. 1988, NASA Special Publication, 497, 160

\bibitem[{{Gooch}(1995)}]{Gooch:1995}
{Gooch}, R. 1995, in Astronomical Society of the Pacific Conference Series,
  Vol.~77, Astronomical Data Analysis Software and Systems IV, ed. R.~A.
  {Shaw}, H.~E. {Payne}, \& J.~J.~E. {Hayes}, 144

\bibitem[{{Gray} {et~al.}(1999){Gray}, {Landecker}, {Dewdney}, {Taylor},
  {Willis}, \& {Normandeau}}]{Gray:1999}
{Gray}, A.~D., {Landecker}, T.~L., {Dewdney}, P.~E., {et~al.} 1999, \apj, 514,
  221

\bibitem[{{Gregory} \& {Neish}(2002)}]{Gregory:2002}
{Gregory}, P.~C., \& {Neish}, C. 2002, \apj, 580, 1133

\bibitem[{{Gregory} {et~al.}(1979){Gregory}, {Taylor}, {Crampton}, {Hutchings},
  {Hjellming}, {Hogg}, {Hvatum}, {Gottlieb}, {Feldman}, \&
  {Kwok}}]{Gregory:1979}
{Gregory}, P.~C., {Taylor}, A.~R., {Crampton}, D., {et~al.} 1979, \aj, 84, 1030

\bibitem[{{Groenewegen} {et~al.}(1989){Groenewegen}, {Lamers}, \&
  {Pauldrach}}]{Groenewegen:1989}
{Groenewegen}, M.~A.~T., {Lamers}, H.~J.~G.~L.~M., \& {Pauldrach}, A.~W.~A.
  1989, \aap, 221, 78

\bibitem[{{Guetter} \& {Vrba}(1989)}]{Guetter:1989}
{Guetter}, H.~H., \& {Vrba}, F.~J. 1989, \aj, 98, 611

\bibitem[{{Han} {et~al.}(2006){Han}, {Manchester}, {Lyne}, {Qiao}, \& {van
  Straten}}]{Han:2006}
{Han}, J.~L., {Manchester}, R.~N., {Lyne}, A.~G., {Qiao}, G.~J., \& {van
  Straten}, W. 2006, \apj, 642, 868

\bibitem[{{Harvey-Smith} {et~al.}(2011){Harvey-Smith}, {Madsen}, \&
  {Gaensler}}]{Harvey:2011}
{Harvey-Smith}, L., {Madsen}, G.~J., \& {Gaensler}, B.~M. 2011, \apj, 736, 83

\bibitem[{{Hasegawa} {et~al.}(1983){Hasegawa}, {Sato}, \&
  {Fukui}}]{Hasegawa:1983}
{Hasegawa}, T., {Sato}, F., \& {Fukui}, Y. 1983, \aj, 88, 658

\bibitem[{{Haverkorn} {et~al.}(2004){Haverkorn}, {Gaensler},
  {McClure-Griffiths}, {Dickey}, \& {Green}}]{Haverkorn:2004}
{Haverkorn}, M., {Gaensler}, B.~M., {McClure-Griffiths}, N.~M., {Dickey},
  J.~M., \& {Green}, A.~J. 2004, \apj, 609, 776

\bibitem[{{Heald}(2009)}]{Heald:2009}
{Heald}, G. 2009, in IAU Symposium, Vol. 259, IAU Symposium, ed. K.~G.
  {Strassmeier}, A.~G. {Kosovichev}, \& J.~E. {Beckman}, 591--602

\bibitem[{{Heyer} \& {Terebey}(1998)}]{Heyer:1998}
{Heyer}, M.~H., \& {Terebey}, S. 1998, \apj, 502, 265

\bibitem[{{Heyer} {et~al.}(1996){Heyer}, {Brunt}, {Snell}, {Howe}, {Schloerb},
  {Carpenter}, {Normandeau}, {Taylor}, {Dewdney}, {Cao}, {Terebey}, \&
  {Beichman}}]{Heyer:1996}
{Heyer}, M.~H., {Brunt}, C., {Snell}, R.~L., {et~al.} 1996, \apjl, 464, L175

\bibitem[{{Hill} {et~al.}(2017){Hill}, {Landecker}, {Carretti}, {Douglas},
  {Sun}, {Gaensler}, {Mao}, {McClure-Griffiths}, {Reich}, {Wolleben}, {Dickey},
  {Gray}, {Haverkorn}, {Leahy}, \& {Schnitzeler}}]{Hill:2017}
{Hill}, A.~S., {Landecker}, T.~L., {Carretti}, E., {et~al.} 2017, \mnras, 467,
  4631

\bibitem[{{Hollenbach} \& {Tielens}(1999)}]{Hollenbach:1999}
{Hollenbach}, D.~J., \& {Tielens}, A.~G.~G.~M. 1999, Reviews of Modern Physics,
  71, 173

\bibitem[{{Howarth} \& {Prinja}(1989)}]{Howarth:1989}
{Howarth}, I.~D., \& {Prinja}, R.~K. 1989, \apjs, 69, 527

\bibitem[{{Jose} {et~al.}(2016){Jose}, {Kim}, {Herczeg}, {Samal}, {Bieging},
  {Meyer}, \& {Sherry}}]{Jose:2016}
{Jose}, J., {Kim}, J.~S., {Herczeg}, G.~J., {et~al.} 2016, \apj, 822, 49

\bibitem[{{Kerton} {et~al.}(2013){Kerton}, {Arvidsson}, \&
  {Alexander}}]{Kerton:2013}
{Kerton}, C.~R., {Arvidsson}, K., \& {Alexander}, M.~J. 2013, \aj, 145, 78

\bibitem[{{Kharchenko} {et~al.}(2005){Kharchenko}, {Piskunov}, {R{\"o}ser},
  {Schilbach}, \& {Scholz}}]{Kharchenko:2005}
{Kharchenko}, N.~V., {Piskunov}, A.~E., {R{\"o}ser}, S., {Schilbach}, E., \&
  {Scholz}, R.-D. 2005, \aap, 438, 1163

\bibitem[{{Kiminki} {et~al.}(2015){Kiminki}, {Kim}, {Bagley}, {Sherry}, \&
  {Rieke}}]{Kiminki:2015}
{Kiminki}, M.~M., {Kim}, J.~S., {Bagley}, M.~B., {Sherry}, W.~H., \& {Rieke},
  G.~H. 2015, \apj, 813, 42

\bibitem[{{Lagrois} \& {Joncas}(2009{\natexlab{a}})}]{Lagrois:2009}
{Lagrois}, D., \& {Joncas}, G. 2009{\natexlab{a}}, \apj, 691, 1109

\bibitem[{{Lagrois} \& {Joncas}(2009{\natexlab{b}})}]{Lagrois:2009b}
---. 2009{\natexlab{b}}, \apj, 693, 186

\bibitem[{{Lagrois} {et~al.}(2012){Lagrois}, {Joncas}, \&
  {Drissen}}]{Lagrois:2012}
{Lagrois}, D., {Joncas}, G., \& {Drissen}, L. 2012, \mnras, 420, 2280

\bibitem[{{Landecker} {et~al.}(2010){Landecker}, {Reich}, {Reid}, {Reich},
  {Wolleben}, {Kothes}, {Uyan{\i}ker}, {Gray}, {Del Rizzo}, {F{\"u}rst},
  {Taylor}, \& {Wielebinski}}]{Landecker:2010}
{Landecker}, T.~L., {Reich}, W., {Reid}, R.~I., {et~al.} 2010, \aap, 520, A80

\bibitem[{{Lestrade} {et~al.}(1999){Lestrade}, {Preston}, {Jones}, {Phillips},
  {Rogers}, {Titus}, {Rioja}, \& {Gabuzda}}]{Lestrade:1999}
{Lestrade}, J.-F., {Preston}, R.~A., {Jones}, D.~L., {et~al.} 1999, \aap, 344,
  1014

\bibitem[{{Macquart} {et~al.}(2012){Macquart}, {Ekers}, {Feain}, \&
  {Johnston-Hollitt}}]{Macquart:2012}
{Macquart}, J.-P., {Ekers}, R.~D., {Feain}, I., \& {Johnston-Hollitt}, M. 2012,
  \apj, 750, 139

\bibitem[{{Mao} {et~al.}(2010){Mao}, {Gaensler}, {Haverkorn}, {Zweibel},
  {Madsen}, {McClure-Griffiths}, {Shukurov}, \& {Kronberg}}]{Mao:2010}
{Mao}, S.~A., {Gaensler}, B.~M., {Haverkorn}, M., {et~al.} 2010, \apj, 714,
  1170

\bibitem[{{Massey} {et~al.}(1995){Massey}, {Johnson}, \&
  {Degioia-Eastwood}}]{Massey:1995}
{Massey}, P., {Johnson}, K.~E., \& {Degioia-Eastwood}, K. 1995, \apj, 454, 151

\bibitem[{{Massi}(2004)}]{Massi:2004b}
{Massi}, M. 2004, \aap, 422, 267

\bibitem[{{Massi} {et~al.}(2017){Massi}, {Migliari}, \&
  {Chernyakova}}]{Massi:2017}
{Massi}, M., {Migliari}, S., \& {Chernyakova}, M. 2017, \mnras, 468, 3689

\bibitem[{{Massi} {et~al.}(2004){Massi}, {Rib{\'o}}, {Paredes}, {Garrington},
  {Peracaula}, \& {Mart{\'{\i}}}}]{Massi:2004}
{Massi}, M., {Rib{\'o}}, M., {Paredes}, J.~M., {et~al.} 2004, \aap, 414, L1

\bibitem[{{Moore} {et~al.}(2007){Moore}, {Bretherton}, {Fujiyoshi}, {Ridge},
  {Allsopp}, {Hoare}, {Lumsden}, \& {Richer}}]{Moore:2007}
{Moore}, T.~J.~T., {Bretherton}, D.~E., {Fujiyoshi}, T., {et~al.} 2007, \mnras,
  379, 663

\bibitem[{{Nakano} {et~al.}(2017){Nakano}, {Soejima}, {Chibueze}, {Nagayama},
  {Omodaka}, {Handa}, {Sunada}, {Kamezaki}, \& {Burns}}]{Nakano:2017}
{Nakano}, M., {Soejima}, T., {Chibueze}, J.~O., {et~al.} 2017, \pasj, 69, 16

\bibitem[{{Navarete} {et~al.}(2011){Navarete}, {Figueredo}, {Damineli},
  {Mois{\'e}s}, {Blum}, \& {Conti}}]{Navarete:2011}
{Navarete}, F., {Figueredo}, E., {Damineli}, A., {et~al.} 2011, \aj, 142, 67

\bibitem[{{Normandeau} {et~al.}(1996){Normandeau}, {Taylor}, \&
  {Dewdney}}]{Normandeau:1996}
{Normandeau}, M., {Taylor}, A.~R., \& {Dewdney}, P.~E. 1996, \nat, 380, 687

\bibitem[{{Normandeau} {et~al.}(1997){Normandeau}, {Taylor}, \&
  {Dewdney}}]{Normandeau:1997}
---. 1997, \apjs, 108, 279

\bibitem[{{Oey} {et~al.}(2005){Oey}, {Watson}, {Kern}, \& {Walth}}]{Oey:2005}
{Oey}, M.~S., {Watson}, A.~M., {Kern}, K., \& {Walth}, G.~L. 2005, \aj, 129,
  393

\bibitem[{{O'Sullivan} {et~al.}(2012){O'Sullivan}, {Brown}, {Robishaw},
  {Schnitzeler}, {McClure-Griffiths}, {Feain}, {Taylor}, {Gaensler},
  {Landecker}, {Harvey-Smith}, \& {Carretti}}]{OSullivan:2012}
{O'Sullivan}, S.~P., {Brown}, S., {Robishaw}, T., {et~al.} 2012, \mnras, 421,
  3300

\bibitem[{{Paredes} {et~al.}(2007){Paredes}, {Rib{\'o}}, {Bosch-Ramon}, {West},
  {Butt}, {Torres}, \& {Mart{\'{\i}}}}]{Paredes:2007}
{Paredes}, J.~M., {Rib{\'o}}, M., {Bosch-Ramon}, V., {et~al.} 2007, \apjl, 664,
  L39

\bibitem[{{Pellegrini} {et~al.}(2007){Pellegrini}, {Baldwin}, {Brogan},
  {Hanson}, {Abel}, {Ferland}, {Nemala}, {Shaw}, \&
  {Troland}}]{Pellegrini:2007}
{Pellegrini}, E.~W., {Baldwin}, J.~A., {Brogan}, C.~L., {et~al.} 2007, \apj,
  658, 1119

\bibitem[{{Purcell} {et~al.}(2015){Purcell}, {Gaensler}, {Sun}, {Carretti},
  {Bernardi}, {Haverkorn}, {Kesteven}, {Poppi}, {Schnitzeler}, \&
  {Staveley-Smith}}]{Purcell:2015}
{Purcell}, C.~R., {Gaensler}, B.~M., {Sun}, X.~H., {et~al.} 2015, \apj, 804, 22

\bibitem[{{Reynolds} {et~al.}(2001){Reynolds}, {Sterling}, \&
  {Haffner}}]{Reynolds:2001}
{Reynolds}, R.~J., {Sterling}, N.~C., \& {Haffner}, L.~M. 2001, \apjl, 558,
  L101

\bibitem[{{Roberts} {et~al.}(1993){Roberts}, {Crutcher}, {Troland}, \&
  {Goss}}]{Roberts:1993}
{Roberts}, D.~A., {Crutcher}, R.~M., {Troland}, T.~H., \& {Goss}, W.~M. 1993,
  \apj, 412, 675

\bibitem[{{Rom{\'a}n-Z{\'u}{\~n}iga} \& {Lada}(2008)}]{Roman:2008}
{Rom{\'a}n-Z{\'u}{\~n}iga}, C.~G., \& {Lada}, E.~A. 2008, {Star Formation in
  the Rosette Complex}, ed. B.~{Reipurth}, 928

\bibitem[{{Rom{\'a}n-Z{\'u}{\~n}iga} {et~al.}(2015){Rom{\'a}n-Z{\'u}{\~n}iga},
  {Ybarra}, {Meg{\'{\i}}as}, {Tapia}, {Lada}, \& {Alves}}]{Roman:2015}
{Rom{\'a}n-Z{\'u}{\~n}iga}, C.~G., {Ybarra}, J.~E., {Meg{\'{\i}}as}, G.~D.,
  {et~al.} 2015, \aj, 150, 80

\bibitem[{{Roshi}(2007)}]{Roshi:2007}
{Roshi}, D.~A. 2007, \apjl, 658, L41

\bibitem[{{Sato}(1990)}]{Sato:1990}
{Sato}, F. 1990, \aj, 99, 1897

\bibitem[{{Savage} {et~al.}(2013){Savage}, {Spangler}, \&
  {Fischer}}]{Savage:2013}
{Savage}, A.~H., {Spangler}, S.~R., \& {Fischer}, P.~D. 2013, \apj, 765, 42

\bibitem[{{Shi} \& {Hu}(1999)}]{Shi:1999}
{Shi}, H.~M., \& {Hu}, J.~Y. 1999, \aaps, 136, 313

\bibitem[{{Sokoloff} {et~al.}(1998){Sokoloff}, {Bykov}, {Shukurov},
  {Berkhuijsen}, {Beck}, \& {Poezd}}]{Sokoloff:1998}
{Sokoloff}, D.~D., {Bykov}, A.~A., {Shukurov}, A., {et~al.} 1998, \mnras, 299,
  189

\bibitem[{{Stil} {et~al.}(2009){Stil}, {Wityk}, {Ouyed}, \&
  {Taylor}}]{Stil:2009}
{Stil}, J., {Wityk}, N., {Ouyed}, R., \& {Taylor}, A.~R. 2009, \apj, 701, 330

\bibitem[{{Sun} {et~al.}(2007){Sun}, {Han}, {Reich}, {Reich}, {Shi},
  {Wielebinski}, \& {F{\"u}rst}}]{Sun:2007}
{Sun}, X.~H., {Han}, J.~L., {Reich}, W., {et~al.} 2007, \aap, 463, 993

\bibitem[{{Sun} {et~al.}(2008){Sun}, {Reich}, {Waelkens}, \&
  {En{\ss}lin}}]{Sun:2008}
{Sun}, X.~H., {Reich}, W., {Waelkens}, A., \& {En{\ss}lin}, T.~A. 2008, \aap,
  477, 573

\bibitem[{{Sun} {et~al.}(2015){Sun}, {Landecker}, {Gaensler}, {Carretti},
  {Reich}, {Leahy}, {McClure-Griffiths}, {Crocker}, {Wolleben}, {Haverkorn},
  {Douglas}, \& {Gray}}]{Sun:2015}
{Sun}, X.~H., {Landecker}, T.~L., {Gaensler}, B.~M., {et~al.} 2015, \apj, 811,
  40

\bibitem[{{Taylor} {et~al.}(1999){Taylor}, {Irwin}, {Matthews}, \&
  {Heyer}}]{Taylor:1999}
{Taylor}, A.~R., {Irwin}, J.~A., {Matthews}, H.~E., \& {Heyer}, M.~H. 1999,
  \apj, 513, 339

\bibitem[{{Taylor} {et~al.}(2009){Taylor}, {Stil}, \& {Sunstrum}}]{Taylor:2009}
{Taylor}, A.~R., {Stil}, J.~M., \& {Sunstrum}, C. 2009, \apj, 702, 1230

\bibitem[{{Taylor} {et~al.}(2003){Taylor}, {Gibson}, {Peracaula}, {Martin},
  {Landecker}, {Brunt}, {Dewdney}, {Dougherty}, {Gray}, {Higgs}, {Kerton},
  {Knee}, {Kothes}, {Purton}, {Uyaniker}, {Wallace}, {Willis}, \&
  {Durand}}]{Taylor:2003}
{Taylor}, A.~R., {Gibson}, S.~J., {Peracaula}, M., {et~al.} 2003, \aj, 125,
  3145

\bibitem[{{Terebey} {et~al.}(2003){Terebey}, {Fich}, {Taylor}, {Cao}, \&
  {Hancock}}]{Terebey:2003}
{Terebey}, S., {Fich}, M., {Taylor}, R., {Cao}, Y., \& {Hancock}, T. 2003,
  \apj, 590, 906

\bibitem[{{Tielens} \& {Hollenbach}(1985)}]{Tielens:1985}
{Tielens}, A.~G.~G.~M., \& {Hollenbach}, D. 1985, \apj, 291, 722

\bibitem[{{Tomisaka}(1990)}]{Tomisaka:1990}
{Tomisaka}, K. 1990, \apjl, 361, L5

\bibitem[{{Tomisaka}(1998)}]{Tomisaka:1998}
---. 1998, \mnras, 298, 797

\bibitem[{{Townsley} {et~al.}(2014){Townsley}, {Broos}, {Garmire}, {Bouwman},
  {Povich}, {Feigelson}, {Getman}, \& {Kuhn}}]{Townsley:2014}
{Townsley}, L.~K., {Broos}, P.~S., {Garmire}, G.~P., {et~al.} 2014, \apjs, 213,
  1

\bibitem[{{Troland} {et~al.}(2016){Troland}, {Goss}, {Brogan}, {Crutcher}, \&
  {Roberts}}]{Troland:2016}
{Troland}, T.~H., {Goss}, W.~M., {Brogan}, C.~L., {Crutcher}, R.~M., \&
  {Roberts}, D.~A. 2016, \apj, 825, 2

\bibitem[{{Vall{\'e}e}(1993)}]{Vallee:1993}
{Vall{\'e}e}, J.~P. 1993, \apj, 419, 670

\bibitem[{Van Der~Walt {et~al.}(2011)Van Der~Walt, Colbert, \&
  Varoquaux}]{van2011numpy}
Van Der~Walt, S., Colbert, S.~C., \& Varoquaux, G. 2011, Computing in Science
  \& Engineering, 13, 22

\bibitem[{{van der Werf} \& {Goss}(1990)}]{vanderWerf:1990}
{van der Werf}, P.~P., \& {Goss}, W.~M. 1990, \aap, 238, 296

\bibitem[{{Van Eck} {et~al.}(2011){Van Eck}, {Brown}, {Stil}, {Rae}, {Mao},
  {Gaensler}, {Shukurov}, {Taylor}, {Haverkorn}, {Kronberg}, \&
  {McClure-Griffiths}}]{vanEck:2011}
{Van Eck}, C.~L., {Brown}, J.~C., {Stil}, J.~M., {et~al.} 2011, \apj, 728, 97

\bibitem[{{Watson} {et~al.}(2008){Watson}, {Povich}, {Churchwell}, {Babler},
  {Chunev}, {Hoare}, {Indebetouw}, {Meade}, {Robitaille}, \&
  {Whitney}}]{Watson:2008}
{Watson}, C., {Povich}, M.~S., {Churchwell}, E.~B., {et~al.} 2008, \apj, 681,
  1341

\bibitem[{{Weaver} {et~al.}(1977){Weaver}, {McCray}, {Castor}, {Shapiro}, \&
  {Moore}}]{Weaver:1977}
{Weaver}, R., {McCray}, R., {Castor}, J., {Shapiro}, P., \& {Moore}, R. 1977,
  \apj, 218, 377

\bibitem[{{West} {et~al.}(2007){West}, {English}, {Normandeau}, \&
  {Landecker}}]{West:2007}
{West}, J.~L., {English}, J., {Normandeau}, M., \& {Landecker}, T.~L. 2007,
  \apj, 656, 914

\bibitem[{{Whiting} {et~al.}(2009){Whiting}, {Spangler}, {Ingleby}, \&
  {Haffner}}]{Whiting:2009}
{Whiting}, C.~A., {Spangler}, S.~R., {Ingleby}, L.~D., \& {Haffner}, L.~M.
  2009, \apj, 694, 1452

\bibitem[{{Wright} {et~al.}(2010){Wright}, {Eisenhardt}, {Mainzer}, {Ressler},
  {Cutri}, {Jarrett}, {Kirkpatrick}, {Padgett}, {McMillan}, {Skrutskie},
  {Stanford}, {Cohen}, {Walker}, {Mather}, {Leisawitz}, {Gautier}, {McLean},
  {Benford}, {Lonsdale}, {Blain}, {Mendez}, {Irace}, {Duval}, {Liu}, {Royer},
  {Heinrichsen}, {Howard}, {Shannon}, {Kendall}, {Walsh}, {Larsen}, {Cardon},
  {Schick}, {Schwalm}, {Abid}, {Fabinsky}, {Naes}, \&
  {Tsai}}]{2010AJ....140.1868W}
{Wright}, E.~L., {Eisenhardt}, P.~R.~M., {Mainzer}, A.~K., {et~al.} 2010, \aj,
  140, 1868

\end{thebibliography}
